\documentclass[11pt,english]{article}
\usepackage{graphicx}
\usepackage{xcolor}
\usepackage{amstext}
\usepackage{caption}
\usepackage{etoolbox}
\usepackage{makeidx}
\usepackage{amsfonts}
\usepackage{authblk} 
\usepackage{sectsty}
\usepackage{amsmath,amssymb,epsfig}
\usepackage{amscd}
\usepackage{amsthm}
\usepackage{mathrsfs}
\usepackage{dsfont}
\usepackage[applemac]{inputenc}
\usepackage[english]{babel}
\usepackage{enumitem} 
\usepackage[]{latexsym}
\usepackage{caption}
\usepackage{subfigure}
\usepackage{hyperref}
\usepackage{blindtext}
\usepackage{verbatim}
\usepackage{cancel}
\usepackage{cases}
\usepackage{cite}%for collecting multiple references

\hypersetup{
	pdftitle={},%
	pdfauthor={},%
	pdfsubject={},%
	pdfkeywords={},%
	colorlinks=true,%
	linkcolor=blue,%
	citecolor=crimson,%
	linktocpage=true,%
	hyperfootnotes=true,%
	pageanchor=true
}

\definecolor{crimson}{rgb}{0.7, 0.08, 0.24}

\newcommand*{\affaddr}[1]{#1}
\newcommand*{\affmark}[1][*]{\textsuperscript{#1}}

\newtheorem*{proof*}{Proof}

\newcommand{\be}{\begin{equation}}

\newcommand{\ee}{\end{equation}}
\def\beqa{\begin{eqnarray}}
\def\eeqa{\end{eqnarray}}
\def\bean{\begin{eqnarray*}}
\def\eean{\end{eqnarray*}}

\newcommand{\dd}{\mathrm{d}}

\newcommand{\der}[2]{\frac{\dd #1}{\dd #2}}
\newcommand{\dder}[2]{\frac{\dd^2 #1}{\dd #2^2}}

\renewenvironment{thebibliography}[1]
         {\section*{References}\frenchspacing\small
          \begin{list}{[\arabic{enumi}]}
         {\usecounter{enumi}\parsep=2pt\topsep 0pt
         \settowidth{\labelwidth}{[#1]}
         \leftmargin=\labelwidth\advance\leftmargin\labelsep
         \rightmargin=0pt\itemsep=1pt\sloppy}}{\end{list}}

 \numberwithin{equation}{section}

\usepackage[normalem]{ulem}

\title{\textbf{\textsf{Effective Quantum Dust Collapse via Surface Matching}}\vspace{0.35cm}}

\author{
\textsf{Johannes M\"unch\affmark[1,2]\footnote{\texttt{johannes.muench@cpt.univ-mrs.fr}}}\\
\affaddr{\affmark[1]\textsf{Aix-Marseille Université, Université de Toulon, CNRS, CPT,}}\\
\affaddr{\textsf{13288 Marseille, France}}\\
\affaddr{\affmark[2]\textsf{Institute for Theoretical Physics, University of Regensburg,}}\\
\affaddr{\textsf{93040 Regensburg, Germany}}\vspace{-0.5cm}
}

\begin{document}

\maketitle

\begin{abstract}
\textsf{The fate of matter forming a black hole is still an open problem, although models of quantum gravity corrected black holes are available. In loop quantum gravity (LQG) models were presented, which resolve the classical singularity in the centre of the black hole by means of a black-to-white hole transition, but neglect the collapse process. The situation is similar in other quantum gravity approaches, where eternal non-singular models are available. In this paper, a strategy is presented to generalise these eternal models to dynamical collapse models by surface matching. Assuming 1) the validity of a static quantum black hole spacetime outside the collapsing matter, 2) homogeneity of the collapsing matter, and 3) differentiability at the surface of the matter fixes the dynamics of the spacetime uniquely. It is argued that these assumptions resemble a collapse of pressure-less dust and thus generalises the Oppenheimer-Snyder-Datt model, although no precise model of the matter has to be assumed. Hawking radiation is systematically neglected in this approach. The junction conditions and the spacetime dynamics are discussed generically for bouncing black hole spacetimes, as proposed by LQG, although the scheme is approach independent. Further, the equations are explicitly solved for the recent model \cite{BodendorferEffectiveQuantumExtended} and a global spacetime picture of the collapse is achieved. The causal structure is discussed in detail and the Penrose diagram is constructed. The trajectory of the collapsing matter is completely constructed from an inside and outside observer point of view. The general analysis shows that the matter is collapsing and re-expanding and crosses the Penrose diagram diagonally. This way the infinite tower of Penrose diagrams, as proposed by several LQG models, is generically not cut out. Questions about different timescales of the collapse for in- and outside observers can be answered.}
\end{abstract}

\section{Introduction}

The collapse of matter can under certain circumstances and according to general relativity (GR) lead to the formation of a black hole and in consequence to a spacetime singularity \cite{PenroseGravitationalCollapseAnd,HawkingPropertiesOfExpanding}, where the matter density and spacetime curvature diverge.
This collapse of extremely dense matter is exactly the scenario of how astrophysical black holes are formed and it remains the question what happens to the matter in the late stages of the collapse, where the GR description fails.
It is commonly believed that only a full (non-perturbative) description of quantum gravity can answer these questions about the fate of matter and spacetime at the Planck scale and at the end of the collapse (see e.g. \cite{BojowaldSingularitiesAndQuantum, NatsuumeTheSingularityProblem}).
As a full theory of quantum gravity is not available, these are still open questions.
In the case of cosmology, symmetry reduced models of loop quantum gravity (LQG), so called loop quantum cosmology (LQC), were able to describe spacetime in the Planck regime \cite{AshtekarQuantumNatureOf,AshtekarMathematicalStructureOf,AshtekarLoopQuantumGravity30Years,DaporCosmologicalEffectiveHamiltonian} (nevertheless note recent criticism \cite{BojowaldCriticalEvaluationof}).
Similar efforts have been taken within the framework of spherically symmetric and static black holes, which are still simple due to a high amount of symmetries.
The field of effective quantum descriptions of black holes in LQG is still active \cite{VakiliClassicalpolymerizationof,CorichiLoopquantizationof,ModestoSemiclassicalLoopQuantum,ModestoSelf-dualBlackHoles,BenAchourPolymerSchwarzschildBlack,BojowaldEffectivelineelements,BoehmerLoopquantumdynamics,AshtekarQuantumExtensionof,AshtekarQuantumTransfigurarationof,BodendorferEffectiveQuantumExtended,BodendorferMassandHorizon,Bodendorferbvtypevariables,AssanioussiPerspectivesonthe,AshtekarPropertiesofarecent,Bouhmadi-LopezAsymptoticnon-flatness,KellyEffectiveloopquantumgravity,KellyBlackholecollapse,GambiniSphericallysymmetricloop,GeillerSymmetriesofthe,SartiniQuantumdynamicsof}.
Besides LQG, also in other approaches of quantum gravity, non-singular solutions of black holes were studied as particular examples show, see e.g. \cite{NicoliniQuantumCorrectedBlackHoles,EassonTheclassicaldouble} in the context of string theory, \cite{AdeifeobaTowardsconditionsfor,PlataniaDynamicalrenormalizationof,MotiOnthequantum} as efforts within asymptotically safe gravity, \cite{NicoliniNoncommutativeBlackHoles,NicoliniRemarksonregular,SmailagicKerrrblackhole} within non-commutative geometry and \cite{BardeenNon-singulargeneral,HaywardFormationandevaporation,DymnikovaVacuumnonsingularblackhole,DymnikovaDeSitter-Schwarzschild,FrolovNotesonnon-singular,FrolovRemarksonnon-singularblackholes} for general regular black hole solutions.

Nevertheless, all the efforts within LQG discuss only spherically symmetric and static black holes, which do not cover the process of the matter collapse.
In fact due to the bouncing character of the LQG inspired models, where the classical singularity is replaced by a transition from a black hole to a white hole interior and a minimal radius is reached, it is not clear how the collapsing matter fits into this picture.
It is simply not included in the analysis, but highly necessary to understand the fate of the matter and to gain a more realistic and complete picture of black holes.  
Another feature of the LQG models is a causal structure, which is an infinite tower of black and white hole regions all connected by a spacelike and regular transition surface.
Most likely, this is not the situation realised in nature and the unphysical regions are cut out by including collapse and Hawking radiation effects, as described in e.g. \cite{AshtekarBlackHoleEvaporation,HaggardBlackHoleFireworks,BianchiWhiteHolesasRemnants,Martin-DussaudEvaporationgblackto}.
Indeed, due to the physical relevance of the black hole collapse and formation, there exist a large literature on quantum models in different approaches as \cite{KieferSingularityavoidancefor,SchmitzTowardsaquantum,PiechockiQuantumOppenheimerSnyder} in Wheeler-de Witt quantum gravity, \cite{ModestoGravitationalcollapsein,HusainCriticalbehaviourin,HossenfelderAmodelfor,BambiNon-singularquantum,BojowaldBlackHoleMass,HaggardBlackHoleFireworks,BarceloThelifetimeproblem,BianchiWhiteHolesasRemnants,Martin-DussaudEvaporationgblackto,ChakrabartyAtoymodelfor} in LQG and other quantum gravity approaches \cite{CasadioHamiltonianformalismfor,BaccettiBlackholeevaporation,HossenfelderConservativesolutionsto,MannaGravitationalCollapsefor,HaywardFormationandevaporation,BonannoRegularblackholes}, see \cite{MalafarinaClassicalcollapseto} for a recent review.
Nevertheless, the situation is still not clear.
Often many assumptions about the character of Hawking radiation needed or no quantum corrected spacetime in the vacuum region is available.
Therefore the full spacetime picture remains unclear as in e.g. \cite{BambiNon-singularquantum,BojowaldBlackHoleMass}.

Already classically is the description of a black hole collapse complicated, depending on the content of matter.
Nevertheless, there exist analytical models \cite{AdlerSimpleAnalyticModels,ChatterjeeMarginallyTrappedSurfaces} and the simplest case of spherically symmetric and homogeneous pressure-less dust is known as Oppenheimer-Snyder-Datt collapse \cite{OppenheimerOnContinuedGravitational,DattOnaclass}.
The strategy is to start with the full time dependent problem, while it turns out that the spacetime dynamics of the matter filled region decouples from the vacuum region.
They influence each other due to the gluing of the boundary surface of the matter region according to the Israel-Darmois junction conditions \cite{IsraelSingularhypersurfacesand,DarmoisLeseuqationsde} and thus fixes the relevant boundary conditions for each region.
The junction conditions ensure the continuity of spacetime and include the stress-energy carried by the surface of the matter.
This way a complete spacetime picture can be obtained.
Assuming that any generalised effective Einstein equations, in any quantum gravity approach, were derived for the dynamical case, the situation would probably be the same: 
These equations have to be solved in the matter filled and vacuum region separately, and afterwards both regions have to be glued according to some generalised junction conditions.
Indeed, this scenario was analysed in recent efforts \cite{BenAchourBouncingcompactobjectsI,BenAchourBouncingcompactobjectsII,BenAchourConsistentblackto}.

Nevertheless, these generalised effective equations do not exist in LQG, though there is recent effort towards this \cite{KellyEffectiveloopquantumgravity,KellyBlackholecollapse,GambiniSphericallysymmetricloop}.
The strategy underlying LQC and the black hole models \cite{CorichiLoopquantizationof,ModestoSemiclassicalLoopQuantum,AshtekarQuantumExtensionof,AshtekarQuantumTransfigurarationof,BodendorferEffectiveQuantumExtended,BodendorferMassandHorizon,Bodendorferbvtypevariables}, so-called polymerisation, heavily uses the high amount of symmetry, i.e. spherical symmetry and homogeneity (staticity for outside the black hole horizon).
This loosening the symmetry restrictions turns the homogeneous black hole interior model, which is effectively 1+0-dimensional, into a 1+1 dimensional field theory.
This additional dependence on space and time makes it hard to identify holonomies, which are the core ingredients of polymerisation.
Additional difficulties come from the restriction of the model to remain diffeomorphism invariant, see \cite{BojowaldCovarianceinmodels,BenAchourPolymerSchwarzschildBlack,BojowaldCriticalEvaluationof,BojowaldAno-goresult,BojowaldBlack-holemodels,ArrugaDeformedGeneralRelatiity}.
Therefore no generalised effective Einstein equations and consequently no derived junction conditions are available.

Due to this, the present paper focusses on reasonable technical junction conditions and interprets their physical meaning later.
This does restrict the physical applicability, but is also not based on some generalised junction conditions and consequently approach independent.
This work builds on recent progress of static black holes models, which provide the quantum corrected spacetime in- and outside the horizon \cite{ModestoSemiclassicalLoopQuantum,AshtekarQuantumTransfigurarationof,AshtekarQuantumExtensionof,BodendorferEffectiveQuantumExtended,BodendorferMassandHorizon,Bodendorferbvtypevariables,KellyEffectiveloopquantumgravity,KellyBlackholecollapse,GambiniSphericallysymmetricloop}.
It will be assumed that a generalised version of Birkhoff's theorem holds, i.e. that vacuum and matter region decouple and that the spacetime remains always static in the vacuum region, allows to use these effective eternal black hole models as vacuum spacetime.
In the same way, any other regular black hole solution could be used.
This will be the first central assumption of the present paper.
Restricting further to a simple case of homogeneous pressure-less dust leads to two further assumptions:
As the matter is then homogeneous and spherically symmetric, the matter spacetime has to have the form of a Friedmann-Lemaître-Robertson-Walker (FLRW) metric, which is the second assumption of the present model.
Finally, as will be discussed in the main text, the assumption of pressure-less dust can be translated into the condition that the spacetime should be at least once continuously differentiable along the matter boundary surface.
This is the third and last assumption.
As it will be shown these assumptions allow to determine the full spacetime dynamics if the exterior metric is known.
The three technical assumptions will be formulated and discussed in the main text.
They are applied to the classical scenario to motivate their physical interpretation as a model of pressure-less dust collapse.
Finally, it will be applied to first generic LQG-inspired black hole modes and generic properties of the causal structure and a solution strategy of the equations is developed.
These generic qualitative features are made explicit by solving the junction conditions for the concrete model \cite{BodendorferEffectiveQuantumExtended}.
This way a full spacetime picture of the collapse can be discussed.
As the vacuum region is assumed to be known this is a contemporary approach to the recent work \cite{BenAchourBouncingcompactobjectsI,BenAchourBouncingcompactobjectsII,BenAchourConsistentblackto}.

The strategy discussed in this paper leads to a minimal model of a quantum collapse.
Following the three central assumptions, there will not be a precise model for the matter region and no concrete stress-energy tensor will be available.
Furthermore, the analysis is fully on the effective level and no statement about the true quantum nature of the collapse in the Planck regime can be made.
Nevertheless, this strategy allows discussing a brought range of black hole models, a priori not restricted to the LQG approach.
Hawking radiation and black hole evaporation are systematically neglected to avoid additional assumptions and to increase complexity step-by-step.
This way all assumptions are under control and a clear full spacetime picture might be obtained.
Although the picture remains minimalistic, important questions about the fate of matter, the causal structure, and different time scales can be addressed.
Having neglected Hawking radiation, the only relevant time scale is the collapse time and it can be discussed if two observers can meet again if one rests outside the black hole horizon and the other crosses it and comparisons to \cite{KieferSingularityavoidancefor,HaggardBlackHoleFireworks} will be addressed.
This simple picture will also already allow comparisons with the LQG proposals \cite{AshtekarBlackHoleEvaporation,Martin-DussaudEvaporationgblackto}.

The paper is organised as follows:
In Sec.~\ref{sec:SurfMatch} the general strategy of surface matching in this context and the underlying assumptions are discussed.
The relevant general equations are derived and it is observed that these matching conditions are sufficient to determine the full spacetime dynamics once a vacuum spacetime is known.
In the following Sec.~\ref{sec:classical}, the classical gravitational collapse is reviewed and the technical assumptions of Sec. \ref{sec:SurfMatch} are interpreted as the scenario of a pressure-less homogeneous dust collapse.
The derived junction conditions are applied to the case of bouncing black hole models as they are inspired by LQG, without specifying a particular model in Sec.~\ref{sec:application}.
This allows to derive generic features of the gravitational collapse of LQG black hole models.
In the following Subsecs.~\ref{sec:modelapplication} and \ref{sec:PenroseDiag} the junction conditions are solved explicitly for the recent model \cite{BodendorferEffectiveQuantumExtended} and the resulting causal structure is analysed in terms of a Penrose diagram.
The results of this approach is compared to finding of other work in Sec.~\ref{sec:comparison}.
The paper closes with a general discussion and outlook in Sec.~\ref{sec:conclusions}. 

\section{Surface Matching}\label{sec:SurfMatch}

The proper approach to the spherically symmetric collapse problem, including the full complexity, would be to set up and solve effective quantum corrected equations of a spherically symmetric but time dependent metric defined on a manifold $M$, i.e.

\begin{equation}\label{eq:ansatz1}
\dd s^2 = -a(t,r) \dd t^2 + 2 B(t,r)^2 \dd r \dd t + N(t,r) \dd r^2 + b(t,r)^2 \dd \Omega_2^2 \;,
\end{equation}

\noindent
coupled to a matter source constrained to a compact spherical region in space.
Here $\dd\Omega_2^2$ is the line element of the round 2-sphere $\mathbb{S}^2$.
Solving the full problem requires a separate analysis of the matter region (non-vanishing matter stress-energy tensor) and the vacuum region (vanishing stress-energy tensor).
Afterwards an analysis of the boundary conditions of the dynamical surface of the matter region is needed to match it to the vacuum region in a physically meaningful sense.
This timelike surface of the collapsing matter, where boundary and matching conditions are analysed, is denoted $\Sigma$ from now on. Further the surface of matter is also referred to as \textit{matching surface} or \textit{junction surface}, as there the matching or junction conditions are applied.
Without specifying the model applied for determining the spacetime in the matter and vacuum region, we focus only on the matching conditions between them and do the following assumptions:

\begin{enumerate}[label=\textnormal{(\arabic*)}]
	\item A generalised version of Birkhoff theorem holds \cite{BirkhoffRelativityandModernPhysics,JebsenOnthegerneral}. We assume that in the case of a spherically symmetric collapse the metric remains stationary in the vacuum region, i.e. has timelike Killing vector field (outside possible horizons). This allows us to choose as vacuum metric 
	\begin{equation}\label{eq:metricext}
	\dd s^2 = -a(b) \dd t^2 + N(b) \dd b^2 + b^2 \dd \Omega_2^2 \;,
	\end{equation}

	\noindent
	in the gauge $B = 0$ and $b(r) = r$. The coordinate $b$ is thus identical with the areal radius. This vacuum region metric is valid up to the surface of the collapsing matter, which is parametrised by $b = R(t)$.
	The matter surface $\Sigma$ can in these coordinates be written as
	
	$$
	\Sigma := \left\{\left(t , R(t)\right) \; \middle| \; t \in \mathbb{R} \right\} \times \mathbb{S}^2\;.
	$$\label{assm:1}
	
	\item We restrict to a homogeneous collapse, i.e. the matter density does not depend on the radial direction. Due to symmetry restrictions the metric in the matter region is spherically symmetric and homogeneous, hence has the form\footnote{Note that, for the following discussion is suffices to demand this form of the metric only locally around the matching surface. For simplicity and possible problems in interpreting the situation physically, we assume the stronger condition that the metric has this form in the whole interior.}
	\begin{equation}\label{eq:metricint}
	\dd s^2 = - \dd \tau ^2 + \frac{S(\tau)^2}{1-k \rho} \dd \rho^2  + S(\tau)^2 \rho^2 \dd \Omega_2^2 \;.
	\end{equation}

	\noindent
	The matter region metric is valid up to the surface of the matter, parametrised by $\rho = \rho_o(\tau)$, i.e. in these coordinates it is
	
	$$
	\Sigma := \left\{\left(\tau, \rho_o(\tau)\right)\;\middle|\; \tau \in \mathbb R\right\} \times \mathbb{S}^2 \;.
	$$ 
	\label{assm:2}
	
	\item The matching between matter and vacuum region is $C^1(M)$, i.e. the first and second fundamental forms coincide \cite{FayosMatchingofthe,FayosInteriorsofVaidya,FayosGeneralmatchingof}\footnote{The superscript $i$ denotes the region inside the matter distribution, while $e$ describes the exterior region, outside the matter distribution. To avoid confusion with in-/outside the horizon or black hole interior /exterior, we refer to the two regions as matter and vacuum regions.}:
	
	\begin{equation}\label{eq:mathcing}
	\left.q^i\right|_{\Sigma} \stackrel{!}{=} 	\left.q^e\right|_{\Sigma} \quad , \quad \left.K^i\right|_{\Sigma} \stackrel{!}{=} 	\left.K^e\right|_{\Sigma}\;.
	\end{equation}
	
	\noindent
	These conditions are a special case of the so-called known as Israel-Darmois junction conditions \cite{IsraelSingularhypersurfacesand,DarmoisLeseuqationsde}. 
	\label{assm:3}
\end{enumerate}

There are a few remarks in order: The Birkhoff theorem can be proven to be true for classical GR, but is a simplifying assumption for the quantum model. For the quantum model violations of classical GR are allowed at the Planck scale, i.e. in the vicinity of the classical singularity. 
These quantum effects might violate Birkhoff's theorem.
Nevertheless, these effects are neglected in the following. 
A mathematical proof or disproof of a generalised version of Birkhoff's theorem requires an effective quantum model of the full collapse problem, which is for the case of LQG not available so far\footnote{In alternative theories of gravity there are results in supporting the validity of Birkhoff's theorem \cite{NziokiJebsen-Birkhofftheorem} (for $f(R)$ gravity), \cite{LiGeneralizedBirkhoff} (generalisation in mimetic gravity), \cite{CavagliaTheBirkhoffTheorem} (Dilaton gravity), but there is also evidence against it \cite{KehmViolationofBirkhoff,DeveciogluOnBirkhoffsTheorem}.}.
The assumptions \ref{assm:2} and \ref{assm:3} are discussed in detail in the next section at the example of classical GR.
In agreement with the Israel-Darmois junction conditions, we will conclude that this model describes collapsing dust.
A concrete model for the interior in terms of a stress-energy tensor and energy density will not be available due to the fact that these are purely technical assumptions.

Assumptions \ref{assm:1} and \ref{assm:2} allow now to compute the first and second fundamental forms for the matter and vacuum region, respectively.
For the matter metric, this is 

\begin{equation}\label{eq:qint}
\left.q^i\right|_{\Sigma} = -\frac{1-k \rho_o^2 - S^2 \dot{\rho_o}^2}{1-k \rho_o^2} \dd \tau^2 + \rho_o^2 S^2 \dd\Omega_2^2 \;,
\end{equation}

\noindent
and 

\begin{equation}\label{eq:Kint}
\left.K^i\right|_{\Sigma} = -\frac{\dot{\rho_o}^2 S \left(k \rho_o  - S \dot{\rho_o} \dot{S}\right) +\left(1-k\rho_o^2\right) \left(S \ddot{\rho_o} +2 \dot{\rho_o} \dot{S}\right)}{\left(1-k\rho_o^2\right)\sqrt{1-k\rho_o^2 - S^2 \dot{\rho_o}^2}} \dd\tau^2 
+ \frac{\rho_o S \left(1-k \rho_o^2 + \rho_o \dot{\rho_o} S \dot{S}\right)}{\sqrt{1-k\rho_o^2 - S^2 \dot{\rho_o}^2}} \dd \Omega_2^2 \;,
\end{equation}

\noindent
where $\left.\rho\right|_{\Sigma} = \rho_o(\tau)$ parametrises the matter surface $\Sigma$ (cfr. assumption \ref{assm:2}) and dots denote derivatives with respect to $\tau$.
For the vacuum region, this is instead 

\begin{equation}\label{eq:qext}
\left. q^e\right|_{\Sigma} = \left(-a + N \der{R}{t}^2\right) \dd t^2 + R^2 \dd \Omega_2^2\;,
\end{equation}

\noindent
as well as 

\begin{equation}\label{eq:Kext}
\left.K^e\right|_\Sigma = \frac{2 N a' \der{R}{t}^2 - a a' -a N' \der{R}{t}^2-2aN\dder{R}{t}}{2 \sqrt{a N \left(a-N\der{R}{t}^2\right)}} \dd t^2 
+ \frac{a R}{\sqrt{aN\left(a-N\der{R}{t}^2\right)}} \dd \Omega_2^2\;,
\end{equation}

\noindent
where $a' = \left.\der{a(r)}{r}\right|_{r = R(t)}$ and analogous for $N'$.
All quantities are evaluated at the matter surface, i.e. $a = a\left(r = R(t)\right)$ and $N=N\left(r = R(t)\right)$ and therefore functions of time.
The matching conditions Eq. \eqref{eq:mathcing} lead to four equations, which are strong enough to determine the four unknown quantities $t(\tau)$, $R(t)$, $\rho_o(\tau)$ and $S(\tau)$, assuming the vacuum region metric as given.
Matching the angular part of Eqs. \eqref{eq:qint} and \eqref{eq:qext} yields the condition

\begin{equation}
R(t(\tau)) = \rho_o(\tau) S(\tau)\;,
\end{equation}

\noindent
while matching the temporal components leads to the equation

\begin{equation}
-a \dd t^2 = -N  \left(\rho_o S\right)\dot{}\;^2 \dd \tau^2 -\frac{1-k \rho_o^2 - S^2 \dot{\rho_o}^2}{1-k \rho_o^2} \dd \tau^2 \;.
\end{equation}

\noindent
This can be re-expressed as 

\begin{equation}\label{eq:times}
\dd t^2 = \frac{1-\frac{S^2 \dot{\rho_o}^2}{1-k\rho_o^2} + N  \left(\rho_o S\right)\dot{}\;^2}{a} \dd \tau^2\;,
\end{equation}

\noindent
which determines how the two time coordinates are related.
Integrating this equation yields the function $t(\tau)$.

From the matching of the angular components of the second fundamental forms Eqs. \eqref{eq:Kint} and \eqref{eq:Kext}, we get

\begin{equation}\label{eq:angularcompts1}
\frac{a R}{\sqrt{aN\left(a-N\dot{R}^2\right)}} = \frac{\rho_o S \left(1-k \rho_o^2 + \rho_o \dot{\rho_o} S \dot{S}\right)}{\sqrt{1-k\rho_o^2 - S^2 \dot{\rho_o}^2}}\;,
\end{equation}

\noindent
which after some algebra leads to 

\begin{equation}\label{eq:Kangularmatch}
0=\left(1-k\rho_o^2 - S^2 \dot{\rho_o}^2 \right) \left(1-k\rho_o^2 -\frac{1}{N} - \rho_o^2 \dot{S}^2 \right) \;.
\end{equation}

\noindent
This equation is obviously satisfied for either the first or the second bracket equal to zero. We assume that $1-k\rho_o^2 - S^2 \dot{\rho_o}^2 \neq 0$, as the opposite leads to degeneracy of the induced metric \eqref{eq:qint}.
This is a special case which corresponds to a matching surface, which is lightlike.
The physical situation hence would be the collapse of massless particles, which we do not discuss here, although this is interesting on its own. 
Demanding thus the second bracket of Eq. \eqref{eq:Kangularmatch} being zero leads to

\begin{equation}\label{eq:rhodS}
\rho_o^2 \dot{S}^2 = 1-k\rho_o^2 -\frac{1}{N} \;.
\end{equation}

\noindent
Using this condition, Eq. \eqref{eq:times} can be rewritten as 

\begin{equation}\label{eq:times2}
\dd t^2 = \frac{N}{(1-k\rho_o^2) a} \left(1-k\rho_o^2 + \rho_o \dot{\rho_o} S \dot{S}\right)^2 \dd \tau^2 \;.
\end{equation}

The last condition follows from matching the temporal components of the extrinsic curvature (Eqs. \eqref{eq:Kint} and \eqref{eq:Kext}), leading to 

\begin{equation}
\frac{2 N a' \der{R}{t}^2 - a a' -a N' \der{R}{t}^2-2aN\dder{R}{t}}{2 \sqrt{aN \left(a-N\der{R}{t}^2\right)}} \dd t^2 = -\frac{\dot{\rho_o}^2 S \left(k \rho_o  - S \dot{\rho_o} \dot{S}\right) +\left(1-k\rho_o^2\right) \left(S \ddot{\rho_o} +2 \dot{\rho_o} \dot{S}\right)}{\left(1-k\rho_o^2\right)\sqrt{1-k\rho_o^2 - S^2 \dot{\rho_o}^2}} \dd\tau^2
\end{equation}

\noindent
Using Eqs. \eqref{eq:times}, \eqref{eq:times2} and Eq. \eqref{eq:rhodS} allows to simplify this expression and gives after a long and tedious, but straight forward computation

\begin{equation}
0= \left( 1-k\rho_o^2 - \dot{\rho_o}^2 S^2 \right) \left( N\frac{\rho_o\dot{\rho_o} \dot{S}^2}{1-k\rho_o^2}\left(k\rho_o^2-\frac{1}{N}\right) + \frac{\rho_o \dot{S}}{2} \left(\frac{a'}{a} + \frac{N'}{N}\right) + \frac{a'}{2a} \frac{\rho_o^2 \dot{\rho_o} S \dot{S}^2}{1-k\rho_o^2} - N k \rho_o \dot{\rho_o} + \frac{N'}{2 N} \dot{\rho_o} S \right)\;.
\end{equation}

\noindent
As we assume the first bracket to be non-zero, vanishing of the second bracket leads to an equation for $\dot{\rho_o}$, i.e.

\begin{equation}
\frac{\dot{\rho_o}}{\rho_o} = \frac{(1-k\rho_o^2)\frac{\rho_o \dot{S}}{2} \left(\frac{a'}{a} + \frac{N'}{N}\right)}{1-\frac{1}{N} + \frac{a'}{2N a} R - \left(1-k\rho_o^2\right) \frac{R}{2}\left(\frac{N'}{N} + \frac{a'}{a}\right)} \;.
\end{equation}

Note that we did not assume any theory or underlying model for the matter region of spacetime and the collapsing matter.
Knowing instead the vacuum metric, i.e $a(r)$ and $N(r)$ determines completely the dynamics of the surface of the collapsing matter as well as the dynamics of the interior metric $S(\tau)$ via the four equations

\begin{subequations}\label{eq:matchingfinal}
	\begin{equation}\label{eq:RrhoS}
	R(t(\tau)) = \rho_o(\tau) S(\tau)\;,
	\end{equation}
	\begin{equation}\label{eq:timefinal}
	\dd t^2 = \frac{1-\frac{S^2 \dot{\rho_o}^2}{1-k\rho_o^2} + N  \left(\rho_o S\right)\dot{}\;^2}{a} \dd \tau^2 = \frac{N}{(1-k\rho_o^2) a} \left(1-k\rho_o^2 + \rho_o \dot{\rho_o} S \dot{S}\right)^2 \dd \tau^2 \;,
	\end{equation}
	\begin{equation}\label{eq:rhodS2}
	\rho_o^2 \dot{S}^2 = 1-k\rho_o^2 -\frac{1}{N} \;,
	\end{equation}
	\begin{equation}\label{eq:drho}
	\frac{\dot{\rho_o}}{\rho_o} = \frac{(1-k\rho_o^2)\frac{\rho_o \dot{S}}{2} \left(\frac{a'}{a} + \frac{N'}{N}\right)}{1-\frac{1}{N} + \frac{a'}{2N a} R - \left(1-k\rho_o^2\right) \frac{R}{2}\left(\frac{N'}{N} + \frac{a'}{a}\right)} \;,
	\end{equation}
\end{subequations}

\noindent
which represents ($a(r)$, $N(r)$ given) a complete set of differential equations.

Let us point out an important quantity in the context of spherically symmetric collapse models, namely the Misner-Sharp mass

$$
M_{MS} = \frac{b^3}{8} R_{\mu \nu \alpha \beta} \epsilon^{\mu \nu} \epsilon^{\alpha \beta}
$$ 

\noindent
where $b$ is the areal radius of the spherically symmetric system and $\epsilon^{\mu \nu} = g^{\mu \alpha} g^{\nu \beta} \epsilon_{\alpha \beta}$ with $\epsilon_{\alpha \beta} \dd x^\alpha \wedge \dd x^{\beta} = b^2 \sin\theta \dd \theta \wedge \dd \phi$ is the volume two-form of a fixed time and radius two-sphere. The Misner-Sharp mass (see \cite{SzabadosQuasi-LocalEnergy} and references therein) is a quasi-local measure of the gravitational mass enclosed in the sphere of a given radius.
Inserting the metrics \eqref{eq:metricext} and \eqref{eq:metricint} leads to

\begin{equation}
M_{MS}^{i}(\tau,\rho) = \frac{1}{2} \rho^3 S \left(k + \dot{S}^2\right)   \quad , \quad M_{MS}^{e}(b) = \frac{b}{2}\left(1-\frac{1}{N(b)}\right)\;.
\end{equation}

\noindent
A physically sensible condition for the matching of the interior and exterior is

\begin{equation}\label{eq:MSmachting}
\left.M_{MS}^{i}\right|_{\Sigma} \stackrel{!}{=} \left.M_{MS}^{e}\right|_{\Sigma} \;,
\end{equation}

\noindent
which leads exactly to Eq. \eqref{eq:rhodS2} and is compatible with the Israel-Darmois junction conditions.

In the following section, we give an interpretation of the conditions \ref{assm:1}-\ref{assm:3} as the collapse of homogeneous pressure-less dust (Oppenheimer-Synder-Datt collapse \cite{OppenheimerOnContinuedGravitational,DattOnaclass}).

\section{Classical General Relativistic Collapse and its Interpretation}\label{sec:classical}

Let us now recall simple models of the homogeneous spherically symmetric collapse of an ideal fluid in classical general relativity and analyse how the equations derived in the previous section fit in this picture. In details we recall the so called Oppenheimer-Snyder-Datt \cite{OppenheimerOnContinuedGravitational,DattOnaclass} dust model and the uniform fluid sphere collapse with linear equation of state (see \cite{AdlerSimpleAnalyticModels} for an overview).

For the spherically symmetric collapse one starts with the ansatz \eqref{eq:ansatz1} with the gauge $a(t,r) = 1$ and $B(t,r) = 0$ for the matter region

\begin{equation}\label{eq:ansatz2i}
\dd s^2 = - \dd \tau^2 + N(\tau,\rho) \dd \rho^2 + b(\tau,\rho)^2 \dd \Omega_2^2\;,
\end{equation}

\noindent
where the coordinates were renamed according to $t \mapsto \tau$ and $r \mapsto \rho$.
For the vacuum region a suitable gauge is $b(r,t) = r$ and $B(r,t) = 0$, leading to

\begin{equation}\label{eq:ansatz2e}
\dd s^2 = -a(b,t) \dd t^2 + N(b,t) \dd b^2 + b^2 \dd \Omega_2^2 \;.
\end{equation}

\noindent
where $b$ is used as a coordinate instead of $r$.
In addition the stress-energy tensor is assumed to be of the form $T_{\mu \nu}(t,b) = 0$ for $b > R(t)$, where $R(t)$ describes the outer boundary of the matter distribution. Assuming an ideal fluid, the stress-energy tensor for $b< R(t)$ is 

$$
T_{\mu \nu}(t,r) = (e+P) \,u_{\mu} u_{\nu} + P g_{\mu \nu}\quad,\quad r < R(t)  \;,
$$

\noindent
where $u_{\mu}$ is the velocity field of the fluid, $e$ the energy density and $P$ the pressure of the fluid.
The coordinate $\rho$ describes the position of the matter constituents, which are assumed to be fixed at this coordinate position. Due to this, the fluid velocity takes the form $u^\mu = (1,0,0,0)^\mu$. 
The condition of homogeneity is $e = e(t)$. 
Further a linear equation of state (eos) is assumed with

\begin{equation}
P(e) = \begin{cases}
0, & \text{Oppenheimer-Snyder-Datt}\\
\alpha e,& \text{uniform sphere with linear eos}
\end{cases} \;,
\end{equation}

\noindent
where $\alpha > 0$. As $e=e(t)$ is a function of $t$ only, the same homogeneity condition holds true for the pressure $P$.
This metric and stress-energy tensor can be inserted into the Einstein equations

\begin{equation}
R_{\mu \nu} - \frac{R}{2} g_{\mu \nu} = \kappa T_{\mu \nu} \;,
\end{equation}

\noindent
with $\kappa = 8\pi$ and $G = c = 1$, which reduce to a set of only a few independent equations. They allow to determine $N(\tau,\rho)$, leading to the matter region metric (see e.g. \cite{fliessbach12,WaldGeneralRelativity,MisnerThorneWheeler}) 

\begin{equation}
\dd s^2 = - \dd \tau^2 + \frac{S(\tau)^2}{1-k \rho^2} \dd \rho^2 + \rho^2 S(\tau)^2 \dd \Omega_2^2 \;,
\end{equation}

\noindent
where $k$ is an integration constant, while  the function $S(\tau)$ can be is determined by the remaining equations, see below. This form of the matter region metric is the only metric consistent with homogeneity and spherically symmetry and reflects exactly the assumption \ref{assm:2} (cfr. Sec. \ref{sec:SurfMatch}). The remaining Einstein equations reduce to 

\begin{subequations}\label{eq:eqsint}
\begin{align}
\frac{4 \pi}{3} S^{3(1+\alpha)}(\tau) e(\tau) &= m \;,\label{eq:energycons}\\
\dot{S}^2 = -k + \frac{8 \pi e S^2}{3} &= -k + \frac{2 m}{S^{1+3 \alpha} } \;,\label{eq:Sclass}
\end{align}
\end{subequations} 

\noindent
where $m$ is a further integration constant.
Eq. \eqref{eq:energycons} determines the dynamics of the matter in dependence of $S(\tau)$ and follows directly from the energy conservation condition $\nabla_\mu T^{\mu \nu} = 0$. 
At this point note that $k > 0$ together with Eq. \eqref{eq:Sclass} leads to a maximal value for $S$, i.e. corresponds to a collapse from a finite maximal size of the matter distribution from rest ($\dot{S} =0$). 
Instead $k = 0$ is the marginally free situation, which corresponds to a collapse from infinity with zero velocity asymptotically and $k > 0$ is a collapse from infinity with non-zero asymptotic velocity. 
Solving Eq. \eqref{eq:Sclass} determines the matter region metric completely.

Still remaining is to determine the vacuum part of spacetime.
Inserting the metric ansatz \eqref{eq:ansatz2e} into the Einstein equation for the exterior region ($b > R(t)$, $T_{\mu \nu} = 0$) leads due to the Birkhoff theorem \cite{BirkhoffRelativityandModernPhysics,JebsenOnthegerneral} to the Schwarzschild metric

\begin{equation}
\dd s^2 = -\left(1- \frac{2 M}{b}\right) \dd t^2 + \frac{1}{1- \frac{2 M}{b}} \dd b^2 + b^2 \dd \Omega_2^2 \;,
\end{equation}
 
\noindent
with mass $M$ as integration constant, which determines the vacuum metric.

By solving Eqs. \eqref{eq:eqsint} the system is fully determined and the last remaining step is to match both metrics at $\rho = \rho_o(\tau)$ and $b = R(t)$ using Eqs. \eqref{eq:matchingfinal}. Identifying $a(b) = 1/N(b) = 1- 2M/b$ leads to

\begin{subequations}\label{eq:matchingfinalclass}
	\begin{equation}\label{eq:Rclass}
	R(t(\tau)) = \rho_o(\tau) S(\tau)\;,
	\end{equation}
	\begin{equation}\label{eq:dtbydtau}
	\dd t^2 = \frac{1-\frac{S^2 \dot{\rho_o}^2}{1-k\rho_o^2} + N  \left(\rho_o S\right)\dot{}\;^2}{a} \dd \tau^2 = \frac{N}{(1-k\rho_o^2) a} \left(1-k\rho_o^2 + \rho_o \dot{\rho_o} S \dot{S}\right)^2 \dd \tau^2 \;,
	\end{equation}
	\begin{equation}\label{eq:rhodS2class}
	\frac{2 m \rho_o^2}{S^{1+3\alpha}(\tau)} = \frac{2 M}{R(\tau)} \;,
	\end{equation}
	\begin{equation}\label{eq:drhoclass}
	\frac{\dot{\rho_o}}{\rho_o} = 0\;,
	\end{equation}
\end{subequations}

\noindent
Let us focus on Eq. \eqref{eq:rhodS2class} first. Using Eq. \eqref{eq:Rclass} this yields 

$$
\frac{\rho_o^{3(1+\alpha)}}{R^{3\alpha}} = \frac{M}{m} \quad \Longleftrightarrow \quad \rho_o^3 = \frac{M}{m} S^{3 \alpha} \;,
$$
\noindent
which fixes the dynamics of the coordinate value of the matter surface $\rho_o(\tau)$. Unfortunately this contradicts Eq. \eqref{eq:drhoclass} as long as $\alpha \neq 0$. As argued in \cite{AdlerSimpleAnalyticModels} the fact that the global metric fails to be $C^1(M)$ corresponds to surface tension of the matter. The metric hence can only be $C^1(M)$ when there is no pressure at all, i.e. $\alpha = 0$ and the matter being pressure-less dust\footnote{In a realistic model of stellar collapse, the matter distribution is of course not homogeneous. Due to this the pressure can depend on the radial direction and can vanish at the surface of the star. If this is true the matching is $C^1(M)$ even if the pressure is non-vanishing in the interior of the star.}. For a model which includes pressure, one would hence not demand the matching of the second fundamental forms, but only the slightly weaker condition of the coincidence of the Misner-Sharp masses \eqref{eq:MSmachting}.
This is discussed in detail in terms of the  Israel-Darmois junction conditions, which take the stress-energy tensor of the matching surface into account by means of a jump of the extrinsic curvatures across this surface \cite{IsraelSingularhypersurfacesand,DarmoisLeseuqationsde}.
This leads to an interpretation of assumption \ref{assm:3}: If the matter has internal pressure, assumption \ref{assm:3} cannot be satisfied. Concluding, assumptions \ref{assm:1}-\ref{assm:3} describe the matching of a homogeneous spherically symmetric distribution of pressure-less dust, which is the simplest imaginable model.

We furthermore can reverse the logic and interpret the condition $\dot{\rho_o} \neq 0$. For pressure-less dust, there are no spacial interactions of neighbouring particles, i.e. their coordinate position $\rho$ remains fixed during the collapse process.
The time $\tau$ in the metric \eqref{eq:ansatz2i} corresponds to the eigentime of a dust particle. Including pressure, the particles start to interact among each other leading to a change of their coordinate $\rho$. From this point of view, we can interpret a dynamical changing of $\rho_o$ as an effect of internal pressure. This will become relevant in the next section, where we apply the surface matching to quantum corrected vacuum spacetime regions.

\section{Application to Effective Black Hole Metrics --- Polymer Models}\label{sec:application}

A common way to study black holes in quantum gravity is to study eternal black holes \cite{CorichiLoopquantizationof,AshtekarQuantumTransfigurarationof,AshtekarQuantumExtensionof,ModestoSemiclassicalLoopQuantum,BodendorferEffectiveQuantumExtended,Bodendorferbvtypevariables,KellyEffectiveloopquantumgravity,KellyBlackholecollapse,GeillerSymmetriesofthe,SartiniQuantumdynamicsof,NicoliniQuantumCorrectedBlackHoles,EassonTheclassicaldouble,AdeifeobaTowardsconditionsfor,PlataniaDynamicalrenormalizationof,MotiOnthequantum,NicoliniNoncommutativeBlackHoles,NicoliniRemarksonregular,SmailagicKerrrblackhole,BardeenNon-singulargeneral,HaywardFormationandevaporation,DymnikovaVacuumnonsingularblackhole,DymnikovaDeSitter-Schwarzschild,FrolovNotesonnon-singular,FrolovRemarksonnon-singularblackholes}, which is of course not the scenario that takes place in nature. Using the matching conditions \eqref{eq:matchingfinal} provides us with a strategy to extend an eternal black hole model to a dust collapse model according to the interpretation of Sec. \ref{sec:classical}. A similar strategy from the matter region point of view was analysed in a recent set of papers \cite{BenAchourBouncingcompactobjectsI,BenAchourBouncingcompactobjectsII,BenAchourConsistentblackto}.

This strategy is in principle applicable to any effectively quantum corrected metric, independent of the particular approach, as long as the initial assumptions are satisfied. In the following we would like to focus on so called poylmer models of black holes, which are motivated by LQG (cfr. \cite{CorichiLoopquantizationof,AshtekarQuantumTransfigurarationof,AshtekarQuantumExtensionof,ModestoSemiclassicalLoopQuantum,BodendorferEffectiveQuantumExtended,Bodendorferbvtypevariables,BodendorferMassandHorizon,KellyEffectiveloopquantumgravity,KellyBlackholecollapse,GeillerSymmetriesofthe,SartiniQuantumdynamicsof}). These effective quantum corrected black hole metrics are typically written in the following form: 

\begin{equation}\label{eq:polymermetricgeneric}
\dd s^2 = -a(r) \dd t^2 + \frac{1}{a(r)} \dd r^2 + b(r)^2 \dd \Omega_2^2 \;, 
\end{equation}

\begin{figure}[t!]
	\centering\includegraphics[scale=0.75]{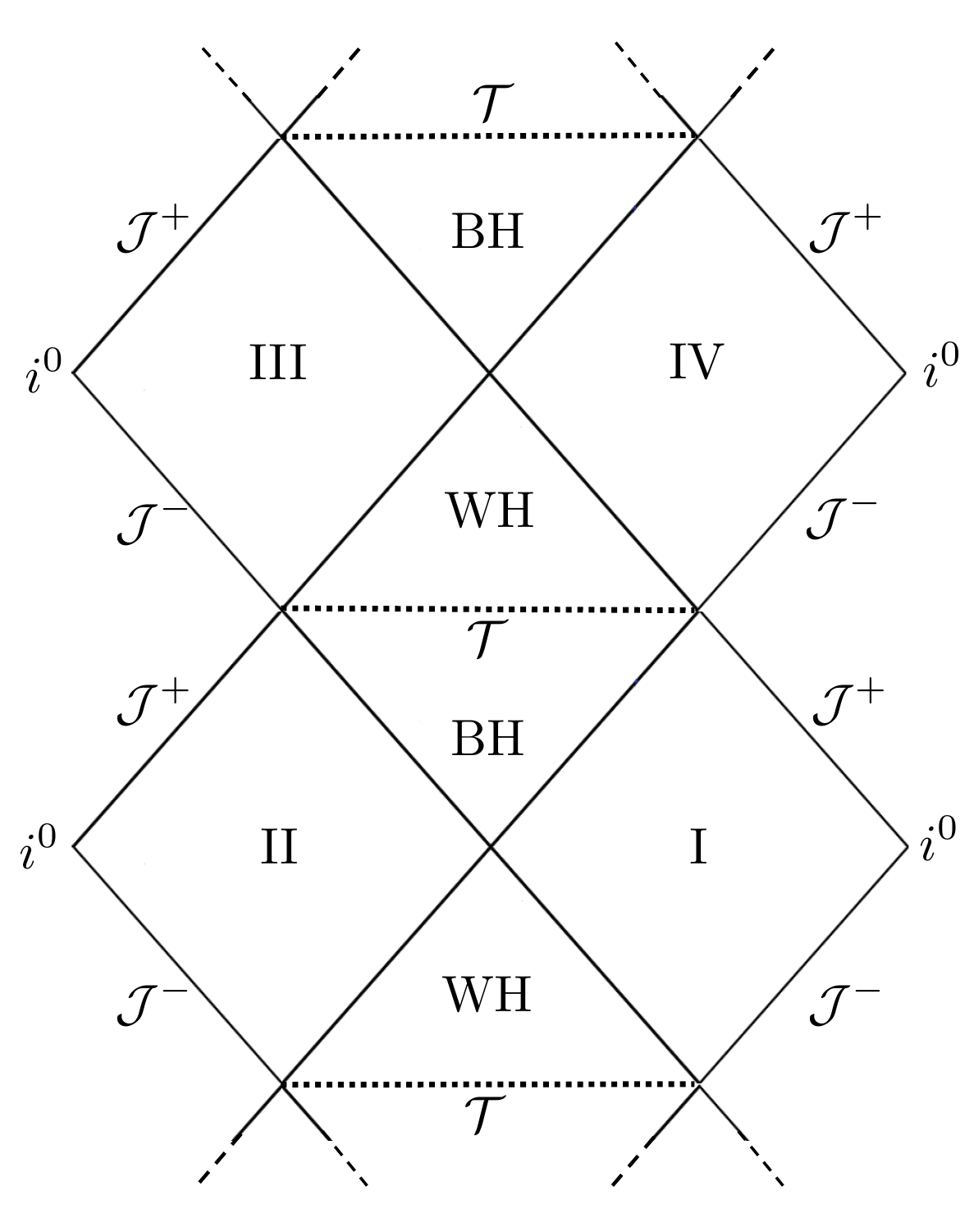}
	\caption{Penrose diagram for the Kruskal extension of the full quantum corrected black hole spacetime according to polymer models as in e.g. \cite{AshtekarQuantumExtensionof,BodendorferEffectiveQuantumExtended,BodendorferMassandHorizon,ModestoSemiclassicalLoopQuantum}}
	\label{Penrosediag2}
\end{figure}

\noindent
Qualitatively, these models are all similar and resolve the classical singularity by means of a regular transition surface. 
At this particular spacelike surface a transition between a trapped and anti-trapped region takes place.
The transition surface has further the minimal reachable areal radius as $b'(r_\mathcal{ T}) = 0$, where $r_\mathcal{ T}$ is the coordinate value of $r$ at which this minimum is reached and $b' = \dd b/\dd r$.
These models usually admit two different horizons, one before the transitions surface and another one afterwards. The first horizon indicates a change from a non-trapped to a trapped region and thus is commonly called black hole horizon. The second one lies behind the transition surface and therefore a change from an anti-trapped to a non-trapped spacetime region happens once this horizon is crossed. Consequently, this horizon is called white hole horizon. Analysing the global causal structure shows that these horizons appear repeatedly.
The causal diagram of such a spacetime is shown in Fig. \ref{Penrosediag2}.
\noindent
The metric \eqref{eq:polymermetricgeneric} can be brought in the form of the above discussion by changing the coordinates via $b = b(r)$, leading to

\begin{equation}
\dd s^2 = -a(b) \dd t^2 + \frac{r'(b)^2}{a(b)} \dd b^2 + b^2 \dd \Omega_2^2 \;, 
\end{equation}

\noindent
where $r'(b) = \frac{d r(b)}{d b}$ and $r(b)$ is the inverse function of $b(r)$. Note that, this inverse function cannot exist uniquely as due to the bouncing behaviour of $b(r)$ there exist a black hole and a white hole branch (cfr. the later discussion of the explicit model). The analysis has to be done for both branches separately. Furthermore, at the transition surface $r = r_{\mathcal{ T}}$ we have $b'(r_{\mathcal{T}}) = 0$, which leads to $r'(b \rightarrow b_{\mathcal T}) \rightarrow \pm \infty$. Inserting this metric in Eqs. \eqref{eq:rhodS2} and \eqref{eq:drho}\footnote{Note that what is $r$ in Eqs. \eqref{eq:rhodS2} and \eqref{eq:drho} is $b$ here, while here $r = r(b)$ is the inverse function of $b(r)$. All expression are evaluated at $R(\tau)$, i.e. $a = a(r(b=R(\tau))) = a(b = R(\tau))$} leads to

\begin{subequations}\label{eq:matchingbounce}
	\begin{equation}
	\rho_o^2 \dot{S}^2 = 1 - k\rho_o^2 - \frac{a(R)}{r'(R)^2} \;,
	\end{equation}
	\begin{equation}
	\frac{\dot{\rho_o}}{\rho_o} =  \frac{(1-k\rho_o^2) \rho_o \dot{S} \frac{r''}{r'}}{1-\frac{a}{r'^2} + \frac{R}{2 r'^2} \left.\der{a}{b}\right|_{b=R`} - \left(1-k\rho_o^2\right) R \frac{r''}{r'}} \;.
	\end{equation}
\end{subequations}

\noindent
Furthermore it turns out to be useful to extract the equation for $R(\tau)$, i.e. (cfr. Eq.~\eqref{eq:RrhoS})

\begin{equation}\label{eq:dRbounce}
\der{R}{\tau} = \rho_o \dot{S} + S \dot{\rho_o} = \rho_o \dot{S} \left( \frac{1-\frac{a}{r'^2} + \frac{R}{2 r'^2} \der{a}{b}}{1-\frac{a}{r'^2} + \frac{R}{2 r'^2} \der{a}{b} - \left(1-k\rho_o^2\right) R \frac{r''}{r'}}  \right) \,.
\end{equation}

\noindent
As mentioned, $r(b)$ has two branches. To avoid tedious computations on both sides separately, it is suitable to rephrase the equations in terms of $r_o(\tau) = b^{-1}(R(\tau))$ and $b(r_o)$ instead. This leads to

\begin{subequations}\label{eq:matchingbounce2}
	\begin{equation}
	\rho_o^2 \dot{S}^2 = 1 - k\rho_o^2 - a b'(r_o)^2 \;,
	\end{equation}
	\begin{equation}
	\frac{\dot{\rho_o}}{\rho_o} =  \frac{-(1-k\rho_o^2) \rho_o \dot{S} \frac{b''}{b'^2}}{1-a b'^2 + \der{a}{r} \frac{b'}{2} b(r_o) + \left(1-k\rho_o^2\right) b(r_o) \frac{b''}{b'^2}} \;.
	\end{equation}
	\begin{equation}\label{eq:ro}
	\der{r_o}{\tau} = \frac{\rho_o \dot{S}}{b'} \left( \frac{1- a b'^2 + \der{a}{r} \frac{b'}{2} b(r_o)}{1- a b'^2 + \der{a}{r}\frac{b'}{2} b(r_o) + \left(1-k\rho_o^2\right) b(r_o) \frac{b''}{b'^2}}  \right) \,.
	\end{equation}
\end{subequations}

\noindent
Note that these equations are functions of $S$ and $r_o$ ($\rho_o = S/b(r_o)$) only and effectively two coupled differential equations. In the case of $k = 0$, there remains only a dependence on $r_o$, which decouples the $r_o$ equation \eqref{eq:ro}.

Before solving these equations for a specific model, we can do some generic analysis of the behaviour around the transition surface.
If the function $b(r)$ is analytic, we can expand the equations around the point $r = r_\mathcal{ T}$.
With $b(r) = b_{\mathcal T} + \frac{b''_{\mathcal{ T}}}{2} (r-r_{\mathcal{ T}}) + \mathcal{O}\left(\left(r-r_{\mathcal{ T}}\right)^3\right)$ , where $b_\mathcal{ T} = b(r_\mathcal{ T})$, we find 

\begin{align}
r' =& \pm \sqrt{\frac{1}{2 b''_{\mathcal{T}}}} \frac{1}{\sqrt{b-b_{\mathcal T}}} \mp \mathcal{O}\left(1\right) \;,
\\
r'' =& \mp \frac{1}{2}\sqrt{\frac{1}{2 b''_{\mathcal{T}}}} \frac{1}{\sqrt{b-b_{\mathcal T}}^3} + \mathcal{O}\left(\frac{1}{\sqrt{b-b_{\mathcal T}}}\right) \;,
\end{align}

\noindent
leading to

\begin{align}
\der{R}{\tau} \simeq&\; 2\text{sign}(\rho_o \dot{S})\frac{R-b_{\mathcal T}}{\sqrt{1-k \rho_o^2} b_{\mathcal T}} + \mathcal O \left(\left(R-b_{\mathcal{ T}}\right)^2\right)\;, \label{eq:dRexp}
\\
\rho_o^2 \dot{S}^2 \simeq&\; 1- k \rho_o^2 - 2 a_{\mathcal T} b''_{\mathcal T} \left(R- b_{\mathcal{T}}\right)+ \mathcal{O}\left(\sqrt{R-b_{\mathcal{ T}}}^3\right)\;, \label{eq:dSexp}
\\
R\frac{\dot{\rho_o}}{\rho_o} \simeq& -\text{sign}(\rho_o \dot{S}) \left( \sqrt{1-k\rho_o^2} - \left(a_{\mathcal T} b''_{\mathcal T} + \frac{2}{b_{\mathcal{ T}}}\right) \frac{R-b_{\mathcal{ T}}}{\sqrt{1-k\rho_o^2}} \right) + \mathcal{O}\left(\sqrt{R-b_{\mathcal{ T}}}^3\right) \;,
\label{eq:drhooexp}
\end{align}

\noindent
where $a_{\mathcal T} = a(b_{\mathcal{T}})$ and $b''_{\mathcal T} = \left.\dder{b}{r}\right|_{r=r_{\mathcal T}}$. At the transition surface, we find the equations to become

\begin{align}
\der{R}{\tau}& \xrightarrow{R \rightarrow b_{\mathcal{T}}} 0 \;,\label{eq:dRlimit}\\
\rho_o^2 \dot{S}^2& \xrightarrow{R \rightarrow b_{\mathcal{T}}} 1- k\rho_o^2 \;, \label{eq:dSbySexp}\\
\frac{\dot{\rho_o}}{\rho_o}& \xrightarrow{R \rightarrow b_{\mathcal{T}}} - \frac{\rho_o \dot{S}}{b_{\mathcal{T}}} = -\text{sign}(\rho_o \dot{S}) \frac{\sqrt{1-k\rho_o^2}}{b_{\mathcal{T}}} \;. \label{eq:drhobyrhoexp}
\end{align}

\noindent
The first equation is intuitively clear and the only consistent possibility. As the matter collapses, it falls into the black hole until the minimal allowed areal radius $b_\mathcal{T}$ -- the transition surface -- is reached. The matter cannot collapse further as the exterior metric does not allow areal radii smaller than $b_{\mathcal{T}}$, i.e. the matter has to re-expand, which corresponds to $\dot{R} = 0$. Eq. \eqref{eq:drhobyrhoexp} instead has an implication on the causal properties of the collapsing surface. Already at this point we can study the causality of the matter surface close to the transition surface. The norm of the tangent vector $\mathcal{V}_{\gamma}$ to the worldline $\gamma^\mu = (\tau, \rho_o(\tau), 0 ,0)^\mu$ of an observer sitting on the surface of the collapsing dust cloud is

\begin{equation}\label{eq:causalitytau}
g(\mathcal{V}_{\gamma},\mathcal{V}_{\gamma}) = -1 + \frac{S^2 \dot{\rho_o}^2}{1-k \rho_o^2} = -1 + \frac{R^2 \frac{\dot{\rho_o}^2}{\rho_o^2}}{1-k \rho_o^2} \xrightarrow{R \rightarrow b_{\mathcal{T}}} 0 \;,
\end{equation}

\noindent
which asymptotically becomes lightlike at the instance the transition surface is passed. We should ensure that the curve remains timelike for the full evolution, which will lead to a constraint on the model parameter (see Sec. \ref{sec:modelapplication}). Note that for this conclusions no specific assumptions about the model were necessary, which makes the asymptotic lightlikeness a generic feature of bouncing exterior black hole models.

Note that the bouncing model is fully compatible with the kinematical constraint of \cite{BenAchourConsistentblackto}, which states that the matter surface can only bounce in a spacetime region where $\theta_+ \theta_- \le 0$. Here $\theta^\pm$ are the in- and outgoing expansions of the spacetime.
In other words, the bounce has to happen in a non-trapped region of spacetime. 
As was shown for several bouncing models \cite{AshtekarQuantumExtensionof,AshtekarQuantumTransfigurarationof,BodendorferEffectiveQuantumExtended,BodendorferMassandHorizon}, the transition surface is characterised by a transition between a trapped and an anti-trapped region and similar results were recently obtained in an analysis of the Raychaudhuri equation for different polymerisation schemes \cite{BlanchetteBlackholesingularity}.
Therefore at the transition surface it is $\theta_+ \theta_- = 0$, consistent with \cite{BenAchourConsistentblackto}.
To be concrete, for a metric of the form Eq.~\eqref{eq:polymermetricgeneric} the expansions read (in the black hole interior, i.e. behind the horizon with $a(r) < 0$)

\begin{equation}\label{eq:expansion}
	\theta_\pm = S^{ab} \nabla_a u_b^\pm = -\sqrt{-2a(r)} \frac{b'(r)}{b(r)} \quad , \quad \theta_+\theta_- = 2|a(r)| \left(\frac{b'(r)}{b(r)}\right)^2 \;.
\end{equation}

\noindent
Here $u_\pm^a$ are in- and outgoing future pointing normals to the constant $t$ and $r$ 2-spheres, which are normalised according to $g(u_\pm, u_\pm) = 0$, $g(u_\pm, u_\mp) = -1$.
Further $S^{ab} = g^{ab} + u^a_+ u^b_- + u_-^a u_+^b$ is the projector on the metric 2-spheres.
See \cite{BodendorferEffectiveQuantumExtended} for details.

We can even solve the equations close to the transition surface. For $k=0$ this yields the asymptotic solution

\begin{equation}
R(\tau) \simeq \left(R_o - b_{\mathcal{T}}\right) \exp\left(\text{sign}(\rho_o \dot{S}) \frac{2}{b_{\mathcal{T}}}\left(\tau - \tau_o\right)\right) + b_{\mathcal{T}} \;,
\end{equation}

\noindent
where the initial conditions $R(\tau_o) = R_o \sim b_\mathcal{ T}$ were chosen.
It shows that it takes infinitely long in $\tau$-time to reach the transition surface and the black and white hole side cannot be described at once by using the coordinate $\tau$. Instead, a more physical time $\sigma$ can be introduced and is defined by

\begin{equation}\label{eq:eigentime}
\dd \sigma = \pm\sqrt{1- \frac{S^2 \dot{\rho_o}^2}{1-k \rho_o^2}} \dd \tau \;.
\end{equation}

\noindent
This is the eigentime of an observer co-moving with the matter surface (cfr. Eq. \eqref{eq:causalitytau}). Indeed, the eigentime until the transition surface is reached is finite (for $k = 0$) as

\begin{align}\label{eq:sigmaoftauatT}
\sigma =& \int_{\tau_o}^{\tau}\sqrt{1- R(\tau')^2 \frac{\dot{\rho_o}(\tau')^2}{\rho_o(\tau')^2}} \dd \tau' 
\notag
\\
\simeq&\, \text{sign}(\rho_o \dot{S})b_{\mathcal T} \sqrt{2 \left( a_{\mathcal{ T}} b''_\mathcal{ T} +\frac{2}{b_\mathcal{ T}}\right)\left(R_o-b_\mathcal{ T}\right) } \left(\exp\left(\text{sign}(\rho_o \dot{S}) \frac{\left(\tau - \tau_o\right)}{b_{\mathcal{T}}}\right)-1\right)
\notag
\\
&\xrightarrow{\tau \rightarrow \pm \infty} -\text{sign}(\rho_o \dot{S})b_{\mathcal T} \sqrt{2 \left( a_{\mathcal{ T}} b''_\mathcal{ T} +\frac{2}{b_\mathcal{ T}}\right)\left(R_o-b_\mathcal{ T}\right) } < \infty\;,
\end{align}

\noindent
where for the positive branch (black hole side) the sign $\text{sign}(\rho_o \dot{S}) = -1$ and $\tau \rightarrow + \infty$ has to be used and the opposite for the negative branch (white hole side).

It is suitable to rephrase Eqs. \eqref{eq:matchingbounce} in terms of $\sigma$, leading to

\begin{align}
\der{r_o}{\sigma} =& \der{r_o}{\tau}(S,r_o) \der{\tau}{\sigma}(S,r_o)\;, \label{eq:equsr-s:r}
\\
\frac{1}{S}\der{S}{\sigma} =& \frac{1}{S}\der{S}{\tau}(S,r_o) \der{\tau}{\sigma}(S,r_o)\;, \label{eq:equsr-s:S}
\\
\frac{1}{\rho_o} \der{\rho_o}{\sigma} =& \frac{1}{\rho_o}\der{\rho_o}{\tau}(S,r_o) \der{\tau}{\sigma}(S,r_o)\;. \label{eq:equsr-s:rho}
\end{align}

\noindent
Note that it is crucial that $1- \frac{S^2 \dot{\rho_o}^2}{1-k \rho_o^2} > 0$ for $\sigma$ and hence the equations being well defined. This leads to a restriction of the parameters of the specific model, as will be discussed below. In the form of Eqs. \eqref{eq:equsr-s:r}-\eqref{eq:equsr-s:rho} the problem is solvable for black and white hole side at once.

To construct the full spacetime picture, it is necessary to re-express the solutions of Eqs. \eqref{eq:equsr-s:r}-\eqref{eq:equsr-s:rho} in terms of the time coordinate $t$ used in the vacuum region metric according to Eq. \eqref{eq:timefinal}. Nevertheless, it is also useful to change the time variable here, as $t$ diverges at the horizons. A suitable time coordinate is the ingoing Eddington-Finkelstein time $v$ defined by 

\begin{equation}\label{eq:v}
v = t + \int^b \sqrt{\frac{N(b')}{a(b')}} \dd b' = t + \int^r \frac{1}{a(r')} \dd r' \;.
\end{equation}

\begin{figure}[t!]
	\centering\includegraphics[width=7.75cm]{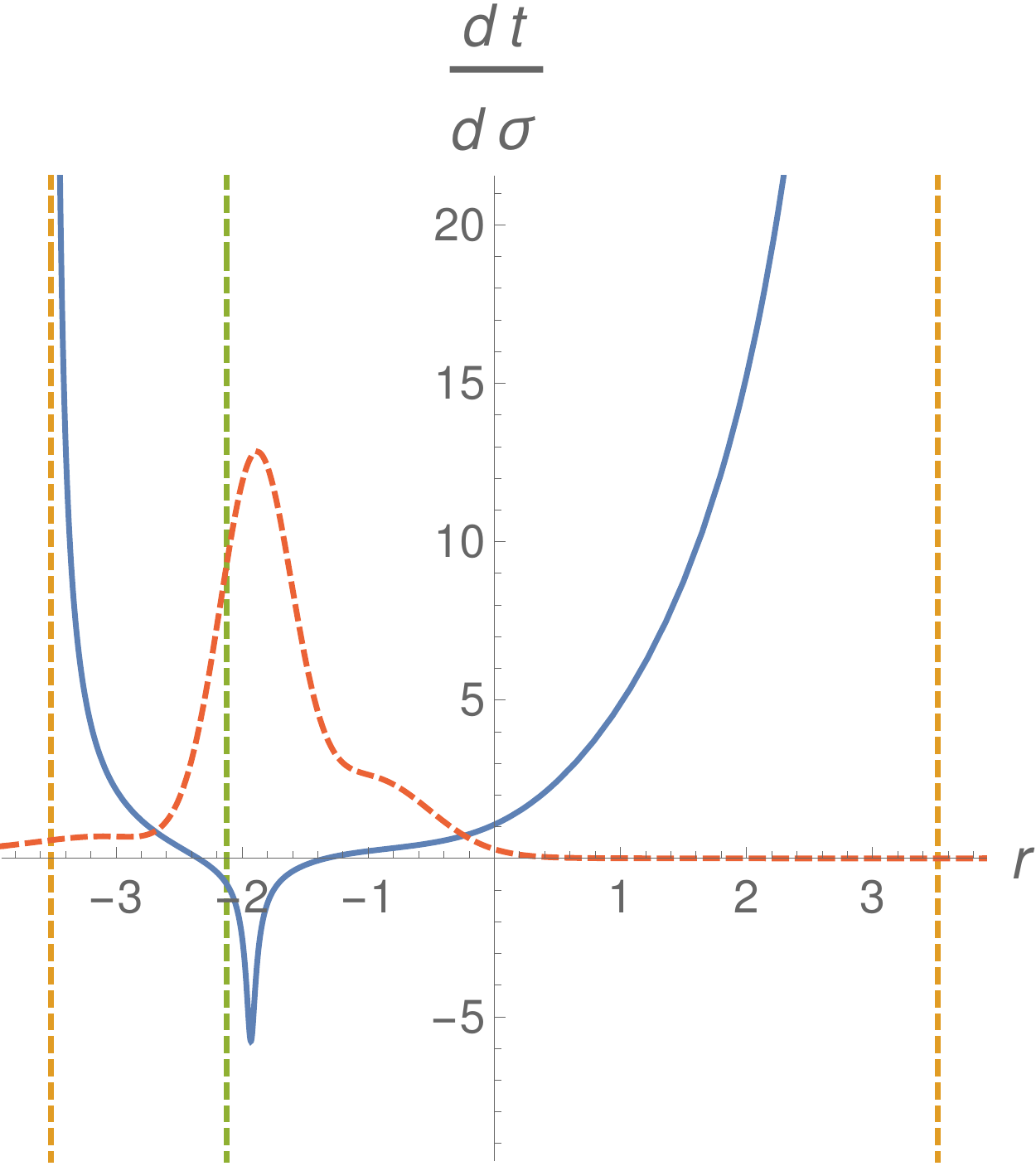}
	\caption{The plot shows the differential relation of $t$ and $\sigma$ for the explicit model introduced below in Sec. \ref{sec:modelapplication} (blue solid line). The yellow dashed lines correspond to the horizons, the green dashed line represents the transition surface. It is visible that $\dd t / \dd \sigma < 0$ around the transition surface. This indicates turning points in the $t$ direction, which is spatial inside the black hole. The red dashed line shows the Kretschmann scalar for reference. The negative part of $\dd t / \dd \sigma$ lies in the high curvature regime and the local minimum is aligned with the curvature maximum. The parameters are chosen as $k = 0$, $M_{BH} = 20$, $M_{WH} = 1$, $\lambda_1 = \lambda_2  =\mathscr L_o=1$.}
	\label{fig:dtbydsigma}
\end{figure}

\noindent
From Eq. \eqref{eq:timefinal} and Eq. \eqref{eq:eigentime} follows ($N = 1/ab'^2$)

$$
\dd t^2 = \frac{1- \frac{S^2 \dot{\rho_o}^2}{1-k\rho_o^2} + N \der{R}{\tau}^2}{a} \dd\tau^2 = \frac{a + \der{r_o}{\sigma}^2}{a^2} \dd\sigma^2 \;,
$$

\noindent
and concluding along $r = r_o(\sigma)$ it is 

\begin{equation}\label{eq:vofsigma}
\dd v = \left(\der{t}{\sigma} + \frac{\der{r_o}{\sigma}}{a}\right)\dd \sigma= \frac{\der{r_o}{\sigma} + \sqrt{\der{r_o}{\sigma}^2+a}}{a} \dd \sigma \;,
\end{equation}

\noindent
where the positive root is chosen as then $\dd v/\dd\sigma > 0$ and the time directions are aligned parallel.  Using again Eq. \eqref{eq:eigentime} and the second equation of Eq. \eqref{eq:timefinal} the following expression can be deduced

$$
a^2 \left(\frac{\dd t}{\dd \sigma}\right)^2 =\der{r_o}{\sigma}^2+a = \frac{\left(1-k\rho_o^2 + S^2 \der{\rho_o}{\sigma}^2 + \rho_o \der{\rho_o}{\sigma} S \der{S}{\sigma}\right)^2}{b'^2 \left(1-k\rho_o^2 + S^2 \der{\rho_o}{\sigma}^2\right)} = \frac{f(S,r_o)^2}{g(S,r_o)} \;,
$$

\noindent 
where $f(S,r_o)^2$ is the numerator and $g(S,r_o)$ the denominator of the fraction. 
Note that, $g(S,r_o) > 0$ but $f(S,r_o)$ can be negative in a region around the transition surface (see Fig. \ref{fig:dtbydsigma}). For getting a smooth function here, we should choose exactly the other sign in front of the square-root for this regime, leading to $\text{sign}(f(S,r_o))\sqrt{\der{r_o}{\sigma}^2+a} = f(S,r_o)/\sqrt{g(S,r_o)}$. This is depicted in Fig. \ref{fig:dtbydsigma}
for the model discussed in the next section. Physically this means that $\dd t/\dd\sigma < 0$, which is no problem as $t$ is a spacelike coordinate in the interior of the black hole and just signals a change of the spacial direction of the dust cloud. $\dd v/ \dd \sigma $ remains positive everywhere, as it should (see Fig. \ref{fig:vofsigma} below). Due to this the function $v(\sigma)$ is invertible, while $t(\sigma)$ is not. This is another reason to use $v$ instead of $t$.

This change of sign in Fig.~\ref{fig:dtbydsigma} will be visible as turning behaviour of the matter surface from the outside observer point of view (see Fig.~\ref{fig:PDextsketch} below).
The physical reason for this behaviour is not clear at this point.
Certainly, it is related to the high curvature (quantum) regime (see the red dashed line in Fig.~\ref{fig:dtbydsigma}), where also the energy conditions (cfr. \cite{BouhmadiAconsistentmodel}) are violated.
Also for different model parameters ($M_{BH}$, $M_{WH}$) as the ones used in the analysis of the following section, a regime in which $\dd t/ \dd \sigma < 0$, remains present.
It would be interesting to understand the physical origin better and also if this is a generic feature of different models.
As the expressions are involved a detailed analysis is complicated and left for future research.

Let us analyse now in more details the behaviour of Eq. \eqref{eq:vofsigma}. Generically for a collapse it is $\dd r_o/\dd\sigma < 0$. Due to this $\der{r_o}{\sigma} + \sqrt{\der{r_o}{\sigma}^2+a} \rightarrow 0$ for $r \rightarrow r_s^{(+)}$, where $r_s^{(\pm)}$ is the location of the black $(+)$ and white $(-)$ hole horizons, i.e. $a(r_s^{(\pm)}) = 0$. As the denominator of Eq. \eqref{eq:vofsigma} also vanishes at the horizon, a detailed analysis of the convergences shows that $\der{v}{\sigma}$ remains finite for $r \rightarrow r_s^{(+)}$. This allows to describe interior and exterior black hole regions simultaneously. Furthermore the same behaviour is true for the white hole horizon $r_s^{(-)}$\footnote{On the white hole side it is $\dd R/\dd \sigma = b'(r_o(\sigma))\cdot \der{r_o}{\sigma} > 0$. As $b'(r) < 0$ on the white hole side, $\der{r_o}{\sigma} < 0$ is still true and $\dd v / \dd \sigma$ remains finite at the white hole horizon.}.
Without making further assumptions about the model the trajectory of the dust cloud can be sketched in the Penrose diagram Fig. \ref{Penrosediag2}. Note that in this diagram constant $v$-lines are $45^\circ$-lines starting in region $\texttt{I}$ and terminating in region $\texttt{III}$. Assuming the collapse begins in region $\texttt{I}$, the dust cloud reaches the horizon connecting BH and region $\texttt{I}$ in Fig. \ref{Penrosediag2} within finite time $v$. As argued above the cloud also reaches the white hole horizon at a later but finite value of $v$. As the constant $v$-lines approach infinity going towards the white hole horizon connecting WH and region $\texttt{IV}$, the dust cloud has to leave the interior towards region $\texttt{III}$. This physically allows an observer in the vacuum region to exit (without touching the dust cloud) the black and white hole interior region though $\texttt{IV}$. From there is it possible to jump into the next BH region and continue along this infinite tower of Penrose diagrams.

As a last point it is possible to check the finiteness of curvature invariants in the matter region. As the vacuum region follows from a polymer black hole model, the singularity is generically avoided and curvature scalars remain finite at the transition surface. The same behaviour is expected for the matter region and can be checked explicitly around the transition surface. The Ricci scalar for the matter region metric Eq. \eqref{eq:metricint} reads

\begin{equation}
R = 6\left(\frac{\dot{S}^2}{S^2} + \frac{k}{S^2} + \frac{\ddot{S}}{S}\right) \;,
\end{equation}

\noindent
which by using Eqs. \eqref{eq:matchingbounce} can be re-written as 

\begin{equation}
\frac{R}{6} = \frac{\dot{S}^2}{S^2} + \frac{k}{S^2} - \frac{\der{a}{r} b'+2 a b''}{2 b} - \left(\frac{\dot{S}^2}{S^2} + \frac{k}{S^2} - \frac{\der{a}{r} b'+2 a b''}{2 b}\right) \frac{S \dot{\rho_o}}{\rho_o \dot{S}} \,.
\end{equation}

\noindent
Indeed, as the expansions Eqs. \eqref{eq:dRlimit}-\eqref{eq:drhobyrhoexp} show, this is finite at the transition surface and gives

\begin{equation}\label{eq:Riccitrans}
\left.R\right|_\mathcal{ T} = \frac{12}{b_\mathcal{ T}^2} \;.
\end{equation}

\subsection{Application for $k=0$}\label{sec:modelapplication}

After the general discussion, let us now analyse a specific model and furthermore restrict to the marginally free case, i.e. $k = 0$.
The application to $k\neq 0$ is possible, but much more involved. It is assumed that the qualitative structure does not change for $k \neq 0$ in the effective quantum regime.

Let us introduce the polymer black hole model of \cite{BodendorferEffectiveQuantumExtended}. According to this model the exterior spacetime is described by the metric 

\begin{equation}\label{eq:metricq1}
\dd s^2 = - a(r) \dd t^2 + \frac{1}{a(r)} \dd r^2 + b(r)^2 \dd \Omega_2^2 \;,
\end{equation}

\noindent
with
\begin{align}
b &= \frac{\mathscr L_o}{\lambda_2}\left( 3 D C^2 \lambda_1^2\right)^\frac{1}{3} \frac{\left(\frac{\lambda_2^6}{16 C^2 \lambda_1^2 \mathscr L_o^6} \left( \frac{\mathscr L_o r}{\lambda_2} + \sqrt{1+\frac{\mathscr L_o^2 r^2}{\lambda_2^2}} \right)^6 +1 \right)^\frac{1}{3}}{\left( \frac{\mathscr L_o r}{\lambda_2} + \sqrt{1+\frac{\mathscr L_o^2  r^2}{\lambda_2^2}} \right)}\; , \label{bquant}\\
a &= \left(\frac{\lambda_2}{\mathscr L_o}\right)^4 \left(1+\frac{\mathscr L_o^2 r^2}{\lambda_2^2}\right) \left( 1 - \frac{3 C D}{2 \lambda_2} \frac{1}{\sqrt{1+\frac{\mathscr L_o^2 r^2}{\lambda_2^2}}} \right) \frac{ \left(\frac{1}{3 D C^2 \lambda_1^2}\right)^{\frac{2}{3}} \left( \frac{\mathscr L_o r}{\lambda_2} + \sqrt{1+\frac{\mathscr L_o^2 r^2}{\lambda_2^2}} \right)^2}{\left(\frac{\lambda_2^6}{16 C^2 \lambda_1^2 \mathscr L_o^6} \left( \frac{\mathscr L_o r}{\lambda_2} + \sqrt{1+\frac{\mathscr L_o^2 r^2}{\lambda_2^2}} \right)^6 +1 \right)^{\frac{2}{3}}} \; , \label{aquant} 
\end{align}

\noindent
where $\mathscr L_o \lambda_1$ and $\lambda_2/\mathscr L_o$ are the polymerisation scales of the model. $\mathscr L_o = \int_{0}^{L_o} \left.\sqrt{a}\right|_{r=r_{\text{ref}}}$ is the physical size of the fiducial cell, which is used as regulator in the model (see \cite{BodendorferEffectiveQuantumExtended} for details) and $C/\mathscr L_o$, $D$ are integration constants related to the black hole and white hole masses as 

\begin{equation}\label{CandD}
\frac{C}{\mathscr L_o} = \frac{\lambda_2^3}{4 \lambda_1 \mathscr L_o^4}\left(\frac{M_{WH}}{M_{BH}}\right)^{\frac{3}{2}} \quad , \quad D = \left(\frac{2 \mathscr L_o}{\lambda_2}\right)^3\left( \frac{2}{3}\left(\frac{\lambda_1 \lambda_2}{3}\right)^3 M_{BH}^3 \left(\frac{M_{BH}}{M_{WH}}\right)^{\frac{9}{2}} \right)^{\frac{1}{4}}\;.
\end{equation}

\noindent
Details about the construction of the model can be found in \cite{BodendorferEffectiveQuantumExtended}. The global structure of this spacetime is depicted as Penrose diagram in Fig. \ref{Penrosediag2}.
The metric is in the limit $r \rightarrow \pm \infty$ well approximated by Schwarzschild spacetime with masses $M_{BH}$ and $M_{WH}$, respectively. At $r_{s}^{(\pm)} = \pm \frac{3CD}{2\sqrt{n}} \sqrt{1- \left(\frac{2 \lambda_2}{3 C D}\right)^2}$ there are two Killing horizons and at 

$$
r_\mathcal{T} = \frac{\lambda_2}{2\mathscr L_o} \left(\left(\frac{\lambda_2^3}{4 C \lambda_1 \mathscr L_o^3}\right)^{-\frac{1}{3}}-\left(\frac{\lambda_2^3}{4 C \lambda_1 \mathscr L_o^3}\right)^{\frac{1}{3}}\right)\;,
$$

\noindent
the minimal value of $b(r)$, the transition surface, is reached.
The areal radius of this surface is given by 

\begin{equation}\label{eq:btransition}
b_\mathcal{T} := b(r_\mathcal{T}) = \left(\frac{3\lambda_1CD}{2}\right)^{\frac{1}{3}}\;.
\end{equation}

This metric is used as input for Eqs.  \eqref{eq:matchingfinal}, which as argued above reduce to \eqref{eq:equsr-s:r}-\eqref{eq:equsr-s:rho}. The strategy now is as follows:

\begin{enumerate}
	\item Solve Eqs. \eqref{eq:equsr-s:r}-\eqref{eq:equsr-s:S} simultaneously, as they form a closed system\footnote{Due to our assumption $k=0$, the Eq. \eqref{eq:equsr-s:r} decouples from the other equations and can be solved separately. The result can be inserted into Eq. \eqref{eq:equsr-s:S}, which can also be integrated.}. The solution for $\rho_o(\sigma)$ is given by $\rho_o(\sigma) = b(r_o(\sigma))/S(\sigma)$.
	\item As argued above $\tau$ does not cover the whole collapse process, i.e. the metric Eq. \eqref{eq:metricint} is not sufficient to describe the whole process. A new choice of coordinates $(\sigma, \mathcal R)$ is discussed, which covers black and white hole side.
	\item Using Eq. \eqref{eq:vofsigma} allows to compute the curve $b(r_o(v))$, which describes the collapse from a outside observer point of view.
	\item This is all needed information to construct the Penrose diagram for the matter region and further to plot the trajectory of the matter surface into the Penrose diagram of the exterior spacetime. This closes the full spacetime picture of the collapse process.
\end{enumerate}

\begin{figure}[h!]
	\centering
	\subfigure[]
	{\includegraphics[width=7.3cm]{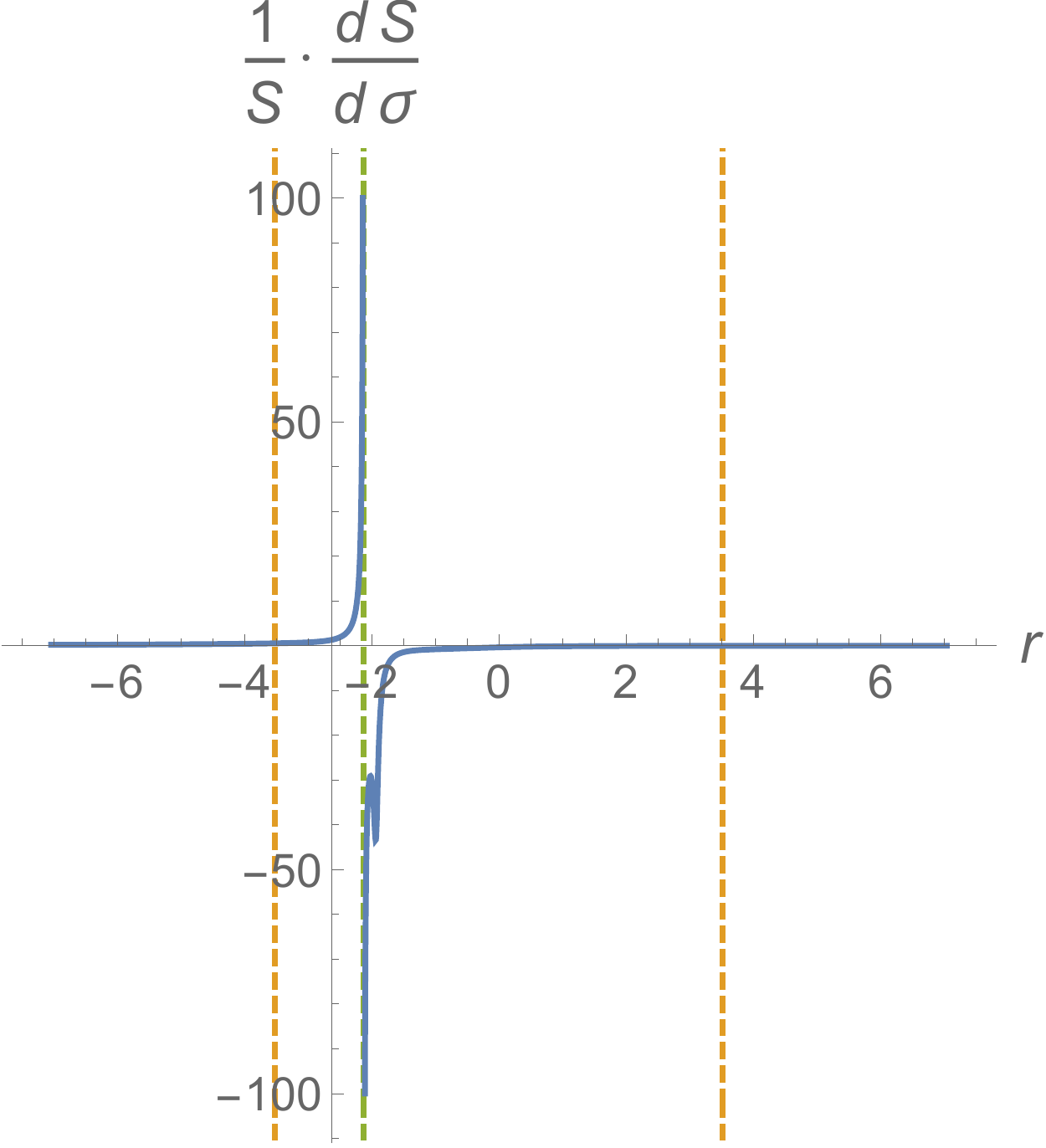}}
	\hspace{2mm}
	\subfigure[]
	{\includegraphics[width=7.3cm]{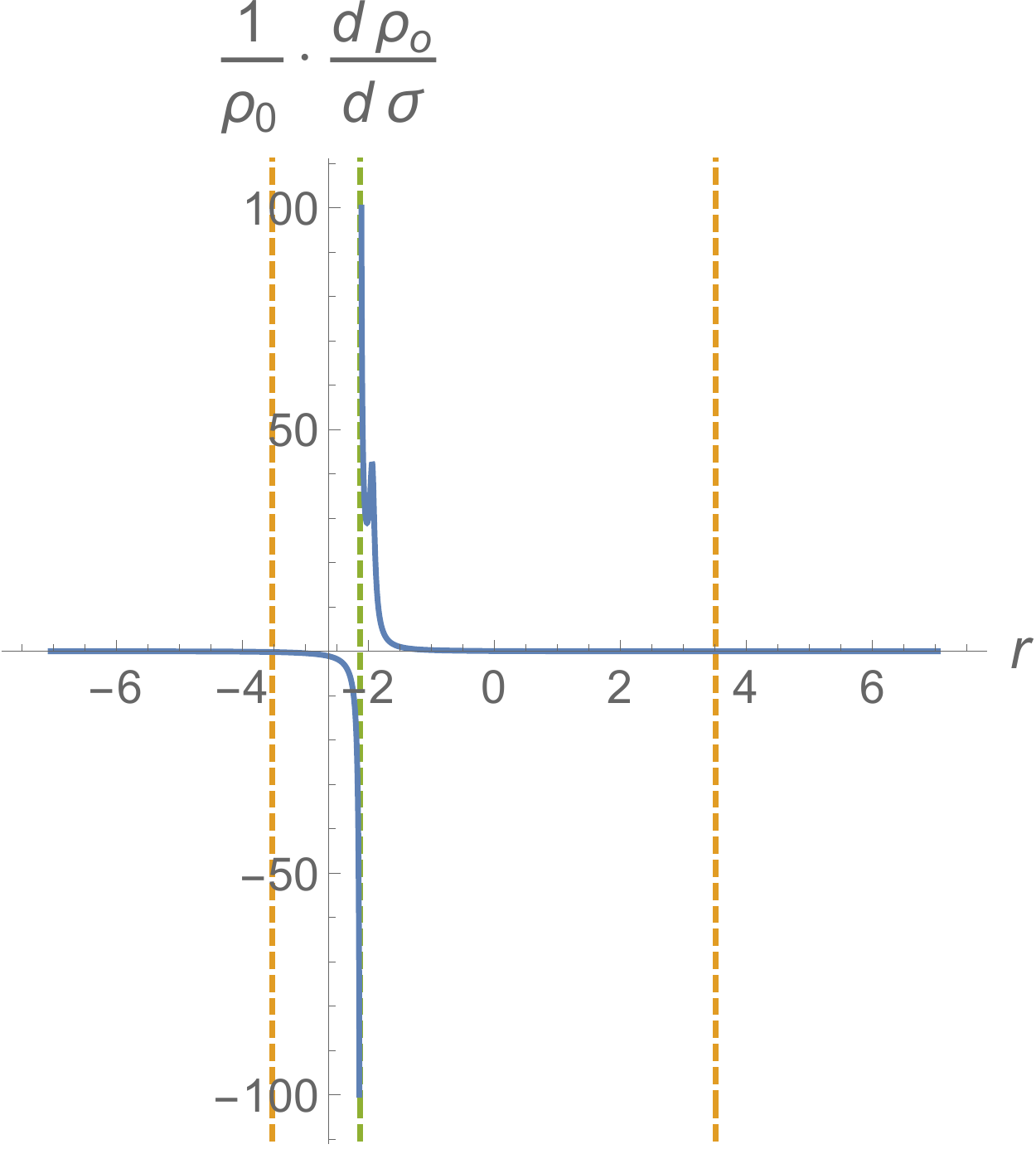}}
	\caption{Functional dependence of $\frac{1}{S} \der{S}{\sigma}$ in (a) and $\frac{1}{\rho_o} \der{\rho_o}{\sigma}$ in (b) of the radial coordinate $r$. The yellow dashed lines indicate the position of the horizons, while the green dashed line represents the transition surface. Both expressions diverge and change their sign at the transition surface. Parameter values are $k = 0$, $M_{BH} = 20$, $M_{WH} = 1$, $\lambda_1 = \lambda_2  =\mathscr L_o=1$.}
	\label{fig:PSSrho}
\end{figure}

\begin{figure}[h!]
	\centering\includegraphics[width=7.35cm]{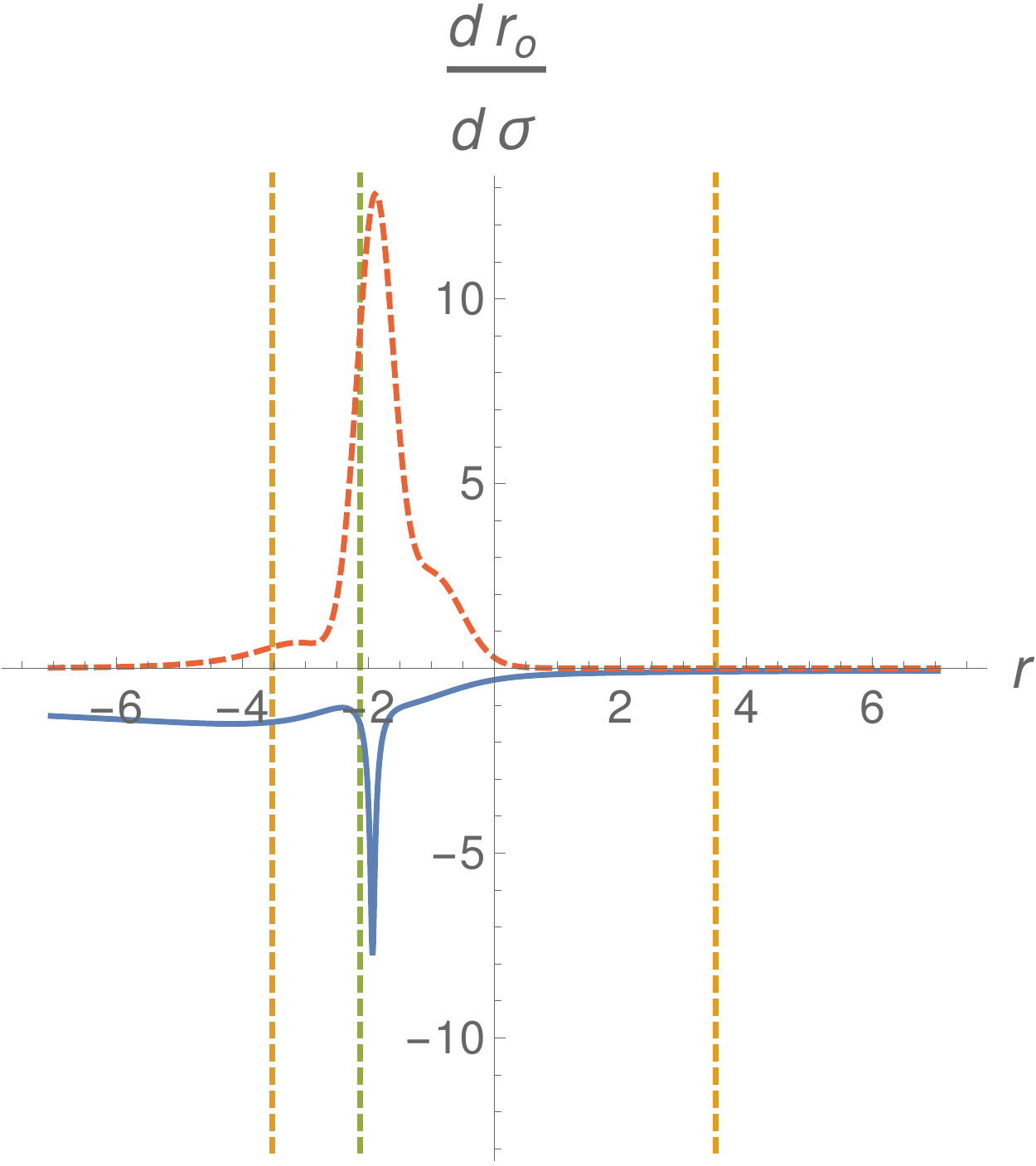}
	\caption{Phasespace plot of $\der{r_o}{\sigma}$ as a function of the radial coordinate $r$ (solid blue). As argued, $\der{r_o}{\sigma}$ remains negative and finite everywhere. The parameters are $k = 0$, $M_{BH} = 20$, $M_{WH} = 1$, $\lambda_1 = \lambda_2  =\mathscr L_o=1$. Yellow dashed lines indicate to the horizons, the green dashed line represents the transition surface. The red dashed line is the Kretschmann scalar as reference. Note that the maximum of the Kretschmann scalar lies slightly before the transition surface and is aligned with the local minimum of $\der{r_o}{\sigma}$.}
	\label{fig:PSr}
\end{figure}

\begin{figure}[h!]
	\centering
	\subfigure[]
	{\includegraphics[width=7.3cm]{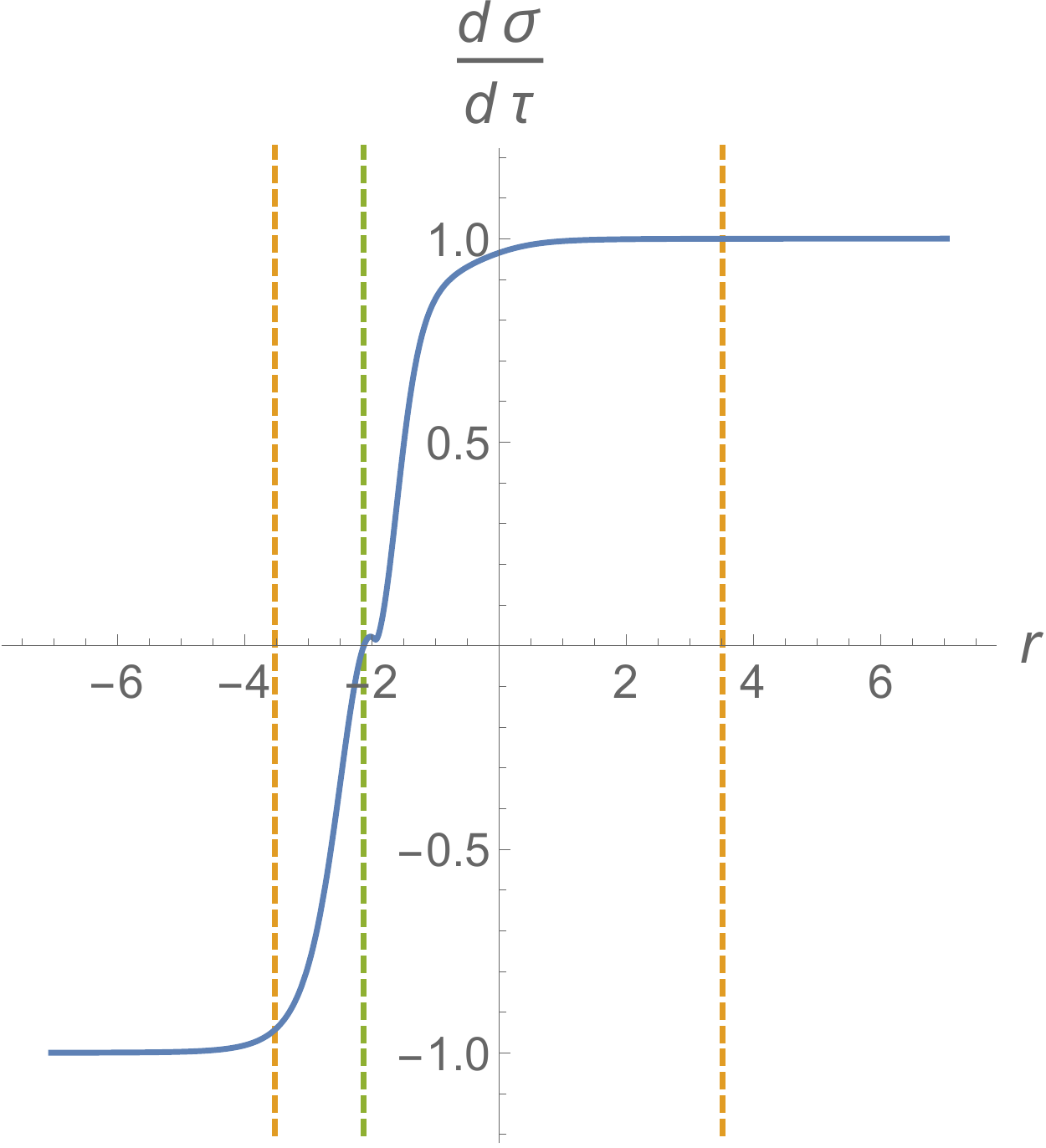}}
	\hspace{2mm}
	\subfigure[]
	{\includegraphics[width=7.3cm]{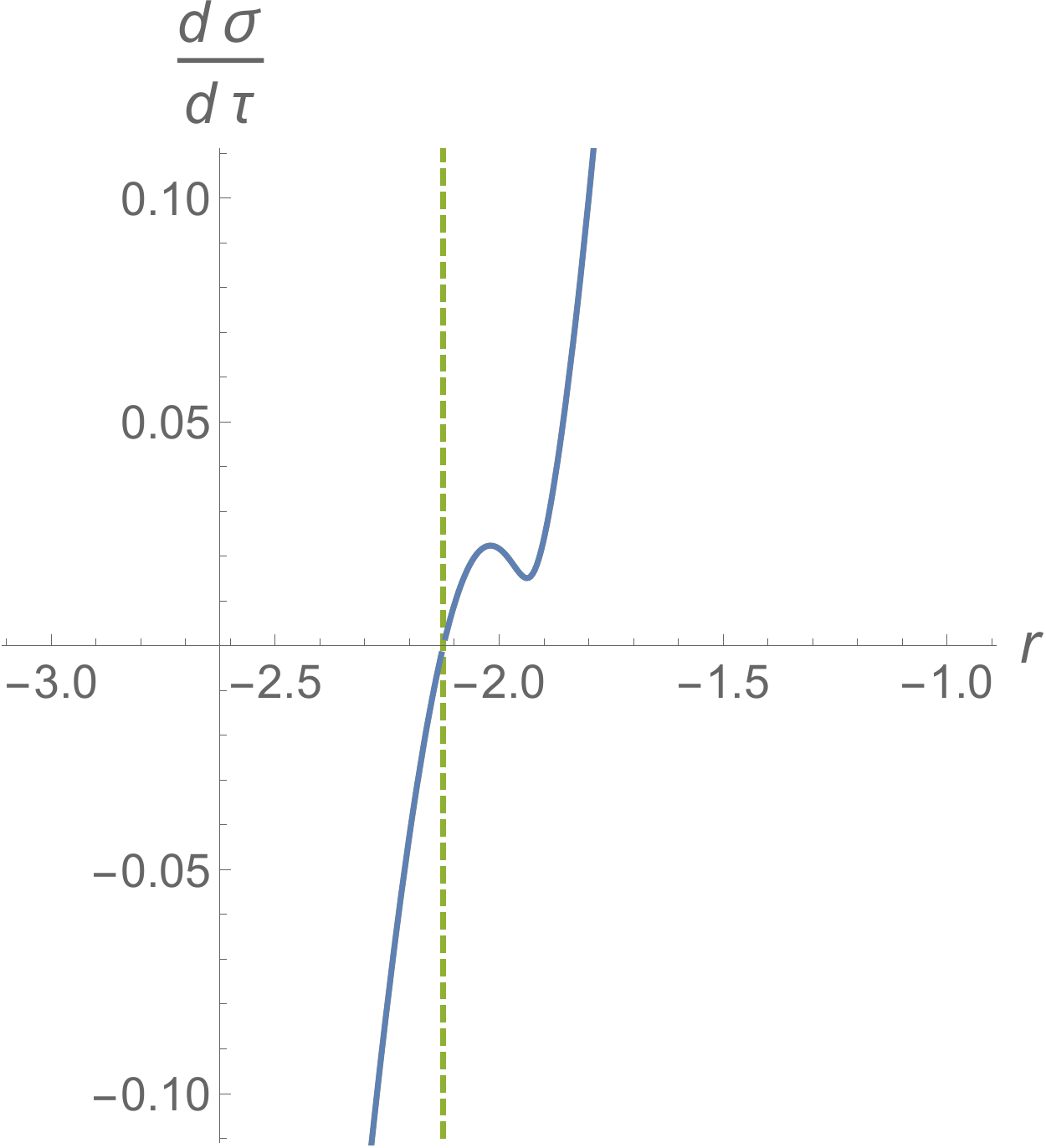}}
	\caption{Plot of the relation \eqref{eq:eigentime} in (a) and on a smaller scale around the transition surface in (b). Yellow dashed lines correspond to the horizons, green dashed line represents the transition surface with $k = 0$, $M_{BH} = 20$, $M_{WH} = 1$, $\lambda_1 = \lambda_2  =\mathscr L_o=1$.}
	\label{fig:PSsigma}
\end{figure}

\begin{figure}[t!]
	\centering
	\subfigure[]
	{\includegraphics[width=7.35cm]{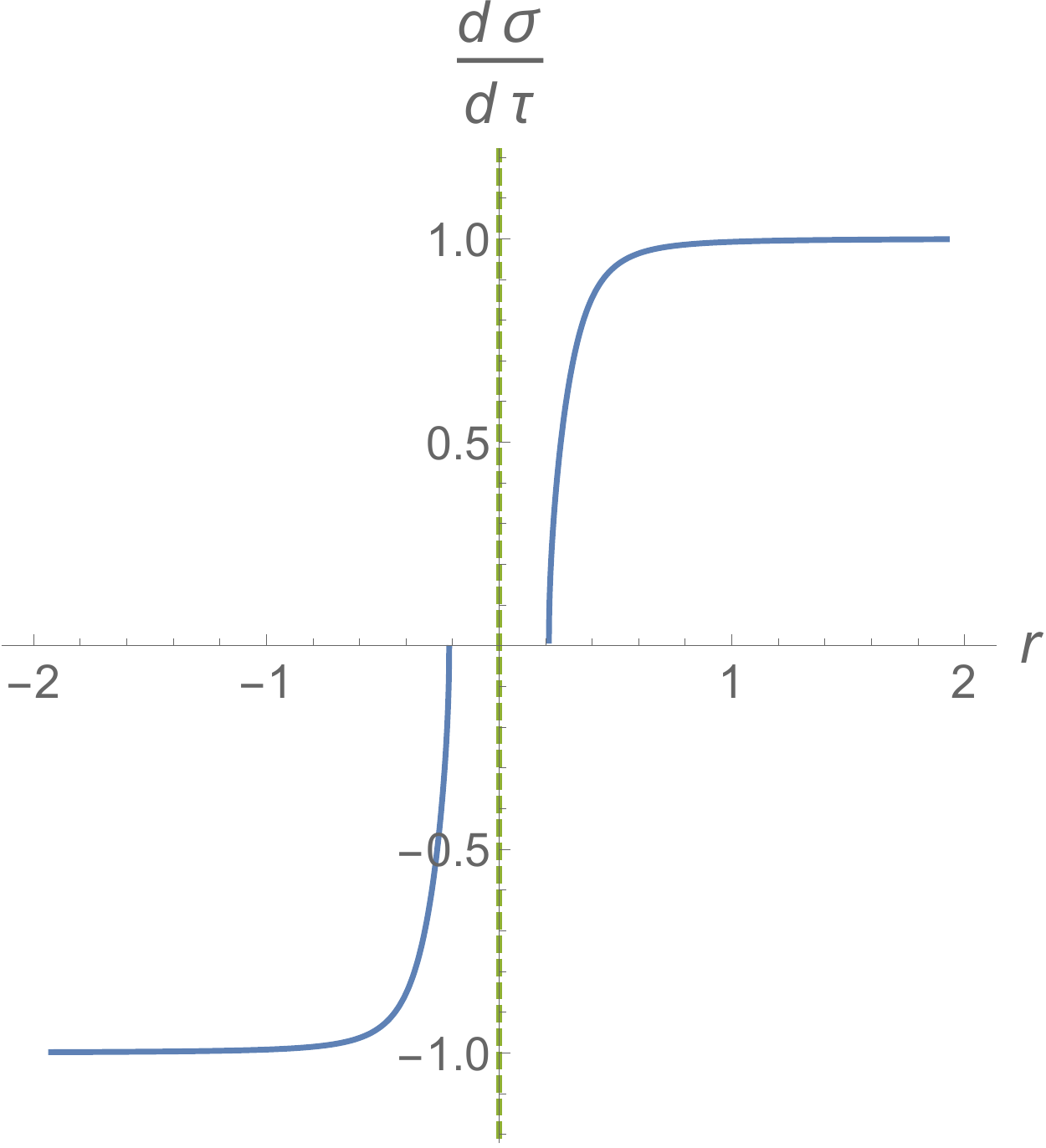}}
	\hspace{2mm}
	\subfigure[]
	{\includegraphics[width=7.35cm]{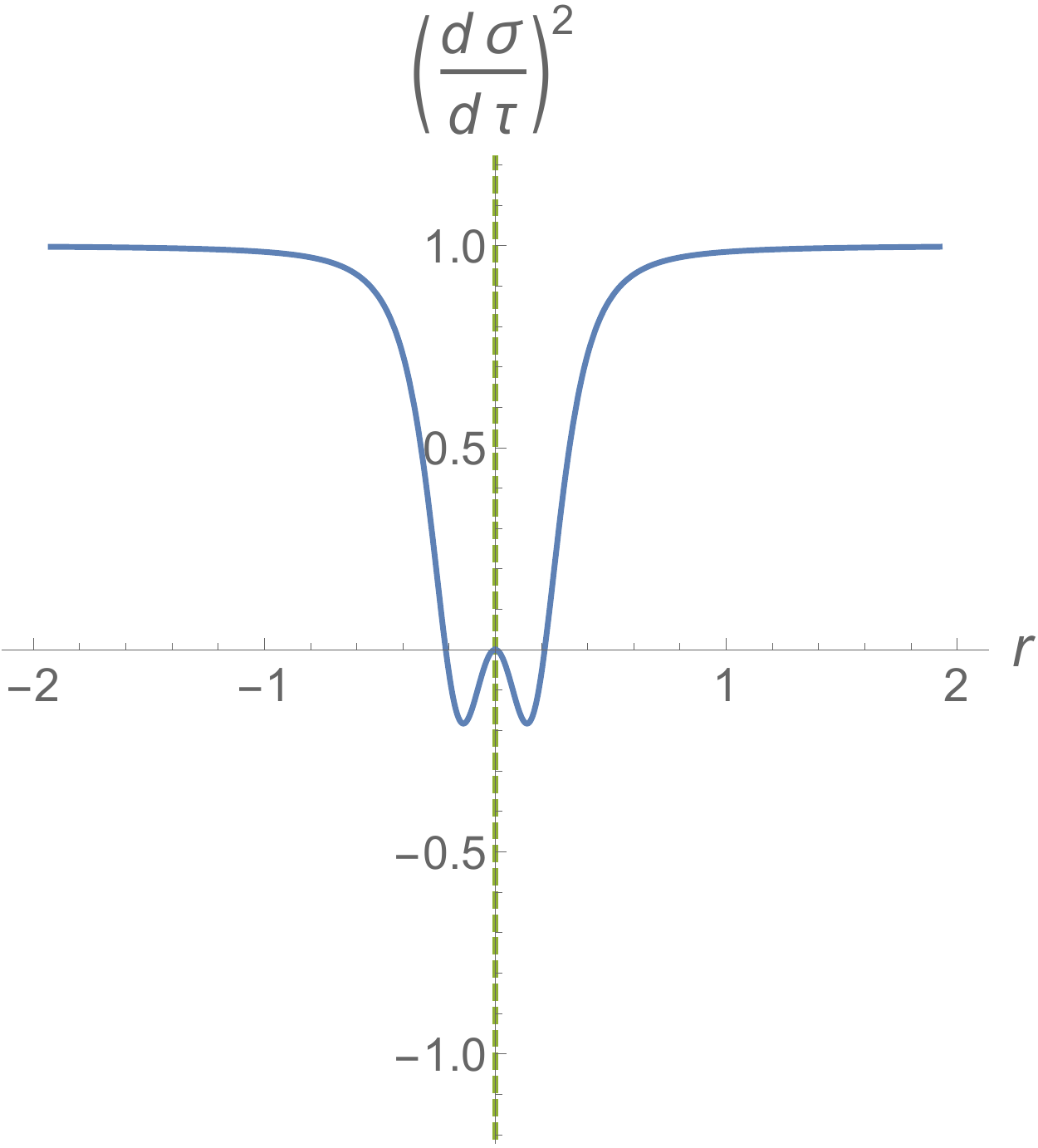}}
	\caption{Plot of the relation \eqref{eq:eigentime} in (a) and squared in (b) for equal masses $M_{BH} = M_{WH} = 5$. Obviously $\left(\dd \sigma / \dd \tau\right)^2$ becomes negative, which reflects that the matter surface trajectory becomes spacelike and non-physical. This restricts the possible choices of masses $M_{BH}$ and $M_{WH}$. Here it is $k = 0$, $M_{BH} = 5$, $M_{WH} = 5$, $\lambda_1 = \lambda_2  =\mathscr L_o=1$. The green dashed line represents the transition surface.}
	\label{fig:PSsigmaSym}
\end{figure}

Recall that in the case of $k=0$, Eqs. \eqref{eq:equsr-s:r}-\eqref{eq:equsr-s:rho} are functions of $r_o$ only. Hence, it is possible to plot them as a function of $r_o$ as depicted in Fig. \ref{fig:PSSrho}-\ref{fig:PSsigma}.
Note that both $\frac{1}{S} \frac{\dd S}{\dd \sigma}$ and $\frac{1}{\rho_o} \der{\rho_o}{\sigma}$ are divergent at the transition surface. Nevertheless, $\der{r_o}{\sigma}$ remains finite and negative everywhere, as expected. Fig. \ref{fig:PSsigma} shows the coordinate transformation Eq. \eqref{eq:eigentime}. The transformation is smooth around the transition surface, although zero is passed. An important observation is, that for the chosen parameters the transformation is valid everywhere. A counterexample is the symmetric bounce $M_{BH} = M_{WH}$, where the transformation becomes negative (cfr. Fig. \ref{fig:PSsigmaSym}). This restricts the possible relations between $M_{BH}$ and $M_{WH}$. A detailed analysis is complicated, and as it is not the main purpose of this paper, left for future investigations.

\begin{figure}[h!]
	\centering
	\subfigure[]
	{\includegraphics[width=6.8cm]{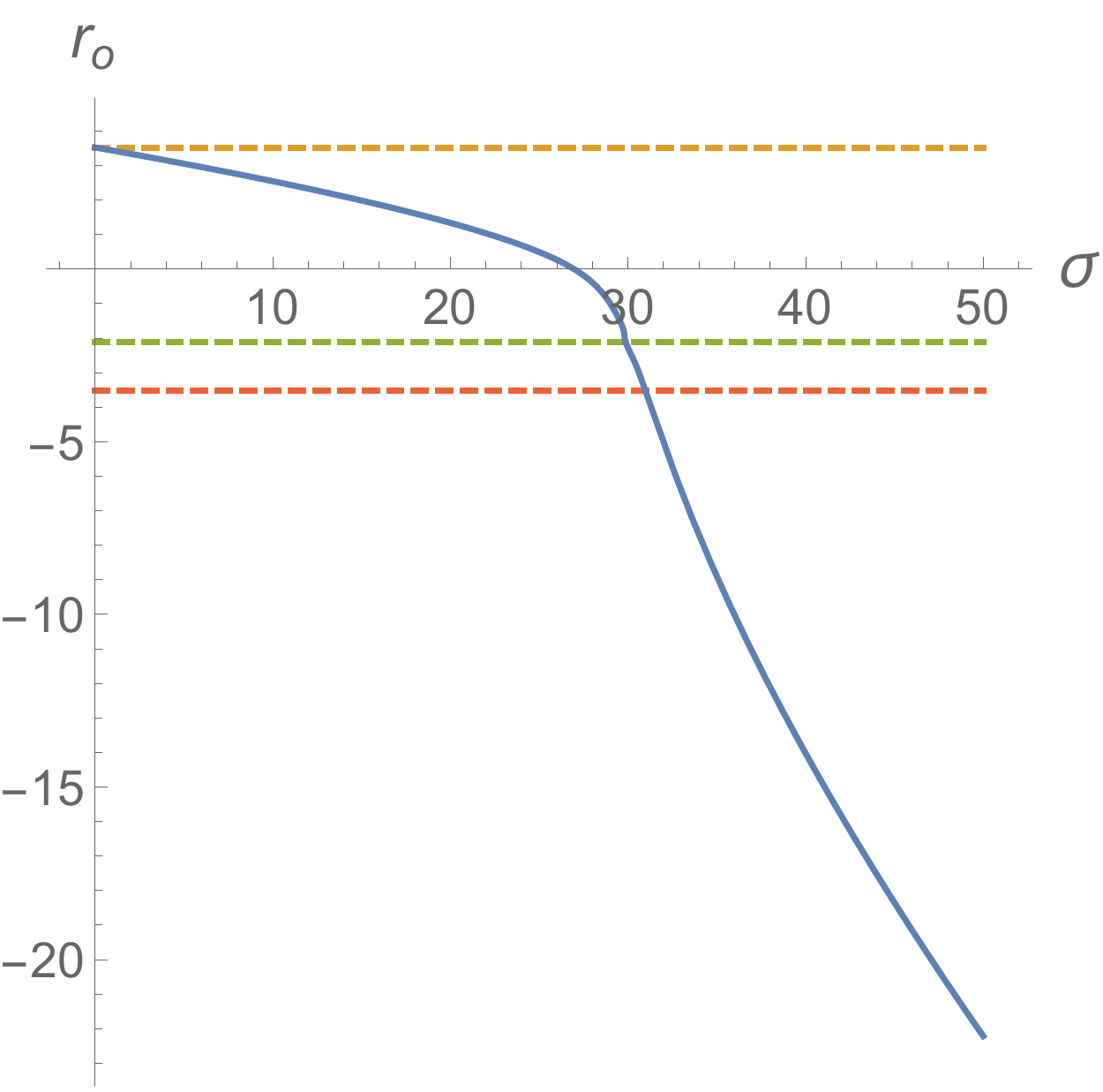}}
	\hspace{2mm}
	\subfigure[]
	{\includegraphics[width=6.8cm]{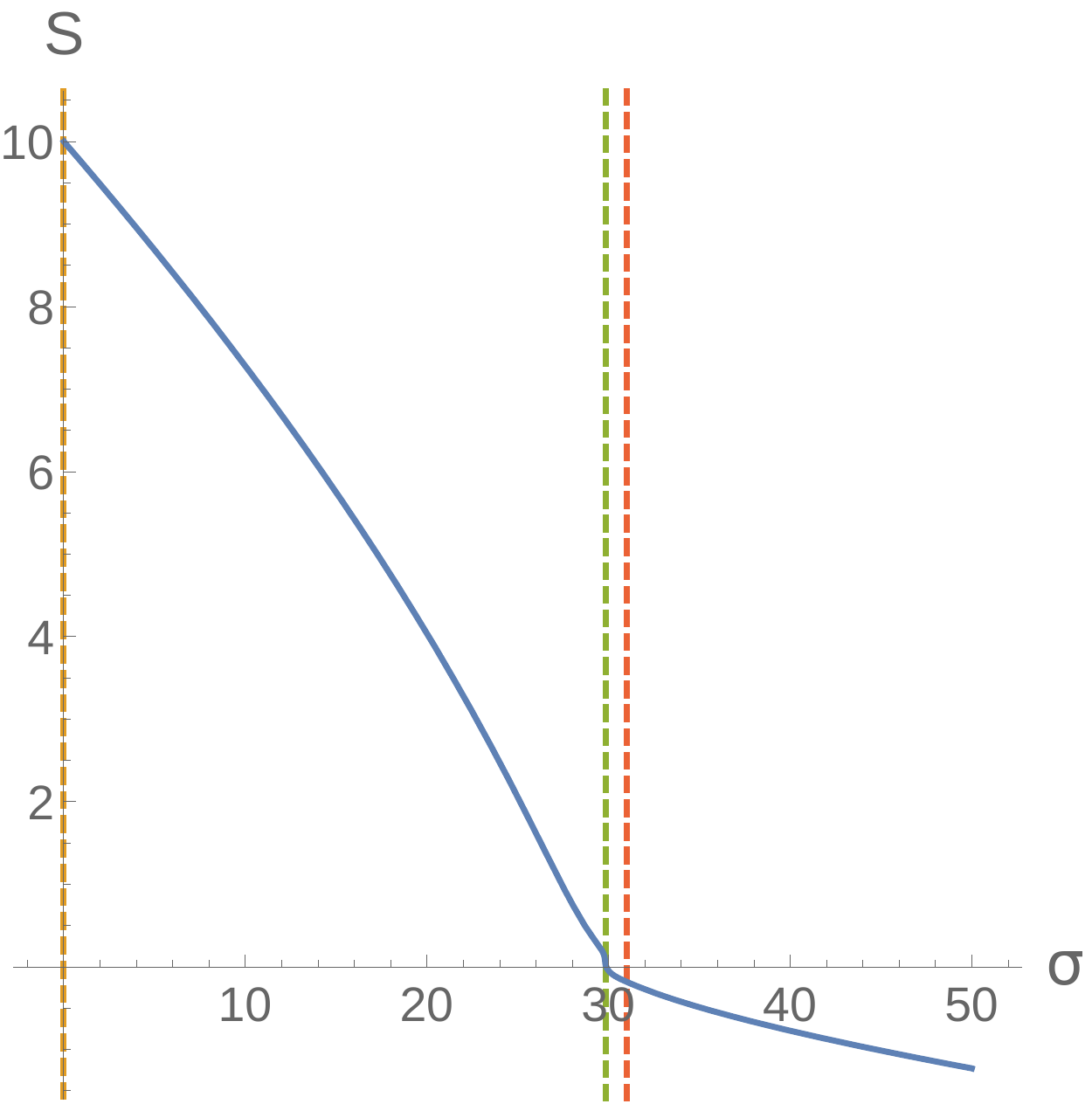}}
	\caption{Solutions of Eqs.~\eqref{eq:equsr-s:r} and \eqref{eq:equsr-s:S} for $r(\sigma)$ in (a) and $S(\sigma)$ in (b). As discussed in the main text $S$ crosses zero at the transition surface. In (a) the yellow dashed line corresponds to the $r$-value of the black hole horizon, the red dashed line to the white hole horizon respectively, and the green dashed line to the transition surface. In (b) the times $\sigma$ are indicated at which the black hole horizon (yellow dashed line), the transition surface (green dashed line), and the white hole horizon (red dashed line) are crossed. The parameters are $k = 0$, $M_{BH} = 20$, $M_{WH} = 1$, $\lambda_1 = \lambda_2  =\mathscr L_o=1$.}
	\label{fig:solr+S}
\end{figure}
\begin{figure}[h!]
	\centering
	\subfigure[]
	{\includegraphics[width=6.85cm]{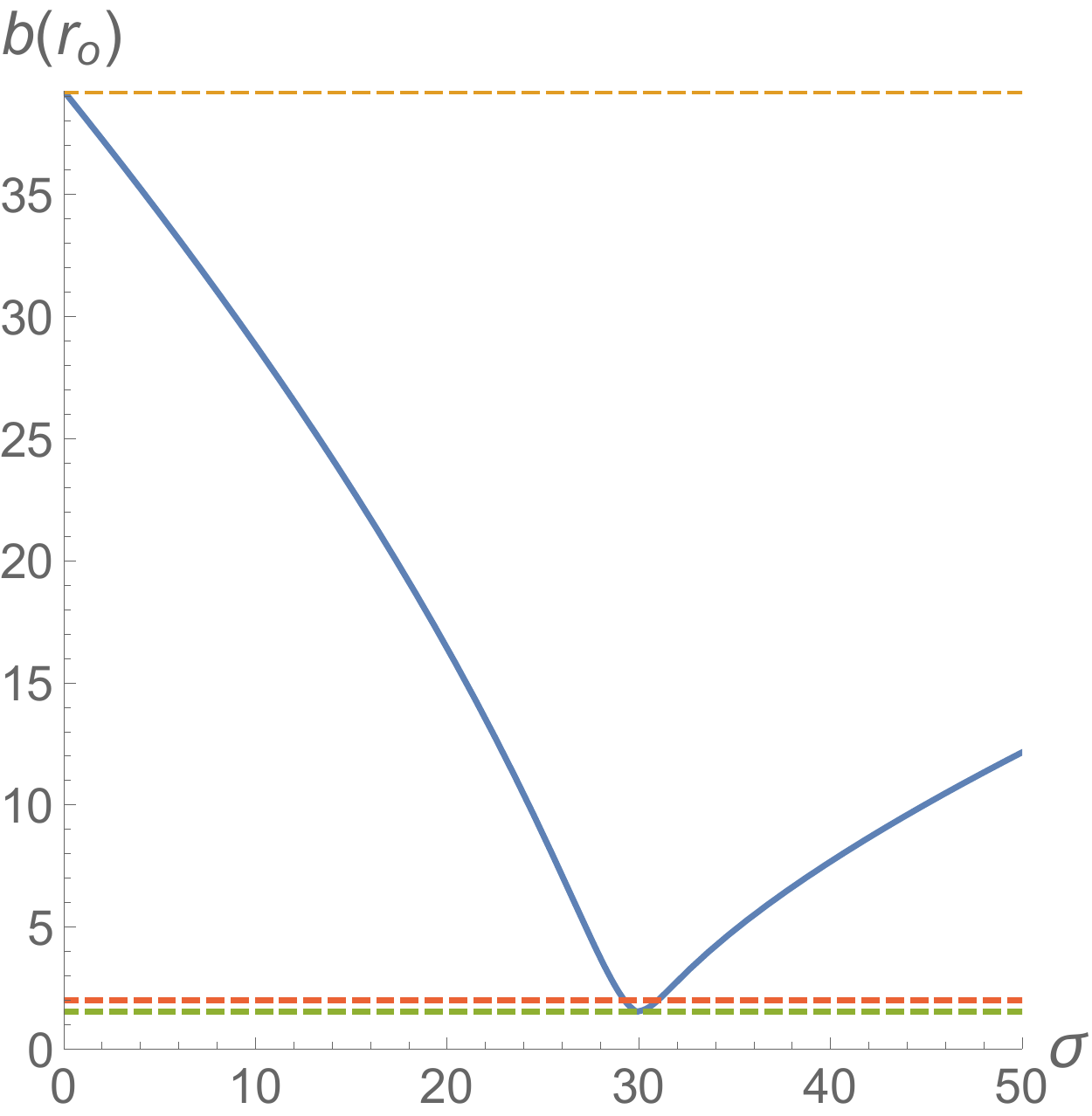}}
	\hspace{2mm}
	\subfigure[]
	{\includegraphics[width=6.85cm]{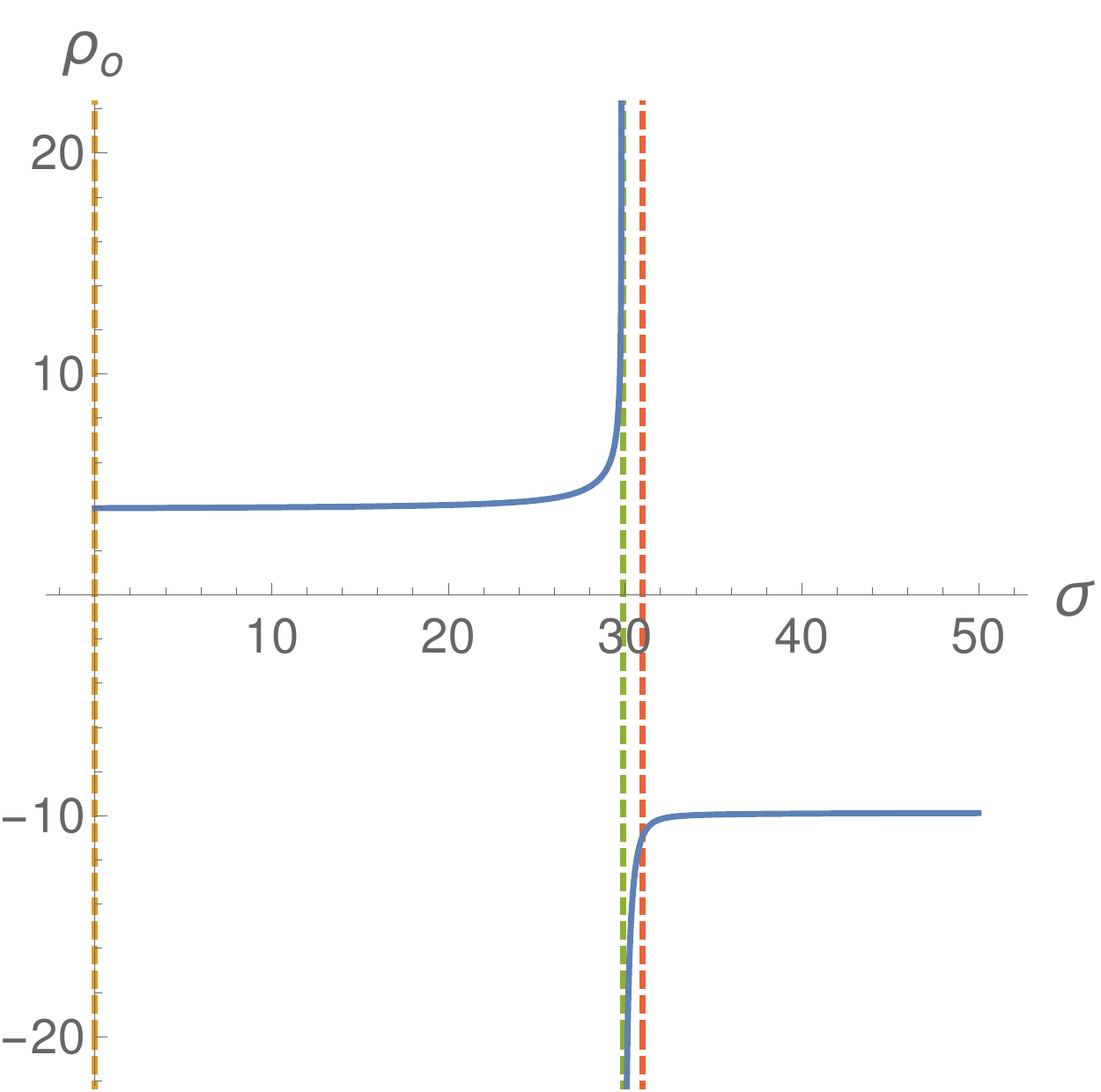}}
	\caption{The solutions for $R(\sigma)=b(r_o(\sigma))$ in (a) and $\rho_o(\sigma)$ (b) can be reconstructed out of $r_o(\sigma)$ and $S(\sigma)$. The bouncing behaviour of the collapsing shell is nicely visible in (a). As $S(\sigma_{\mathcal T}) = 0$, $\rho_o$ has to diverge at the transition surface to yield a finite radius $R(\sigma_{\mathcal{ T}}) = b_{\mathcal{ T}}$, see (b). In (a) the yellow dashed line corresponds to the $b$-value of the black hole horizon, the red dashed line to the white hole horizon respectively, and the green dashed line to the transition surface. In (b) the times $\sigma$ are indicated at which the black hole horizon (yellow dashed line), the transition surface (green dashed line), and the white hole horizon (red dashed line) are crossed. The parameters are $k = 0$, $M_{BH} = 20$, $M_{WH} = 1$, $\lambda_1 = \lambda_2  =\mathscr L_o=1$.}
	\label{fig:solR+rho}
\end{figure}
\newpage 
Solving Eqs. \eqref{eq:equsr-s:r} and \eqref{eq:equsr-s:S} numerically leads to the result depicted in Fig. \ref{fig:solr+S}.
Out of this the functions $R(\sigma) = b(r_o(\sigma))$ and $\rho_o(\sigma)$ can be reconstructed, see Fig. \ref{fig:solR+rho}. Note that $S(\sigma)$ passes zero and becomes negative afterwards. This corresponds to a change of the orientation of the chosen reference frame. Compatibly also $\rho_o(\sigma)$ changes its sign at the transition surface.
As $S$ vanishes and $\rho_o$ diverges at the transition surface, the quantities $\frac{1}{S} \der{S}{\sigma}$ and $\frac{1}{\rho_o} \der{\rho_o}{\sigma}$ diverge as depicted in Fig. \ref{fig:PSSrho}. Nevertheless, the derivatives $\der{S}{\sigma}$ and $\der{b(r_o)}{\sigma}$ remain finite at the transition surface. As the only physically relevant quantity is $b(r_o)$, divergences in $S$ and $\rho_o$ are not relevant. Although $S$ vanishes, the Ricci curvature remains finite  everywhere (cfr. Eq. \eqref{eq:Riccitrans}).

An important next step is to find a chart, which describes black and white hole side simultaneously (at least in a neighbourhood of the transition surface). As argued above (cfr. Eq. \eqref{eq:sigmaoftauatT}) in $\tau$-time it takes infinitely long to reach the transition surface. Hence, the metric in the form Eq. \eqref{eq:metricint} and $\tau, \rho$ coordinates can only cover one half of the process. Therefore let us introduce the new coordinates $\sigma(\tau)$ given by Eq. \eqref{eq:eigentime}, which is equivalently given by (for $k=0$)

\begin{align}\label{eq:tauofsigma}
\tau = \int_{\sigma_{ref}}^{\sigma} \text{sign}(r_o(\sigma')-r_{\mathcal T}) \sqrt{1+S^2(\sigma') \der{\rho_o}{\sigma}^2(\sigma')} \dd\sigma' \quad , \quad \dd\sigma = \pm\sqrt{1- S^2 \der{\rho_o}{\tau}^2} \dd\tau \;,
\end{align}

\noindent
and $\mathcal R(\sigma,\rho)$ defined by

\begin{align}\label{eq:defR}
\mathcal R = S(\sigma) \cdot \rho \quad ,\quad \dd\mathcal R = \frac{\dd S}{\dd \sigma} \rho \dd \sigma + S(\sigma) \dd \rho\;.
\end{align}

As shown in Fig. \ref{fig:PSsigma} the sign change in Eq. \eqref{eq:tauofsigma} is needed to make $\der{\sigma}{\tau}$ a differentiable function. This is no necessary requirement, but simplifies the equations. The new coordinate $\mathcal R$ can be viewed as the physical distance from the centre of the dust distribution.
The surface $\mathcal{R} = b(r_o(\sigma))$, i.e. $\rho = \rho_o(\sigma)$, describes the matter surface.
Inserting this transformation into Eq. \eqref{eq:metricint} leads to 

\begin{equation}\label{eq:metricintboth}
\dd s^2 = -\left(1+ S^2(\sigma) \left(\der{\rho_o}{\sigma}\right)^2- \frac{\mathcal R^2}{S^2(\sigma)} \left(\der{S}{\sigma}\right)^2 \right)\dd\sigma^2 - 2 \frac{\mathcal R}{S(\sigma)} \der{S}{\sigma} \dd\sigma \dd \mathcal R + \dd \mathcal R^2 + \mathcal{R}^2 \dd \Omega_2^2 \;.
\end{equation}

\noindent
This metric describes the full spacetime inside the dust cloud on both the black and white hole side. 
Each point in the interior can uniquely be described by its areal radius $\mathcal R$ and the instance of matter surface eigentime $\sigma$.
An important check is now if this metric is at least $C^1(M)$ across the transition surface. To check this, we compute the induced metric and the extrinsic curvature of the $\sigma = const.$-surfaces, leading to

\begin{align}
\left.\dd s \right|_{\sigma = const.} =&\, \dd \mathcal R^2 + \mathcal R^2 \dd \Omega_2^2 \;, \label{eq:dssigma}
\\
\left. K \right|_{\sigma=const.} =& -\der{S}{\sigma} \frac{1}{\sqrt{S^2+ S^4 \left(\der{\rho_o}{\sigma}\right)^2}}\left(\dd \mathcal R^2 + \mathcal R^2 \dd \Omega_2^2\right)\;.\label{eq:Ksigma}
\end{align}

\begin{figure}[t!]
	\centering
	\subfigure[]
	{\includegraphics[width=7.5cm]{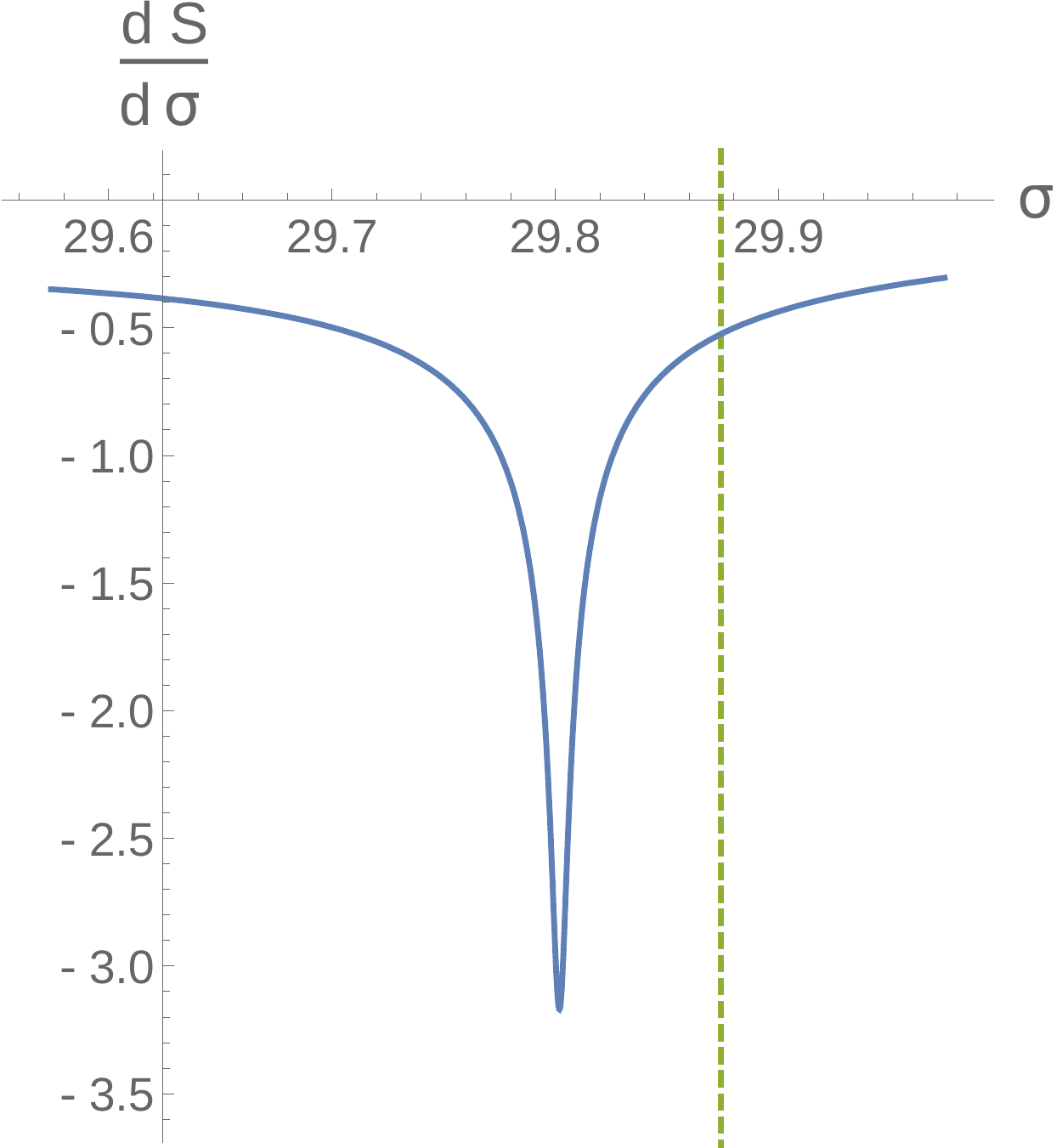}}
	\hspace{2mm}
	\subfigure[]
	{\includegraphics[width=7.5cm]{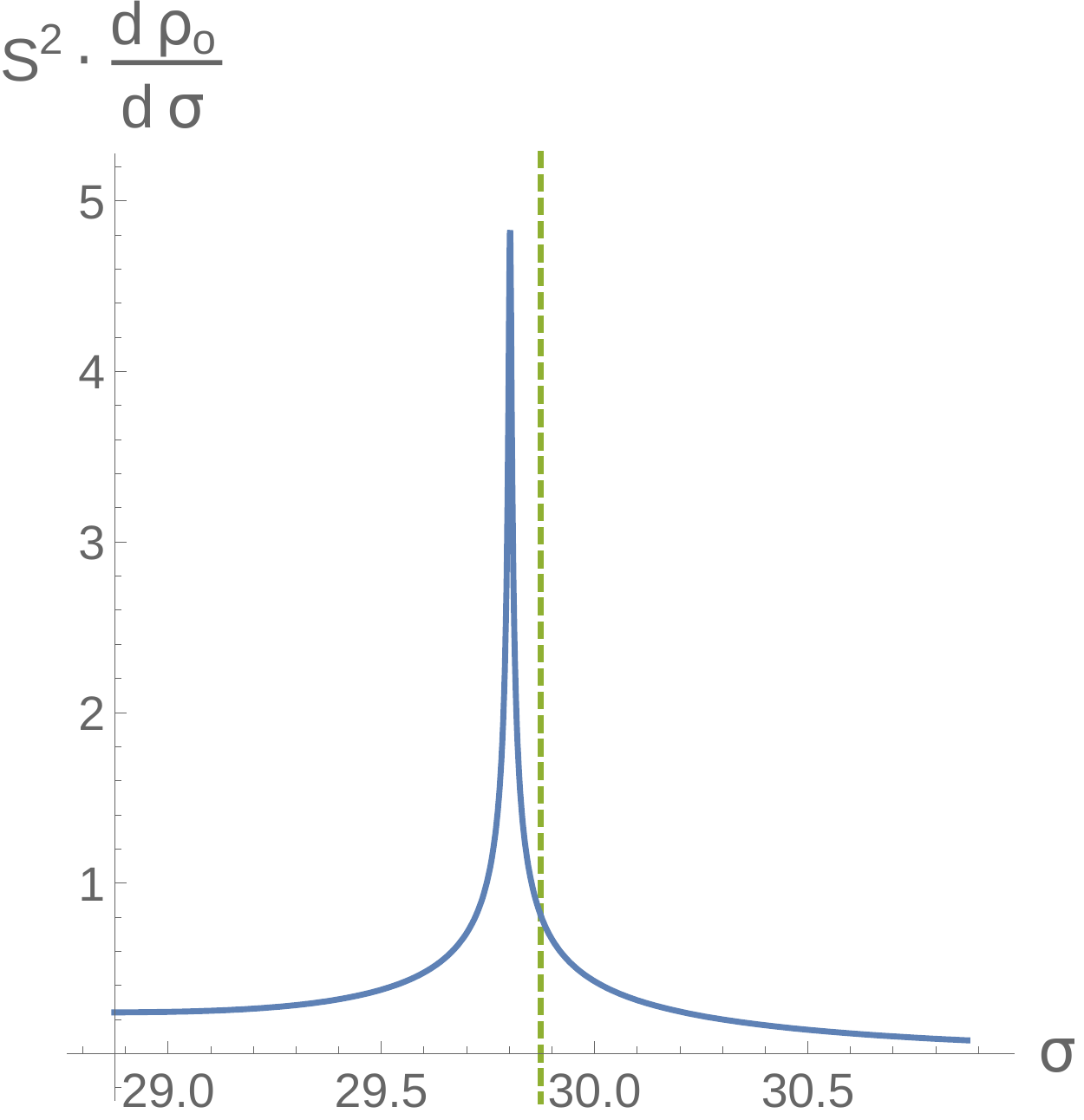}}
	\caption{The quantities $\dd S/\dd \sigma $ (a) and $S^2\cdot\dd \rho_o/\dd \sigma $ (b) stay finite at the transition surface (green dashed line). Paramters are $k = 0$, $M_{BH} = 20$, $M_{WH} = 1$, $\lambda_1 = \lambda_2  =\mathscr L_o=1$.}
	\label{fig:Ssqrhoo}
\end{figure}

\noindent
The induced metric \eqref{eq:dssigma} is continuous at the transition surface.
Also the extrinsic curvature is continuous for $\sigma \rightarrow \sigma_{\mathcal T}$ as $\der{S}{\sigma}$  and $S^2 + S^4 \left(\der{\rho_o}{\sigma}\right)^2$ are both continuous and finite at the transition surface (see Fig. \ref{fig:Ssqrhoo}).
This can also be seen by rewriting the global factor in Eq. \eqref{eq:Ksigma} as $\pm \frac{1}{S} \der{S}{\tau}$ by using Eq. \eqref{eq:tauofsigma}.
This is continuous at the transition surface as Eq. \eqref{eq:dSexp} shows.

\begin{figure}[t!]
	\centering
	\subfigure[]
	{\includegraphics[width=7.5cm]{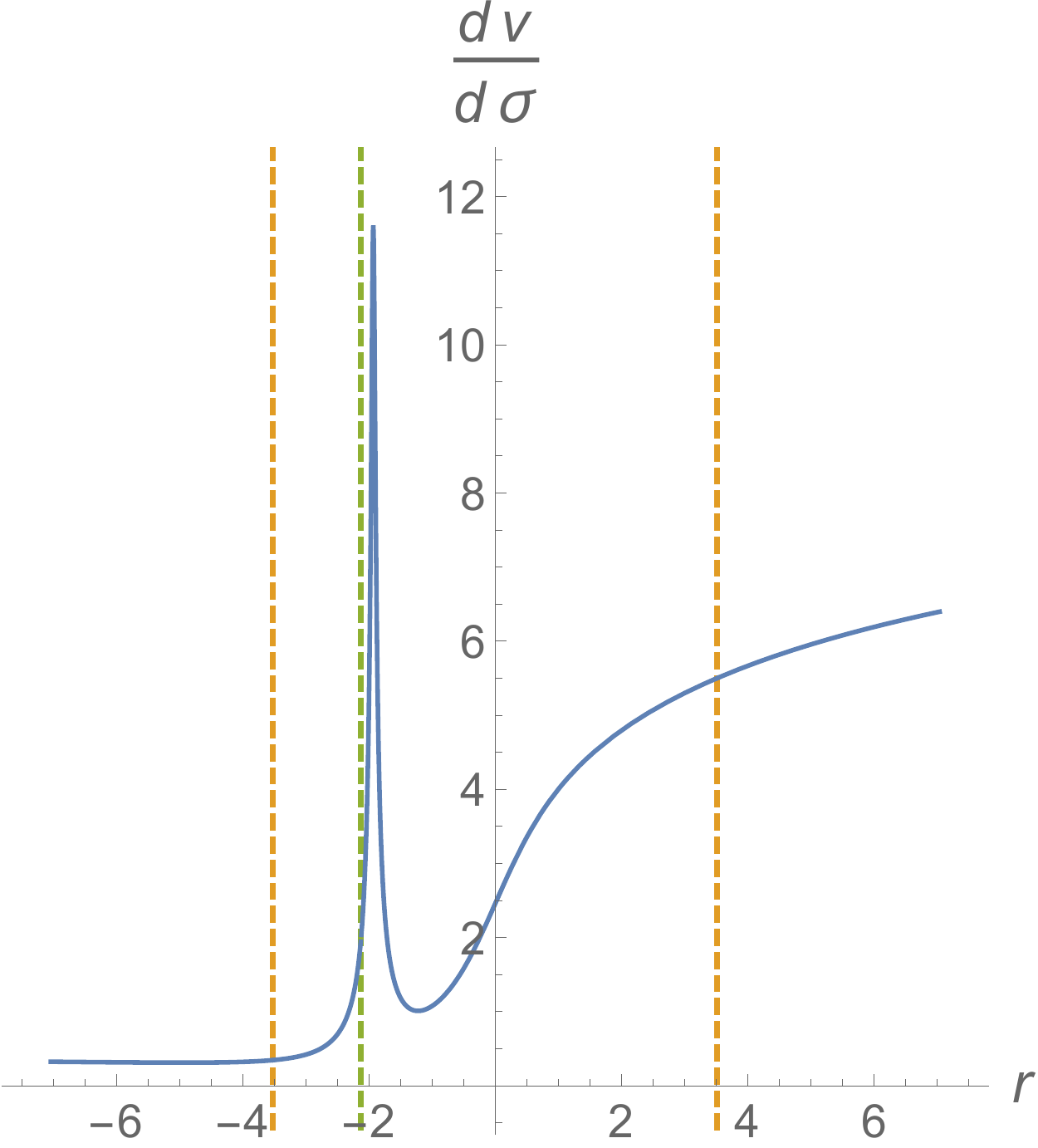}}
	\hspace{2mm}
	\subfigure[]
	{\includegraphics[width=7.5cm]{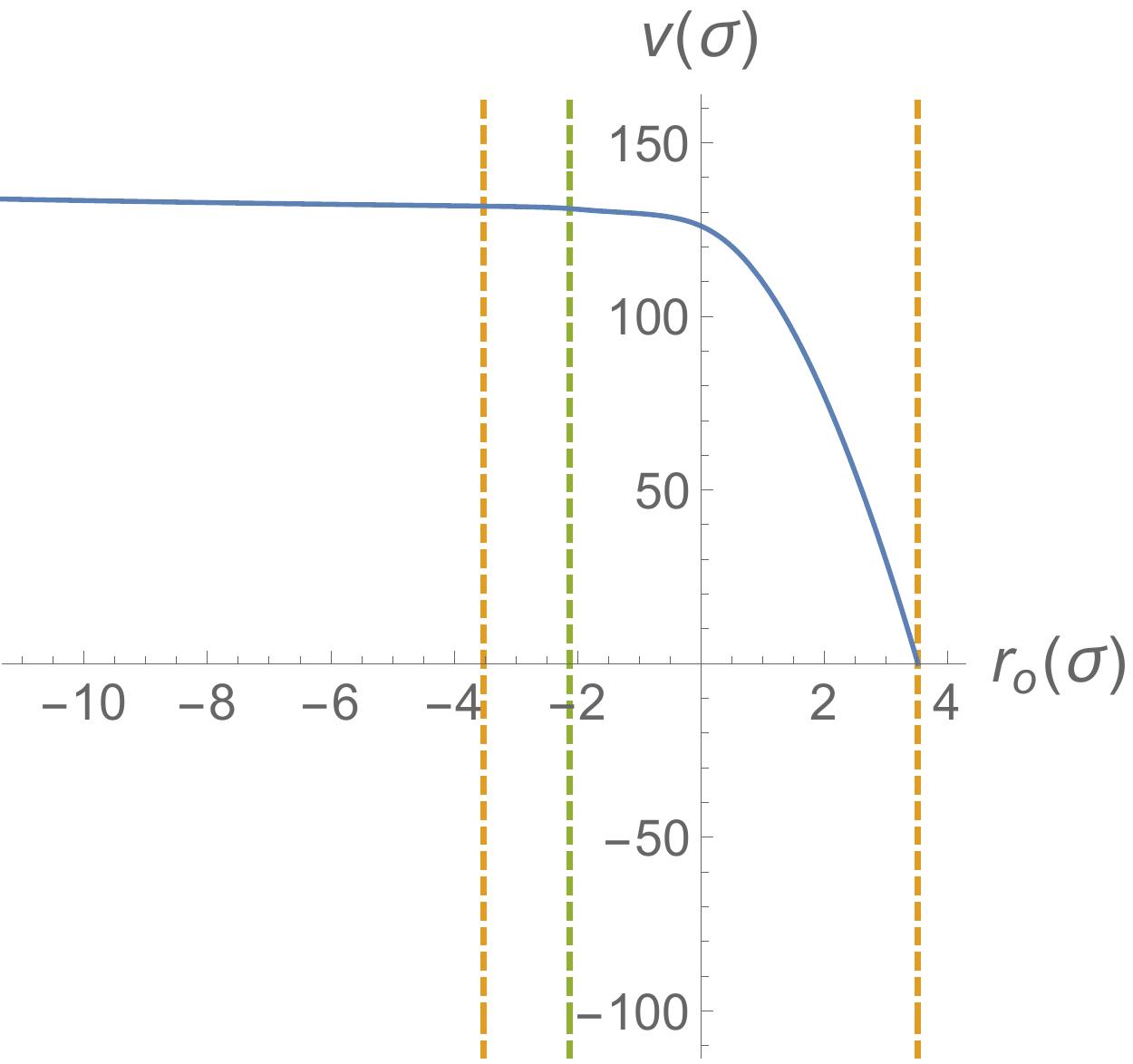}}
	\caption{$\der{v}{\sigma}$ as a function of $r$ in (a) and the solution $v(\sigma)$ in a parametric plot against $r_o(\sigma)$ in (b). The function is finite everywhere and invertible. The parameters are $k = 0$, $M_{BH} = 20$, $M_{WH} = 1$, $\lambda_1 = \lambda_2  =\mathscr L_o=1$. The yellow dashed lines correspond to the horizons, green dashed line represents the transition surface.}
	\label{fig:vofsigma}
\end{figure}

Let us now move to the vacuum observer point of view. The last missing piece is the solution for $v(\sigma)$, which can be obtained from integrating Eq. \eqref{eq:vofsigma} and is depicted in Fig. \ref{fig:vofsigma}.
As argued above it is important to notice, that $v(\sigma)$ remains finite everywhere. From the fact that $v$ is finite at the white hole horizon shows that the collapsing shell is moving from region $\mathtt{I}$ to region $\mathtt{III}$ in Fig. \ref{Penrosediag2}.
With this result the spacetime is completely determined from the matter and vacuum observer point of view.

At this point, it is also possible to make a rough estimation on the collapse time scale.
For this assume that $R(\tau_o) = R_o$ is a certain reference radius at which the clock starts to count.
The time to reach the transition surface is given by

\begin{equation}
\sigma_{\mathcal{ T}} = \int_{\tau_o}^{\infty} \sqrt{1-S^2 \der{\rho_o}{\tau}^2}\, \dd \tau \,.
\end{equation}

\noindent
Assume now further that $\tilde{\tau}$ is a time in the regime where quantum effects become relevant but are not yet dominant.
Therefore, the above integral can be split as follows without making large errors:\footnote{The classical result can be found for example in \cite{fliessbach12}.}

\begin{align}
\sigma_{\mathcal{ T}} \approx& \underbrace{\int_{\tau_o}^{\tilde{\tau}} \,\dd \tau}_{\text{class.}} + \underbrace{\int_{\tilde{\tau}}^{\infty} \sqrt{1-S^2 \der{\rho_o}{\tau}^2}\,\dd \tau}_{\text{quant.}}
\notag
\\
\stackrel{\text{Eq.~\eqref{eq:sigmaoftauatT}}}{\approx}& \frac{\pi}{2} \sqrt{\frac{R_o^3}{2 M_{BH}}} + b_{\mathcal T} \sqrt{2 \left( a_{\mathcal{ T}} b''_\mathcal{ T} +\frac{2}{b_\mathcal{ T}}\right)\left(R(\tilde{\tau})-b_\mathcal{ T}\right) }
\end{align}

\noindent
The estimation works analogously for the white hole side and therefore the full collapse time from $R_o^{BH}$ on the black hole side to $R_o^{WH}$ on the white hole side is

\begin{equation}\label{eq:collapsetime1}
\sigma_{\text{collapse}} \approx \frac{\pi}{2} \sqrt{\frac{\left(R_o^{BH}\right)^3}{2 M_{BH}}} + \frac{\pi}{2} \sqrt{\frac{\left(R_o^{WH}\right)^3}{2 M_{WH}}} + 2b_{\mathcal T} \sqrt{2 \left( a_{\mathcal{ T}} b''_\mathcal{ T} +\frac{2}{b_\mathcal{ T}}\right)\left(R(\tilde{\tau})-b_\mathcal{ T}\right)}.
\end{equation}

\noindent
When choosing $R_o^{BH} \propto M_{BH}$ and $R_o^{WH} \propto M_{WH}$ and neglecting the quantum contribution, this gives

\begin{equation}\label{eq:collapsetime2}
\sigma_{\text{collapse}} \propto M_{BH} + M_{WH} \;.
\end{equation}

\noindent
Note that, as was argued above, the two masses are not completely independent due to the demand that the shell remains timelike everywhere.
For the choice of parameters used in this section it is furthermore $M_{BH} \gg M_{WH} \sim M_{\text{Planck}}$.

\subsection{Penrose Diagram and Global Structure}\label{sec:PenroseDiag}

With all these computations done, we can construct the Penrose diagram for the spacetime in the matter region (inside the dust cloud) and for the vacuum region in the following.

\subsubsection{Matter Region}

Before constructing the Penrose diagram, there are some general remarks on the interior metric Eq.~\eqref{eq:metricintboth} in order.
Although this metric is $C^1(M)$ for $\sigma = const.$ slices, the metric coefficients $g_{\sigma \sigma}$ and $g_{\sigma \mathcal R}$ diverge.
Passing for example along $\mathcal R = const.$ trough the transition surface, takes infinite time as $g_{\sigma \sigma}$ is singular.
This is avoided by the surface of the matter satisfying $\mathcal R = S(\sigma) \rho_o(\sigma)$ and thus moves along a trajectory, which becomes null at $\sigma_{\mathcal T}$.
Due to this, it is only possible to pass the transition surface in finite time if the speed of light is approached fast enough.
As the surface of the matter behaves like this and due to the homogeneity assumption, it seems plausible that also the other (smaller) dust shells show this behaviour.
Therefore, a freely falling observer should be dragged along this ``matter flow'' and pass trough the transition surface.
This is only speculative at this stage of the model, as no precise matter model underlies this matching strategy.
Presumably, repeating this analysis with a clear interior picture, in terms of a stress-energy tensor would shed light into this behaviour.
Another subtle property is that the eigentime of a curve can only remain finite, if the trajectory satisfies $\mathcal R(\sigma_{\mathcal T}) = b_{\mathcal{ T}}$.
As a consequence, if the whole cloud of collapsing dust passes the transition surface, all of it has to be localised at $\mathcal R = b_\mathcal{ T}$ at time $\sigma_{\mathcal T}$.
This would correspond to a shell-crossing singularity.
Again, the model is too minimal to make computations and the exact behaviour cannot be described.

For constructing the Penrose diagram in the interior of the dust cloud, let us start with the metric in $(\sigma,\mathcal R)$-coordinates as in Eq. \eqref{eq:metricintboth}. 
Radial lightlike geodesics satisfy 

\begin{align}
0 =&  -\left(1+ S^2(\sigma) \left(\der{\rho_o}{\sigma}\right)^2- \frac{\mathcal R^2}{S^2(\sigma)} \left(\der{S}{\sigma}\right)^2 \right)\dd\sigma^2 - 2 \frac{\mathcal R}{S(\sigma)} \der{S}{\sigma} \dd\sigma \dd \mathcal R + \dd \mathcal R^2 
\notag
\\
&\Leftrightarrow
\notag
\\
\der{\mathcal R}{\sigma} =& \frac{\mathcal R \der{S}{\sigma} \pm \sqrt{S^2 + S^4 \left(\der{\rho_o}{\sigma}\right)^2}}{S} \;.
\label{eq:LLgeoseq}
\end{align}

\begin{figure}[t!]
	\centering\includegraphics[width=7.5cm]{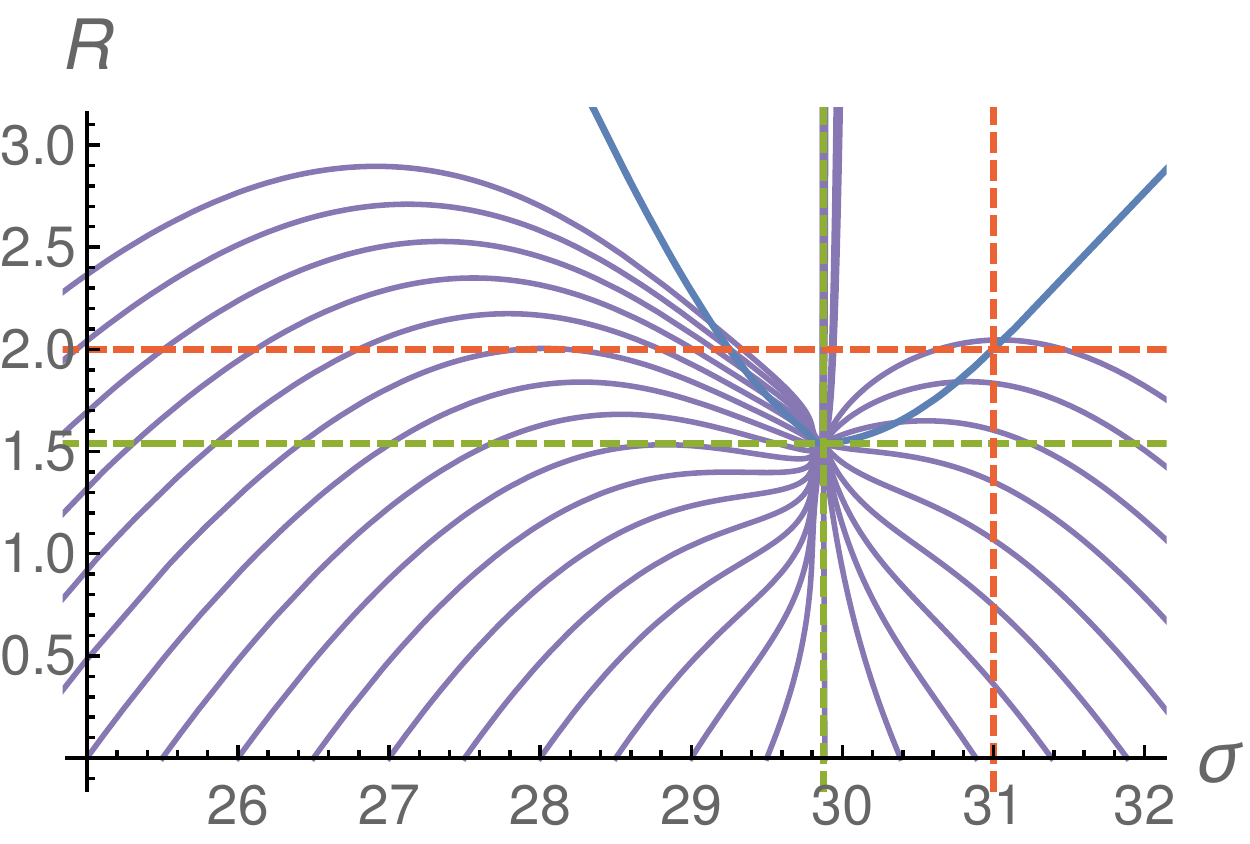}
	\caption{Outgoing radial lightlike geodesics in the vicinity of the transition surface. The geodesics are only valid up to the blue line, which depicts the surface of the matter. From there onwards their evolution is determined by the exterior metric. All geodesics pass through $\mathcal R = b_\mathcal{ T}$ at $\sigma = \sigma_{\mathcal T}$.
	Parameters are $k = 0$, $M_{BH} = 20$, $M_{WH} = 1$, $\lambda_1 = \lambda_2  =\mathscr L_o=1$.}
	\label{fig:LLgeos}
\end{figure}

Note that this equation is in general divergent at the transition surface.
While $\der{S}{\sigma}$ and $S^2 \der{\rho_o}{\sigma}$ are finite there, $S$ vanishes and causes a divergence.
The derivative  remains finite only for one particular geodesic with $\mathcal R(\sigma_{\mathcal T}) = b_\mathcal{ T}$ at $\sigma_{\mathcal T}$ and the negative sign.
It is possible to write a formal solution to this equation, given by

\begin{equation}
R(\sigma) = R_i \frac{S(\sigma)}{S(\sigma_i)} \pm S(\sigma) \left(\eta(\sigma)-\eta(\sigma_i)\right) \;,
\end{equation}

\noindent
where $\eta$ is the conformal time defined by

\begin{equation}\label{eq:defeta}
\eta = \int_{\tau_{ref}}^{\tau} \,\frac{\dd \tau'}{S(\tau')} = \pm \int_{\sigma_{ref}}^{\sigma} \,\frac{\sqrt{1+ S^2 \left(\der{\rho_o}{\sigma}\right)^2}}{S}  \, \dd \sigma' \quad , \quad \dd\eta = \frac{\sqrt{1+ S^2 \left(\der{\rho_o}{\sigma}\right)^2}}{S} \dd \sigma \;,
\end{equation}

\noindent
and $\sigma_i$ is the initial time at which the initial radius $R_i$ is reached.
The reference times $\tau_{ref}$ and $\sigma_{ref}$ can be chosen arbitrarily.
The sign has to be chosen depending on which side (before or after the transition surface) $\sigma_{ref}$ is located.
Fig. \ref{fig:LLgeos} shows the result of the explicit numerical integration.
It is sufficient to study the outgoing geodesics ($+$ sign in Eq. \eqref{eq:LLgeoseq}), as the ingoing light rays hit $\mathcal{R} = 0$ and turn into outgoing ones, only with a parity change in the angles of the $\mathbb{S}^2$ part.
The geodesics are only valid up to the matter surface $\mathcal R = b(r_o(\sigma))$ and have to be continued from there on by means of the exterior geodesic equation.
Note that $\eta(\sigma)$ diverges for $\sigma \rightarrow \sigma_{\mathcal T}$.
According to Eq. \eqref{eq:dSexp}, it is $S(\tau) = S_o \exp(-(\tau-\tau_o)/b_\mathcal{ T})$ in the vicinity of the transition surface, i.e. $\tau,\tau_o \gg 1$.
Generically it is

$$
S(\tau)\cdot\eta(\tau) = S(\tau)\cdot \int_{\tau_{ref}}^{\tau} \, \frac{\dd \tau'}{S(\tau')} \simeq S(\tau)\cdot \underbrace{\int_{\tau_{ref}}^{\tau_o} \, \frac{\dd \tau'}{S(\tau')}}_{\text{finite}}  + S(\tau)\cdot \int_{\tau_o}^{\tau} \, \frac{e^{\frac{\tau'-\tau_o}{b_\mathcal{ T}}}}{S_o}\dd \tau' \longrightarrow b_\mathcal{ T}. 
$$
Due to this (and $S(\sigma_{\mathcal{ T}}) = 0$), all geodesics go trough $b_\mathcal{ T}$ at $\sigma_{\mathcal T}$ and form a caustic.
There is one particular geodesic for which $\der{R}{\sigma} = 0$ at $\sigma = \sigma_{\mathcal T}$.
All earlier geodesics reach the surface of the matter before the transition surface.
This limiting geodesic remains trapped inside the dust.
All later geodesics hit the matter surface exactly at the transition surface, i.e. $\mathcal R = b_\mathcal{ T}$ at $\sigma = \sigma_{\mathcal T}$.
There is an additional geodesic satisfying $\sigma = const. = \sigma_{\mathcal T}$.
The properties worked out in this analysis should be represented in the Penrose diagram. 

For the construction of the Penrose diagram, let us introduce in- and out-going lightcone coordinates by 

\begin{align}
u &= \frac{\mathcal R}{S(\sigma)} -\eta(\sigma)\;,
\\
v &= \frac{\mathcal R}{S(\sigma)} + \eta(\sigma)\;.
\end{align}

\noindent
Outgoing light rays are described by $u = const.$ while ingoing ones are represented by $v = const.$
Note that the involved integral in $\eta$ diverges going across the transition surface.
For $\sigma_{ref} < \sigma_{\mathcal T}$ in Eq. \eqref{eq:defeta}, $u,v \rightarrow \pm\infty$ at the transition surface.
This shows that, depending how $\sigma_{ref}$ is chosen, the chart $(u,v)$ only covers the black hole ($\sigma_{ref} < \sigma_{\mathcal T}$) or the white hole side ($\sigma_{ref} > \sigma_{\mathcal T}$), respectively.
Due to this, it is only possible to represent the interior spacetime by means of two different Penrose diagrams.
Spatial and temporal coordinates can be reconstructed out of $u$ and $v$ by adding and subtracting them, i.e.

\begin{align}
\frac{v+u}{2} &= \frac{\mathcal{R}}{S(\sigma)} = \rho \;,
\\
\frac{v-u}{2} &= \eta(\sigma) \;,
\end{align}

From this, it is straight forward to extract constant $\rho$ and constant $\sigma$ surfaces in the Penrose diagram, which are simply given by 

\begin{align}
\frac{v+u}{2} &= const. \quad,\quad \text{$\rho$ constant} \;,
\\
\frac{v-u}{2} & = const. \quad , \quad \text{$\sigma$ constant} \;.
\end{align}

\noindent
The lightcone coordinates can be compactified by means of 

\begin{align}
\tilde{u} =& \arctan\left(u\right) \;,\\
\tilde{v} =& \arctan\left(v\right) \;.
\end{align}

\noindent
The physical part of the Penrose diagram belongs to $\rho \in \left[0,\rho_o(\sigma)\right)$ on the black hole and $\rho \in \left(\rho_o(\sigma),0\right]$ on the white hole side (where $\rho_o(\sigma)$ is negative). We have the following characteristics for the Penrose diagram of the black hole side

\begin{itemize}
\item $\sigma = \sigma_{\mathcal T}$ corresponds to $\frac{v-u}{2} \rightarrow +\infty$, i.e. $\tilde{v} = \frac{\pi}{2}$ or $\tilde{u} = -\frac{\pi}{2}$.
\item $\sigma = -\infty$ corresponds to $\frac{v-u}{2} = -\infty$, i.e. $\tilde{v} = -\frac{\pi}{2}$ or $\tilde{u} = \frac{\pi}{2}$.
\item $\rho = 0$ is given by $u = - v$, i.e. $\tilde{u} = - \tilde{v}$.
\end{itemize}

Similarly, for the white hole side, we find that

\begin{itemize}
\item $\sigma = \sigma_{\mathcal T}$ corresponds to $\frac{v-u}{2} \rightarrow -\infty$, i.e. $\tilde{v} = -\frac{\pi}{2}$ or $\tilde{u} = \frac{\pi}{2}$.
\item $\sigma = +\infty$ corresponds to $\frac{v-u}{2} = +\infty$, i.e. $\tilde{v} = \frac{\pi}{2}$ or $\tilde{u} = -\frac{\pi}{2}$.
\item $\rho = 0$ is given by $u = - v$, i.e. $\tilde{u} = - \tilde{v}$.
\end{itemize}

\begin{figure}[t!]
	\centering
	\subfigure[]
	{\includegraphics[width=7.5cm]{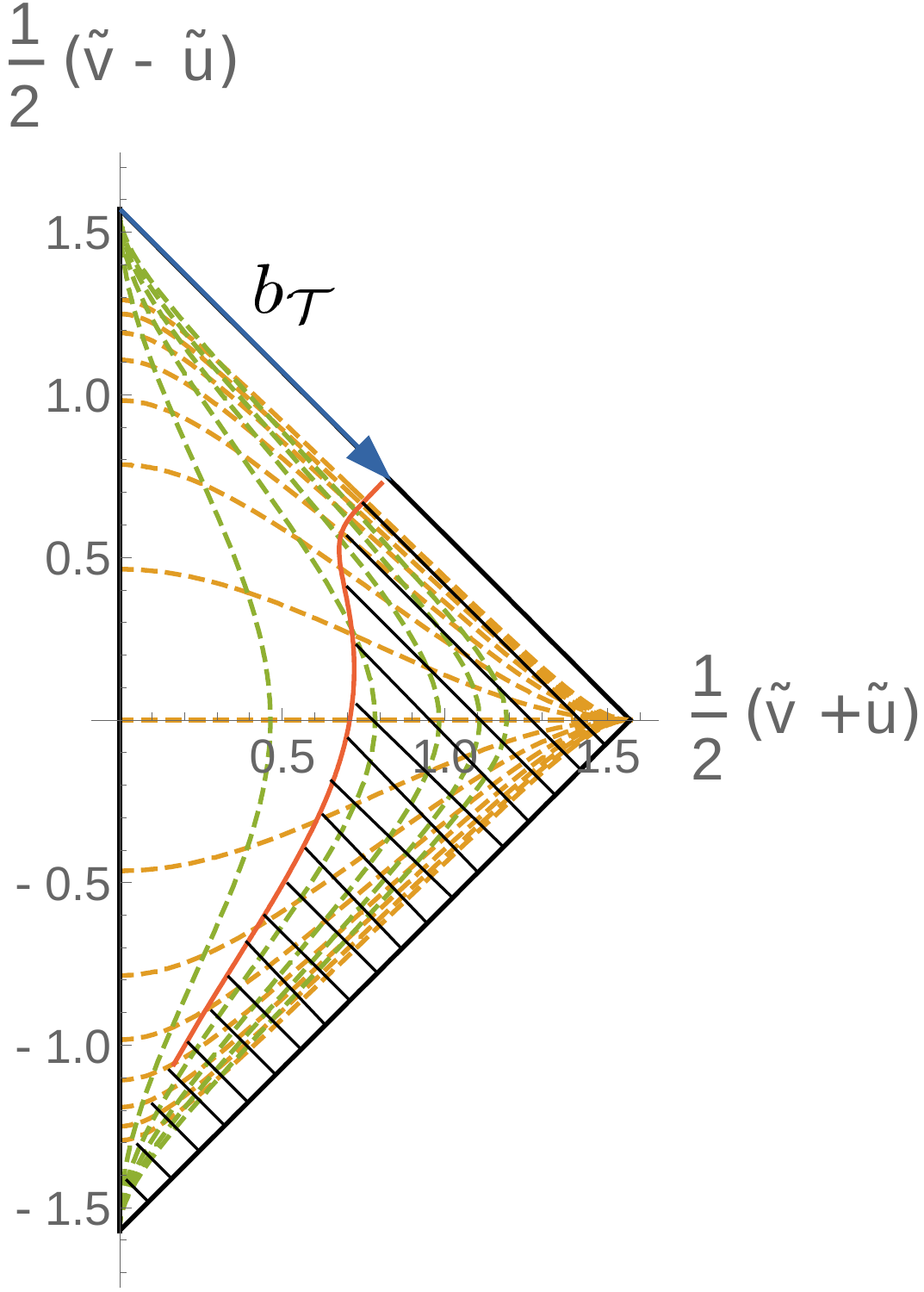}}
	\hspace{2mm}
	\subfigure[]
	{\includegraphics[width=7.5cm]{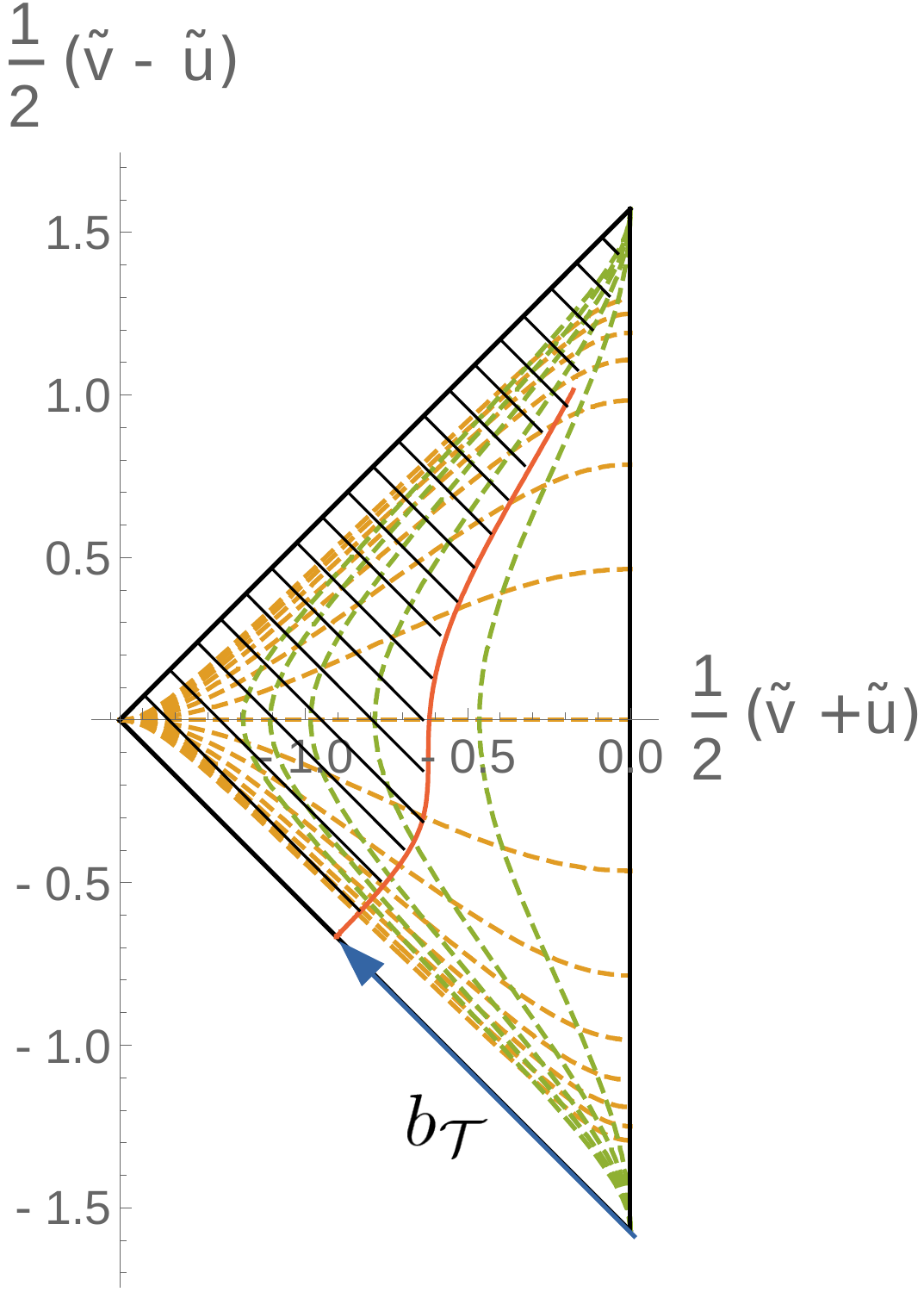}}
	\caption{Penrose diagram patches of the spacetime inside the dust. In (a) the black hole patch is shown, in (b) the white hole part, respectively. Green dashed lines correspond to $\rho = const.$, yellow dashed lines to $\eta = const.$ surfaces. The transition surface is located at the upper diagonal boundary in (a) and the lower one in (b), both times indicated by the blue arrow. The red lines correspond to the surface of the matter $\rho = \rho_o(\sigma)$. The red line reaches from the future/past timelike infinities $i^\pm$ to the transition surface. As the numerics stops at a finite value the red line does not reach $i^{\pm}$. The spacetime is only valid from $\rho = 0$ (vertical boundary of the diagram) up to this surface, the non-physical part is shaded out. Note that, the blue arrows in (a) and (b) need to be identified. The arrows point towards the outside of the matter. An identification needs to take the opposite orientations into account. 
	The parameters are $k = 0$, $M_{BH} = 20$, $M_{WH} = 1$, $\lambda_1 = \lambda_2  =\mathscr L_o=1$.}
	\label{fig:PDint}
\end{figure}

The matter surface is simply given by $v+u = 2 \rho(\sigma)$ and $v-u = 2 \eta(\sigma)$, which can be plotted numerically as parametrised curve.
Fig. \ref{fig:PDint} shows the two Penrose diagrams for both sides, including several constant $\sigma$, i.e. $\eta$, and $\rho$ surfaces. We used $\sigma_{ref} \approx \sigma_{\mathcal T} - 5$ for the black hole side and $\sigma_{ref} \approx \sigma_{\mathcal T} + 2$ for the white hole side, respectively.

The Penrose diagram shows nicely the features already discussed above.
Let us focus first on the black hole side Fig. \ref{fig:PDint} (a).
It shows that all light rays terminate at $\sigma = \sigma_{\mathcal T}$ at the transition surface.
Recall that outgoing light rays are described by $45^\circ$ lines, i.e. $\tilde{v}-\tilde{u} = (\tilde{v}+\tilde{u}) + const.$
It further shows that there is exactly one light ray, which is approached by $\rho = \rho_o(\sigma)$ and becomes tangent at the transition surface.
All light rays emitted earlier than this limiting one, reach the matter surface and escape the matter region.
Later light rays remain trapped inside the matter and have to reach the transition surface.
Timelike curves can leave the diagram only when they converge fast enough to a particular outgoing light ray.
The diagram makes obvious that this trajectory can only exit through the shell $\sigma = \sigma_{\mathcal T}$ and $\mathcal{R} = b_\mathcal{ T}$.
Similar considerations are true for the white hole side Fig. \ref{fig:PDint} (b), only time reversed.
Due to the general discussion in Sec. \ref{sec:application} the qualitative behaviour is not expected to change for other polymer models.

\subsubsection{Vacuum Region}

The Penrose diagram for the spacetime outside the dust was already constructed in \cite{BodendorferEffectiveQuantumExtended}. We use the same coordinates and add the collapsing shell into the picture. For this purpose, we define the coordinates 

\begin{equation}\label{KScoordinates}
T^2-X^2=\exp\left[\left(\left.\frac{\dd a}{\dd r}\right|_{r=r_s^{(\pm)}}\right) r_*(b)\right]\quad , \quad \frac{T}{X} = \begin{cases}
\tanh\left(\frac{t}{2}\left(\left.\frac{\dd \bar{a}}{\dd r}\right|_{r=r_s^{(\pm)}}\right)\right) & -1<\frac{T}{X}<1 \\
\coth\left(\frac{t}{2}\left(\left.\frac{\dd \bar{a}}{\dd r}\right|_{r=r_s^{(\pm)}}\right)\right) & -1<\frac{X}{T}<1 \\
\end{cases} \;,
\end{equation}

\noindent
with 

\begin{equation}\label{rstar}
r_*(r)=\int_{r_\mathcal T}^{r}\dd r\,\frac{1}{a(r)}\;.
\end{equation}

\noindent
These implicit equations can be solved for $X(t,r)$ and $T(t,r)$ leading to

\begin{align}
X^{(\pm)}(t,r) = \begin{cases}
\sinh\left(\frac{t}{2} \left.\frac{\dd a}{\dd r}\right|_{r=r_s^{(\pm)}}\right) \sqrt{\exp\left[\left(\left.\frac{\dd a}{\dd r}\right|_{r=r_s^{(\pm)}}\right)r_*(r)\right]} & \text{inside the horizon}\\
\cosh\left(\frac{t}{2} \left.\frac{\dd a}{\dd r}\right|_{r=r_s^{(\pm)}}\right) \sqrt{\exp\left[\left(\left.\frac{\dd a}{\dd r}\right|_{r=r_s^{(\pm)}}\right)r_*(r)\right]} & \text{outside the horizon}\\
\end{cases}
\\
T^{(\pm)}(t,r) = \begin{cases}
\cosh\left(\frac{t}{2} \left.\frac{\dd a}{\dd r}\right|_{r=r_s^{(\pm)}}\right) \sqrt{\exp\left[\left(\left.\frac{\dd a}{\dd r}\right|_{r=r_s^{(\pm)}}\right)r_*(r)\right]} & \text{inside the horizon}\\
\sinh\left(\frac{t}{2} \left.\frac{\dd a}{\dd r}\right|_{r=r_s^{(\pm)}}\right) \sqrt{\exp\left[\left(\left.\frac{\dd a}{\dd r}\right|_{r=r_s^{(\pm)}}\right)r_*(r)\right]} & \text{outside the horizon}\\
\end{cases}
\end{align}

Note that $(X^{(+)},T^{(+)})$ covers the full Kruskal diagram including the transition surface and the negative horizon $r_s^{(-)}$. Following the analysis of \cite{BodendorferEffectiveQuantumExtended}, we find

\begin{itemize}
	
	\item $r=r_\mathcal T$ corresponds to $\left(T^{(+)}\right)^2-\left(X^{(+)}\right)^2=1$.
	
	\item $r=r_s^{(+)}$ corresponds to $\left(T^{(+)}\right)^2-\left(X^{(+)}\right)^2=0$.
	
	\item $r\to +\infty$ corresponds to $\left(T^{(+)}\right)^2-\left(X^{(+)}\right)^2\to-\infty$.
	
	\item $r=r_s^{(-)}$ corresponds to $\left(T^{(+)}\right)^2-\left(X^{(+)}\right)^2\to-\infty$.
	
\end{itemize}

\noindent
Hence, opposed to what was stated in \cite{BodendorferEffectiveQuantumExtended}, the $(X^{(+)},T^{(+)})$ covers the full black and white hole interior. The analogous analysis can be done for the chart $(X^{(-)},T^{(-)})$. The collapsing shell follows the curve $(X^{(\pm)}(t(\sigma),r_o(\sigma)),T^{(\pm)}(t(\sigma),r_o(\sigma)))$, where $t(\sigma)$ is the solution of Eq. \eqref{eq:timefinal} or can be extracted out of $v(\sigma)$, which was computed in Sec. \ref{sec:modelapplication}. We introduce now the following light cone coordinates

\begin{align}
U^{(\pm)} =& T^{(\pm)}-X^{(\pm)} \;, \\
V^{(\pm)} =& T^{(\pm)}+X^{(\pm)} \;,
\end{align}

\noindent
which can be compactified by means of

\begin{align}
\tilde{U}^{(\pm)} =& \arctan\left(U^{(\pm)}\right)\;,\\
\tilde{V}^{(\pm)} =& \arctan\left(V^{(\pm)}\right)\;, \\
\tilde{X}^{(\pm)} =& \tilde{V}^{(\pm)} -\tilde{U}^{(\pm)} \;,\\
\tilde{T}^{(\pm)} =& \tilde{V}^{(\pm)}+\tilde{U}^{(\pm)} \;.
\end{align}

\begin{figure}[t!]
	\centering
	\subfigure[]
	{\includegraphics[width=7.5cm]{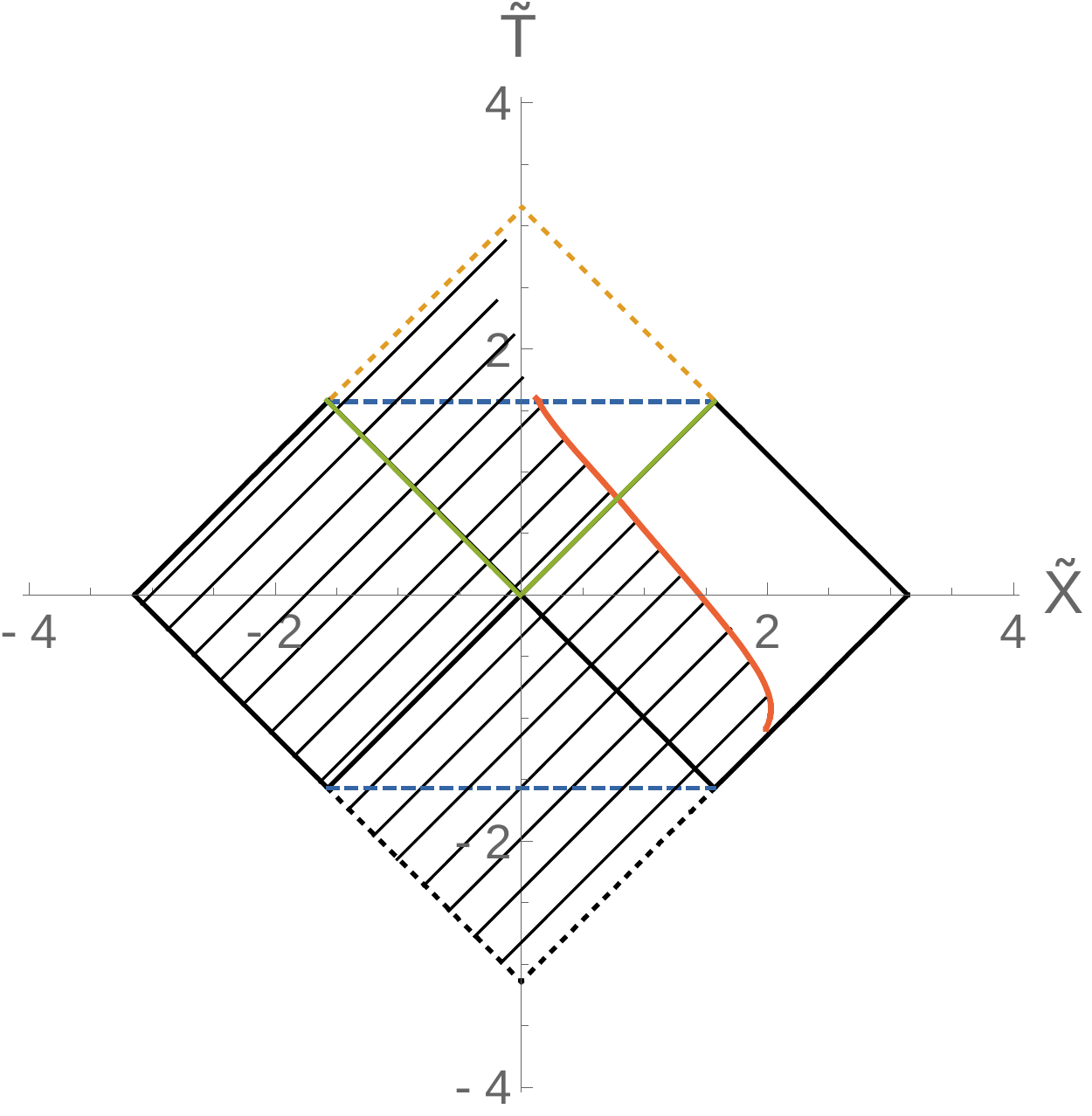}}
	\hspace{2mm}
	\subfigure[]
	{\includegraphics[width=7.5cm]{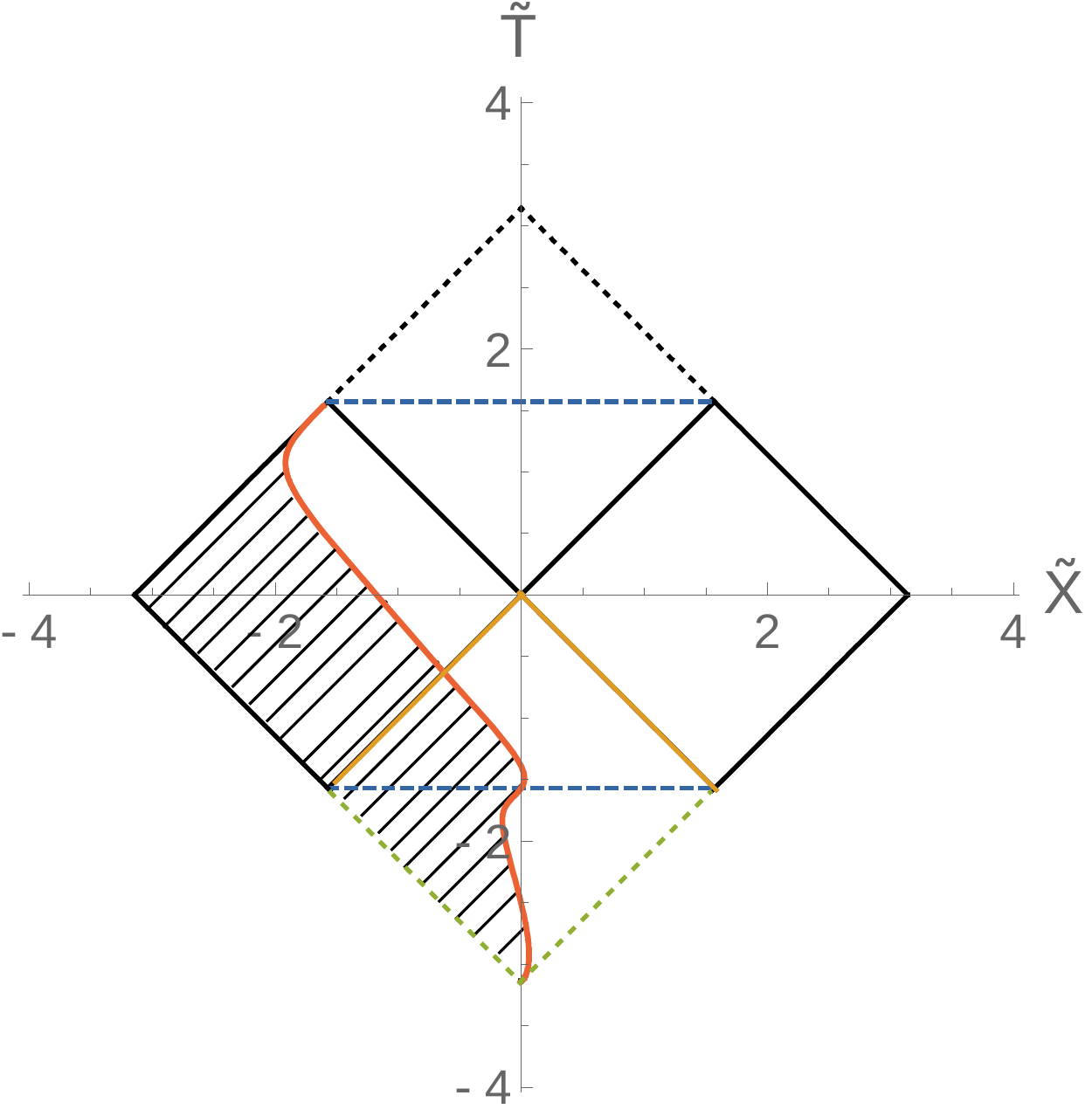}}
	\caption{Penrose diagram of the collapse process from the outside observer point of view. In (a) the black hole side is depicted, while (b) shows the white hole side, respectively. The red line shows the trajectory of the matter surface. The diagram is valid only outside this line, i.e. the non-shaded region. The red line in (a) has to be identified with the red line in Fig.~\ref{fig:PDint} (a) and similar the red line in (b). The transition surface is the blue dashed line. In (a) the red line continues up to the white hole horizon (diagonal dashed yellow lines), which cannot be computed due to numerical uncertainties. The yellow and green horizons of (a) and (b) have to be identified. Parameters are chosen to be $k = 0$, $M_{BH} = 20$, $M_{WH} = 1$, $\lambda_1 = \lambda_2  =\mathscr L_o=1$.}
	\label{fig:PDext}
\end{figure}
\begin{figure}[t!]
	\centering\includegraphics[width=7.5cm]{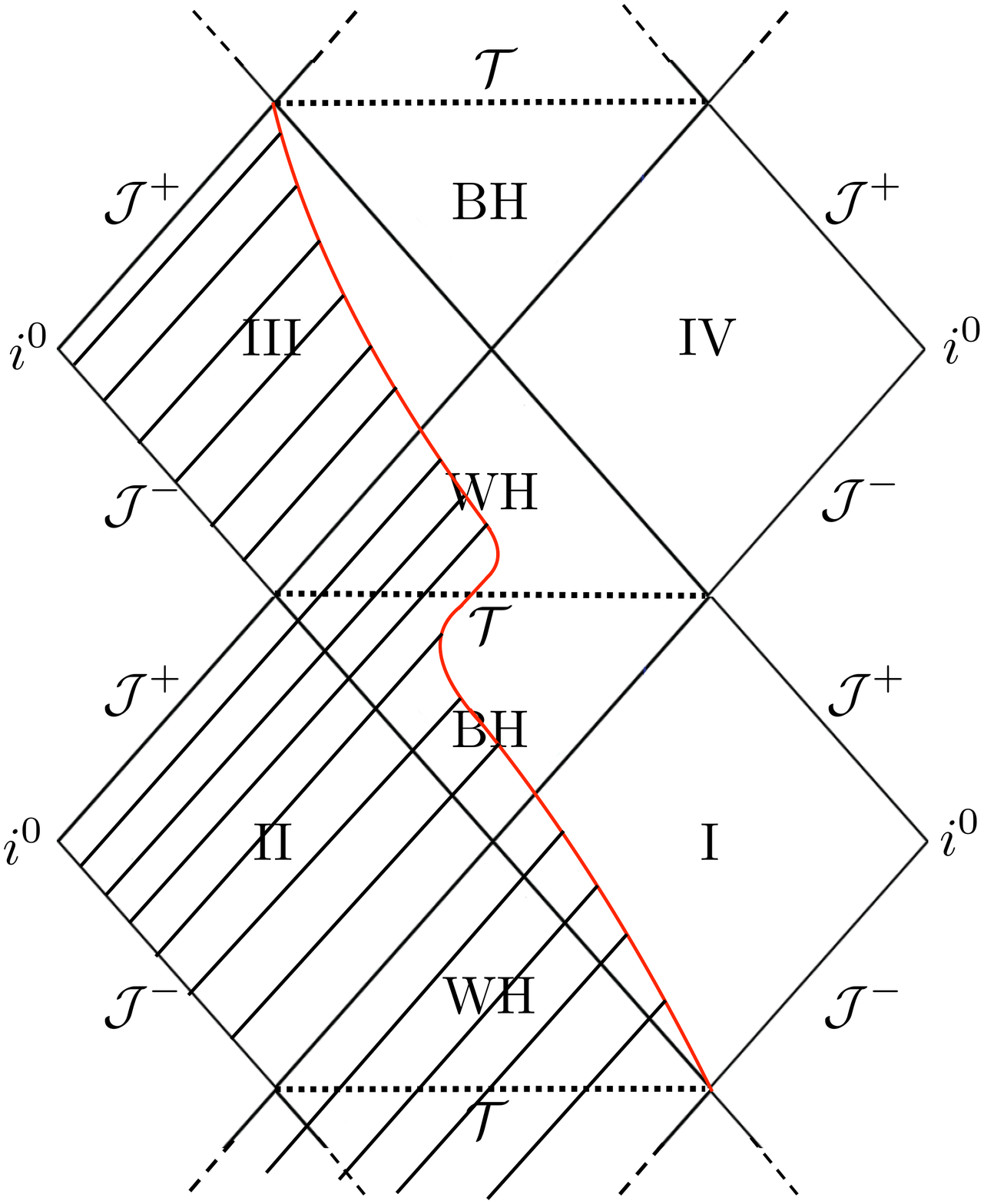}
	\caption{Sketch of the Penrose diagram for the spactime region outside the dust. The spacetime is valid in the non-shaded region and is glued to the interior metric at the red line. As the matter collapses from region \texttt{I} to \texttt{III}, the infinite tower of black and white hole regions still remains.}
	\label{fig:PDextsketch}
\end{figure}

Fig. \ref{fig:PDext} shows $(\tilde{X}^{(\pm)}(t(\sigma),r_o(\sigma)),\tilde{T}^{(\pm)}(t(\sigma),r_o(\sigma)))$ for both patches.
The plot represents the spacetime only in the region outside the red line (non-shaded region).
It is clearly visible what was discussed above: The collapsing dust moves from region $\texttt{I}$ to region $\texttt{III}$. Concluding, an observer could reach region $\texttt{IV}$ without crossing the matter and possibly jump into the black hole again following an infinite tower of black to white hole transitions. As also argued above, this seems to be a generic property of effective polymer black hole models, which are used as exterior spacetime. This behaviour is plausible as a lightlike shell or massive geodesics follow a trajectory from region $\texttt{I}$ to region $\texttt{III}$. The collapsing dust should, in the limit of becoming lightlike, approach the lightlike shell. This limit would be discontinuous, if our result would be a collapse from from region $\texttt{I}$ to region $\texttt{IV}$. Nevertheless, this situation is physically questionable and should be analysed further in future work.
A sketch of the full outside Penrose diagram is shown in Fig.~\ref{fig:PDextsketch}.

Interesting is also the curved behaviour of the collapsing surface (red curve in Figs.~\ref{fig:PDext} and \ref{fig:PDextsketch}).
As argued before this corresponds to the fact that $\dd t/\dd \sigma < 0$ in the vicinity of the transition surface (see Fig.~\ref{fig:dtbydsigma}).
As $t$ is timelike in the black hole exterior this corresponds to a turning point in the spacial direction as $\sigma$-time continues.
Exactly this is visible in Figs.~\ref{fig:PDext} and \ref{fig:PDextsketch}\footnote{Note that it is not nicely visible in Fig.~\ref{fig:PDext} (a). The reason for this is the choice of $r_{\mathcal{ T}}$ as lower integration boundary in Eq.~\eqref{rstar}. This choice ensures that the transition surface is a straight horizontal line in the Penrose diagram. As the transition surface and the black hole horizon are more distant in as the transition surface and the white hole horizon, the $r$-resolution in Fig.~\ref{fig:PDext} (a) is too bad for showing the precise behaviour nicely. A different choice of reference in Eq.~\eqref{rstar} would change this at the cost of the transition surface becoming a curved line.}.
As mentioned, the physical reason is not determined and the analysis is complicated.
The effect maintains for different choices of $M_{BH}$ and $M_{WH}$.
It would be interesting to analyse if this is a generic feature of effective bouncing black hole models.

There is another property directly visible from the Penrose diagram.
Imagine two friends living in region \texttt{I}. One of them jumps into the black hole, the other one remains there.
An important question is: After which time do they meet again \cite{KieferSingularityavoidancefor,HaggardBlackHoleFireworks}?
The Penrose diagram shows immediately that the answer is ``never''.
The exterior spacetime is not modified, i.e. the black hole horizon is still a causal horizon.
Hence, once something or someone going beyond that point becomes causally disconnected from region \texttt{I}, by the definition of a causal horizon.
The only possibility for the two observers to meet again is that the black hole evaporates, which is not modelled here.
This shows that the time scale of the collapse is irrelevant for the understanding of the black hole life cycle as from an outside point of view Hawking evaporation will be always relevant first.
A merging of region \texttt{I} and \texttt{IV} and thus the possibility for the two friends meeting again can only happen after the black hole evaporation.
In turn, this means that a picture as proposed in \cite{AshtekarBlackHoleEvaporation,HaggardBlackHoleFireworks,BianchiWhiteHolesasRemnants,Martin-DussaudEvaporationgblackto} is only possible if Hawking radiation and black hole evaporation are included.

\section{Comparison to Related Work}\label{sec:comparison}

In an independent recent line of research \cite{BenAchourBouncingcompactobjectsI,BenAchourBouncingcompactobjectsII,BenAchourConsistentblackto}, a similar strategy was applied to study the full dynamical black hole setup by means of surface matching.
In this section we would like to summarise similarities and differences in the setup of  \cite{BenAchourBouncingcompactobjectsI,BenAchourBouncingcompactobjectsII,BenAchourConsistentblackto} and the present paper.

\begin{itemize}
	\item In both approaches, \cite{BenAchourBouncingcompactobjectsI,BenAchourBouncingcompactobjectsII,BenAchourConsistentblackto} and the present paper, the Israel-Darmois juction conditions between two spherically symmetric spacetimes are studied.
	In both approaches these spacetimes are effective quantum models of cosmology and black holes.
	
	\item In the present paper, the vacuum spacetime is assumed to be given. 
	Using the assumptions \ref{assm:1}-\ref{assm:3} the matter spacetime, i.e. $S(\tau)$ and the dynamics of the matching surface ($\rho_o(\tau)$) can be determined.
	The reason why this is possible is the fact that assumption \ref{assm:3} demands the energy $\sigma$ and pressure $p$ of the matching surface to be zero.
	This is here interpreted as the Oppenheimer-Snyder-Datt scenario of pressure-less dust.
	Due to this restriction, no model for the matter region of spacetime is needed\footnote{This is an advantage from the model building point of view, in the sense that no additional assumptions are needed. Nevertheless, it would be very useful to have a concrete matter model to interpret the results more easily.}.
	
	In contrast, in \cite{BenAchourConsistentblackto} the matter and the vacuum region of spacetime are assumed to be given.
	The Israel-Darmois junction conditions then allow to determine the dynamics of the matching surface and its energy $\sigma$ and pressure $p$, which are both assumed to be zero in the current approach.
	Therefore the main difference in the approaches is a modification of assumption \ref{assm:3}.
	This way \cite{BenAchourConsistentblackto} can treat more general collapse models, but the matter spacetime has to be known.
	
	\item In \cite{BenAchourConsistentblackto} a kinematic constraint for the case of generic bouncing black hole models (see Eqs.~(2.12) and (2.13) in \cite{BenAchourConsistentblackto}) were derived.
	They can be summarised as 
	\begin{equation}\label{eq:benachourkinematic}
	R_{\text{bounce}} \ge 2 M_{MS}(T_{\text{bounce}}, R_{\text{bounce}}) \quad \Longrightarrow \quad \theta_+ \theta_- \le 0 \;.
	\end{equation}

	This is consistent with the results found here.
	Applied to generic polymer black hole models (see Sec.~\ref{sec:application}), we found that the bounce, i.e. $\der{R}{\sigma} = 0$, happens at the transition surface of the vacuum spacetime.
	It can be generically be shown, that at the transition surface for any metric of the type \eqref{eq:polymermetricgeneric}, the product of the expansions vanishes there (cfr. Eq.~\eqref{eq:expansion})
	\begin{equation}
		\theta_+\theta_- \propto b'(r_{\mathcal{ T}}) = 0 \;.
	\end{equation}
	Further it can be shown that the Misner-Sharp mass $M_{MS}$ is simply $b_{\mathcal{ T}}/2$ at the transition surface.
	Therefore the bounce of the present model satisfies the inequalities of Eq.~\eqref{eq:benachourkinematic}, as the bounce happens at $R(\sigma_{\text{bounce}}) = b_\mathcal{ T}$.
	
	\item Additionally, in \cite{BenAchourConsistentblackto} a dynamical constraint for the energy content of the matching surface was found (cfr. Eqs.~(3.27)-(3.39)).
	The matching of given matter and vacuum spacetimes gives restrictions on $\sigma$ and $p$ of the shell.
	In order to solve these conditions an additional requirement on the spacetime can be derived (see Eq.~(3.29) in \cite{BenAchourConsistentblackto}).
	
	The situation here is similar, although more hidden.
	First of all, independent of the polymer model in the present paper it was found that the matching surface has to approach lightlikeness (cfr. Eq.~\eqref{eq:causalitytau}) at the bounce, which might be interpreted as dynamical constraint.
	Moreover, requiring that the shell is always timelike leads to a constraint for the black hole model parameters
	(see the discussion below Eq.~\eqref{eq:equsr-s:rho} and around Fig.~\ref{fig:PSsigmaSym}).
	As in the present approach, there is no energy and pressure of the matching surface, the dynamical constraint of \cite{BenAchourConsistentblackto} directly translates to a constraint of the model parameter $M_{BH}$ and $M_{WH}$.
	Along the conclusions of \cite{BenAchourConsistentblackto}, this is model dependent and there might or might not be model parameters satisfying this timelikeness condition.
	A clear analysis of this condition was not feasible within the present approach, but it is interpreted as the equivalent of the dynamical constraint of \cite{BenAchourConsistentblackto}.
	
\end{itemize}

It would be interesting to further understand the relation between both approaches to obtain a clearer picture of the dynamical black hole collapse.

\section{Discussion and Outlook}\label{sec:conclusions}

In this paper, a strategy was presented to generalise models of eternal quantum black holes to a dynamical collapse situation of pressure-less dust.
The idea is to match the part of spacetime, which is filled with matter, to the vacuum region, which can be described by a quantum black hole model.
The approach is related to recent work \cite{BenAchourBouncingcompactobjectsI,BenAchourBouncingcompactobjectsII,BenAchourConsistentblackto}.
There are three main assumptions: First, the exterior spacetime is known and spherically symmetric and static eternal model.
This means it is static outside of the black hole horizon and homogeneous in the interior.
This is equivalent to assume that Birkhoff theorem holds even going beyond classical GR, which has to be justified and it is not clear if this is the case.
Second, it is assumed that the matter part of spacetime is homogeneous, i.e. the matter distribution is itself homogeneous, which is the simplest assumable scenario.
The last assumption is, that the spacetime is at least once differentiable at the matching surface between matter and vacuum region.
As argued in Sec.~\ref{sec:classical} and as a consequence of the Israel-Darmois junction conditions \cite{IsraelSingularhypersurfacesand,DarmoisLeseuqationsde} this corresponds to the scenario of collapsing homogeneous dust, i.e. the analogue situation of an Oppenheimer-Snyder-Datt collapse \cite{OppenheimerOnContinuedGravitational,DattOnaclass}.
A surprising observation is that these three assumptions and the following junction conditions fix uniquely the dynamics of the collapsing surface and also of the matter spacetime as long as the vacuum solution is given.
Due to this, the presented strategy allows in principle to generalise any eternal black hole model to a homogeneous dust collapse model.
Moreover, it gives access to a global picture of the collapsing process, which was not possible in previous attempts as in e.g. \cite{BambiNon-singularquantum,BojowaldBlackHoleMass}.

In this paper special focus has been put on effective polymer models as symmetry reduced LQG models \cite{VakiliClassicalpolymerizationof,CorichiLoopquantizationof,ModestoSemiclassicalLoopQuantum,BoehmerLoopquantumdynamics,BenAchourPolymerSchwarzschildBlack,AshtekarQuantumExtensionof,AshtekarQuantumTransfigurarationof,BodendorferEffectiveQuantumExtended,BodendorferMassandHorizon,Bodendorferbvtypevariables,AssanioussiPerspectivesonthe,KellyEffectiveloopquantumgravity,KellyBlackholecollapse,GambiniSphericallysymmetricloop,GeillerSymmetriesofthe,SartiniQuantumdynamicsof}.
Without specifying a concrete model, it was already possible to deduce the qualitative properties of the collapsing surface.
As these models predict a minimal possible radius, the collapse of the matter has to stop at the transition surface and is generically converted into an expansion at the transition surface.
The junction conditions demand that the matter surface itself has to become asymptotically lightlike at this turning point.
This is a surprising result, which should be understood better in future research.
At the current stage, it is not clear if this is physically plausible or the assumptions are too simple to cover the valid physics at the transition surface.
Possibly a full quantum treatment, i.e. leaving the effective approximation would resolve this issue.
As was checked all curvature scalar for the matter and vacuum region remain finite throughout the process and singularities are avoided.
Furthermore, it was shown that the collapsing surface travels diagonally across the Penrose diagram Fig.~\ref{Penrosediag2} from region \texttt{I} to \texttt{III}.
Therefore, the infinite tower of Penrose diagrams remains and the spacetime contains infinitely many black and white holes.
This makes sense from a model building point of view as this is the same behaviour of lightlike and massive geodesics.
Nevertheless, physically this persisting tower of black hole spacetimes seems to be not plausible and it is not clear how contact to other models as \cite{AshtekarBlackHoleEvaporation,HaggardBlackHoleFireworks,BianchiWhiteHolesasRemnants,Martin-DussaudEvaporationgblackto} can be made.
Eventually, black hole evaporation, which was systematically neglected in the present approach, is a crucial ingredient to recover a finite Penrose diagram. 
This property should be studied in future research.
It is remarkable that this is a result of the general analysis and all bouncing polymer models lead to the same qualitative infinite Penrose diagram and a lightlike shell at the transition surface.

As the last part of the paper, the matching strategy was applied to the specific polymer black hole model \cite{BodendorferEffectiveQuantumExtended}.
In the case of $k=0$, i.e. an asymptotically free collapse, the junction conditions were numerically solved and the spacetime globally determined.
An interesting feature of the model \cite{BodendorferEffectiveQuantumExtended} is the fact that there exist two independent mass observables for the black and white hole respectively.
In this paper, it was shown that the junction conditions, together with the demand of a timelike collapse give a constraint on the relation of these to masses.
It turns out that the mass relations found in \cite{BodendorferEffectiveQuantumExtended} are not compatible with the constraint coming from matching the two different spacetimes.
Nevertheless, a detailed analysis of this constraint within these junction conditions is involved and challenging in the current formulation and therefore left for future research.

Having a concrete model at hand it was possible to construct the full spacetime picture and the corresponding Penrose diagrams for the matter and vacuum regions.
Still puzzling is the causal structure of the matter region as it can only be represented in terms of two different diagrams.
Reason for this is that the collapsing surface becomes lightlike at the transition surface and only observers or particles, which accelerate to the speed of light can pass the transition surface.
Furthermore, all lightlike geodesics meet at the transition surface at the same time, i.e. form a caustic.
The physical consequences are not clear and the model is too restricted to gain concrete insights.
A clear matter model, where the stress-energy tensor is given, would possibly lead to further insights.
Finally, it was also possible to explicitly compute the Penrose diagram from the outside point of view.
This allows to clearly pose and answers questions about time scales of different observers.
As a consequence of the model, two observers from which one jumps into the black hole and the other rests outside will never meet each other again.
It becomes clear from the Penrose diagram, that they can only see each other when black hole evaporation is taken into account (cfr. \cite{AshtekarBlackHoleEvaporation}).
This supplements the analysis of other pure quantum collapse models as e.g. \cite{KieferSingularityavoidancefor,SchmitzTowardsaquantum,PiechockiQuantumOppenheimerSnyder}.
The collapse time scale from a radius outside the black hole horizon to the radius outside the white hole horizon can be roughly estimated to be proportional to $M_{BH} + M_{WH}$, where $M_{WH}$ is strongly constrained by the assumption of a timelike matching surface.
In the analysed case here it is $M_{WH} \sim M_{\text{Planck}} \ll M_{BH}$.

The presented strategy is generic and can be applied to all kinds of black hole models, as long as the vacuum metric is fully known.
It would be interesting to analyse in future work how non-bouncing models as \cite{NicoliniQuantumCorrectedBlackHoles,EassonTheclassicaldouble,AdeifeobaTowardsconditionsfor,PlataniaDynamicalrenormalizationof,MotiOnthequantum,NicoliniNoncommutativeBlackHoles,NicoliniRemarksonregular,SmailagicKerrrblackhole,BardeenNon-singulargeneral,HaywardFormationandevaporation,DymnikovaVacuumnonsingularblackhole,DymnikovaDeSitter-Schwarzschild,FrolovNotesonnon-singular,FrolovRemarksonnon-singularblackholes} behave in this context.
It would be important to understand the structure of the matter region better, the property that the collapsing surface approaches lightlikeness at the transition surface and the causal structure.
Therefore, application to other models would supplement the present work.
Further, it would be interesting to find a concrete matter model where details of the stress-energy tensor are known, which are compatible with the matter region dynamics.
This would allow to understand and interpret better the results found here.

To understand the physics of black holes completely, these models have to be generalised and Hawking radiation and black hole evaporation have to be taken into account.
This way comparisons to other models as e.g. \cite{HaggardBlackHoleFireworks,BianchiWhiteHolesasRemnants,Martin-DussaudEvaporationgblackto,AshtekarBlackHoleEvaporation} could be made and consequences for the information loss paradox be worked out.
This is not covered in the present work and should be faced in future research.

Further limitations come from the fact that this strategy relies on effective spacetimes rather than quantum ones.
The physics close to the transition surface certainly lies in the quantum regime and it should be thought about possibilities to extend the presented strategy to quantum spacetimes.
As mentioned this might give more insight about the behaviour of the shell at the transition surface and the condition that it should be asymptotically lightlike there.

In future work, it might also be studied if the strategy could be turned around.
Instead of assuming a metric for the vacuum region and determining the dynamics of the matter region, a cosmological spacetime as given by effective LQC \cite{AshtekarQuantumNatureOf,AshtekarMathematicalStructureOf,AshtekarLoopQuantumGravity30Years,DaporCosmologicalEffectiveHamiltonian} could be used to determine the exterior black hole spacetime, which would be similar to \cite{BenAchourBouncingcompactobjectsI,BenAchourBouncingcompactobjectsII,BenAchourConsistentblackto}.
To which extend this is possible and the results reliable is part of future research.

\section*{Acknowledgements}

The author would like to thank Norbert Bodendorfer for fruitful discussions and support during this project.
The author further thanks Jibril Ben Achour for useful feedback and discussions.
This publication was made possible through the support of the ID\# 61466 grant from the John Templeton Foundation, as part of the “The Quantum Information Structure of Spacetime (QISS)” Project (qiss.fr). The opinions expressed in this publication are those of the author(s) and do not necessarily reflect the views of the John Templeton Foundation.
The author was in the initial phase of the project supported by an International Junior Research Group grant of the Elite Network of Bavaria at the University of Regensburg.

%\bibliographystyle{utphysmendeley}
%\bibliography{/home/work/library/library}

\end{document}